\documentstyle[12pt]{article}
\catcode`\@=11

\@addtoreset{equation}{section}
\def\theequation{\arabic{section}.\arabic{equation}}

\catcode`\@=11

\def\section{\@startsection{section}{1}{\z@}{3.5ex plus 1ex minus
   .2ex}{2.3ex plus .2ex}{\large\bf}}

\def\eqnarray{\stepcounter{equation}\let\@currentlabel=\theequation
    \global\@eqnswtrue
    \global\@eqcnt\z@\tabskip\@centering\let\\=\@eqncr
    $$\halign to \displaywidth\bgroup\@eqnsel\hskip\@centering
      $\displaystyle\tabskip\z@{##}$&\global\@eqcnt\@ne
       \hfil${{}##{}}$\hfil
      &\global\@eqcnt\tw@ $\displaystyle\tabskip\z@{##}$\hfil
       \tabskip\@centering&\llap{##}\tabskip\z@\cr}
\def\lefteqn#1{\hbox to 4\arraycolsep{$\displaystyle #1$\hss}}
\def\thesection{\arabic{section}}

\def\appendix{\setcounter{section}{0}
        \def\thesection{Appendix.}
        \def\theequation{\Alph{section}.\arabic{equation}}}
\long\def\@makefntext#1{\parindent 0cm\noindent
\hbox to 1em{\hss$^{\@thefnmark}$}#1}
\def\IR{{\hbox{{\rm I}\kern-.2em\hbox{\rm R}}}}
\def\IH{{\hbox{{\rm I}\kern-.2em\hbox{\rm H}}}}
\def\IC{{\ \hbox{{\rm I}\kern-.6em\hbox{\bf C}}}}
\def\IZ{{\hbox{{\rm Z}\kern-.4em\hbox{\rm Z}}}}

\newcommand{\beq}{\begin{equation}}
\newcommand{\eeq}{\end{equation}}
\newcommand{\NPB}[1]{{\sl Nucl.~Phys.}~{\bf B#1}}

\newcommand{\CMP}[1]{{\sl Commun.~Math.~Phys.}~{\bf #1}}

\newcommand{\PRL}[1]{{\sl Phys.~Rev.~Lett.}~{\bf #1}}

\newcommand{\CQG}[1]{{\sl Class.~Quant.~Grav.}~{\bf #1}}
\newcommand{\PRD}[1]{{\sl Phys.~Rev.}~{\bf D#1}}

\newcommand{\JMP}[1]{{\sl J.~Math.~Phys.}~{\bf #1}}
\begin{document}
%
%
%
%
\def\citen#1{%
\edef\@tempa{\@ignspaftercomma,#1, \@end, }
\edef\@tempa{\expandafter\@ignendcommas\@tempa\@end}%
\if@filesw \immediate \write \@auxout {\string \citation {\@tempa}}\fi
\@tempcntb\m@ne \let\@h@ld\relax \let\@citea\@empty
\@for \@citeb:=\@tempa\do {\@cmpresscites}%
\@h@ld}
%
\def\@ignspaftercomma#1, {\ifx\@end#1\@empty\else
   #1,\expandafter\@ignspaftercomma\fi}
\def\@ignendcommas,#1,\@end{#1}
%
%
\def\@cmpresscites{%
 \expandafter\let \expandafter\@B@citeB \csname b@\@citeb \endcsname
 \ifx\@B@citeB\relax 
    \@h@ld\@citea\@tempcntb\m@ne{\bf ?}%
    \@warning {Citation `\@citeb ' on page \thepage \space undefined}%
 \else
    \@tempcnta\@tempcntb \advance\@tempcnta\@ne
    \setbox\z@\hbox\bgroup 
    \ifnum\z@<0\@B@citeB \relax
       \egroup \@tempcntb\@B@citeB \relax
       \else \egroup \@tempcntb\m@ne \fi
    \ifnum\@tempcnta=\@tempcntb 
       \ifx\@h@ld\relax 
          \edef \@h@ld{\@citea\@B@citeB}%
       \else 
          \edef\@h@ld{\hbox{--}\penalty\@highpenalty \@B@citeB}%
       \fi
    \else   
       \@h@ld \@citea \@B@citeB \let\@h@ld\relax
 \fi\fi%
 \let\@citea\@citepunct
}
%
\def\@citepunct{,\penalty\@highpenalty\hskip.13em plus.1em minus.1em}%
%
%
\def\@citex[#1]#2{\@cite{\citen{#2}}{#1}}%
%
%
\def\@cite#1#2{\leavevmode\unskip
  \ifnum\lastpenalty=\z@ \penalty\@highpenalty \fi 
  \ [{\multiply\@highpenalty 3 #1
      \if@tempswa,\penalty\@highpenalty\ #2\fi 
    }]\spacefactor\@m}
\let\nocitecount\relax  
%

\begin{titlepage}
\vspace*{.5in}
\begin{center}
{\Large\bf
Angular momentum and an invariant quasilocal energy\\[1ex]\vspace{.0in}
in general relativity}\\
\vspace{.3in}
R{\sc ichard}~J.~E{\sc pp}\footnote{\it E-mail address:
epp@rri.ernet.in}\\
       {\small\it Raman Research Institute}\\
       {\small\it Bangalore 560 080}\\{\small\it India}\\
\end{center}

\vspace{.6in}
\begin{center}
\begin{minipage}{6in}
\begin{center}
{\large\bf Abstract}
\end{center}
{\small
A key feature of the Brown and York definition of quasilocal energy is
that under local boosts of the fleet of observers measuring the
energy, the quasilocal energy surface density transforms as one would
expect based on the equivalence principle, namely, like $E$ in the
special relativity formula: $E^2-\vec{p}^{\;2}=m^2$.  In this paper I will
provide physical motivation for the general relativistic analogue of this
formula, and thereby arrive at a geometrically natural definition of an
`invariant quasilocal energy', or IQE.  In analogy with the invariant mass
$m$, the IQE is invariant under local boosts of the fleet of observers on
a given two-surface $S$ in spacetime.  A reference energy subtraction
procedure is required, but in contrast to the Brown-York procedure, $S$ is
isometrically embedded in a {\it four}-dimensional reference spacetime of
one's choosing.  For example, it is well known that {\it any} sphere,
round or not, can always be isometrically embedded into Minkowski space,
even if its scalar curvature is not everywhere positive.  So rather than
embeddability being a concern, the problem now is that such embeddings are
not unique, leading to an ambiguity in the reference IQE.  However, in
this codimension-two setting there are two curvatures associated with
$S$: the curvature of its tangent bundle, and the curvature of its normal
bundle.  Taking advantage of this fact I will suggest a possible way to
resolve the embedding ambiguity, which at the same time will be seen to
incorporate angular momentum into the energy at the quasilocal level.  I
will analyze the IQE in the following cases: both the spatial and future
null infinity limits of a large sphere in asymptotically flat spacetimes; 
a small sphere shrinking to a point along either spatial or null directions; 
and finally, in asymptotically anti-de Sitter spacetimes.  The last case
reveals a striking similarity between the reference IQE and a certain
counterterm energy recently proposed in the context of the conjectured
AdS/CFT correspondence.
}
\end{minipage}
\end{center}
\end{titlepage}
\addtocounter{footnote}{-1}

\section{Introduction}
\noindent

It is generally agreed that gravitational energy exists, but because of
the equivalence principle it cannot be localized.  The notion of
quasilocal energy is currently one of the most promising descriptions of
energy in the context of general relativity, and can be characterized
simply as follows.  The total energy, including both matter and
gravitational contributions, contained in a finite spatial volume $\Sigma$
can be defined only as the integral of some energy surface density over
its two-surface boundary, $S=\partial\Sigma$.  This implies that, strictly
speaking, there is no such thing as a local energy volume density, except
that which arises from the small $S$ limit of quasilocal energy.\footnote{A 
notable exception is the Tolman density, which integrates to the Komar
mass~\cite{Wald}.  But it can be defined only when the spacetime possesses
special properties, namely a timelike Killing vector field and an
asymptotically flat spatial infinity, and so tells us little about
the nature of energy in a general context.}  And even this local notion is
not truly local because it cannot be integrated over a finite volume
unless one is willing to ignore effects due to gravity.  In short, energy
is associated with closed spacelike two-surfaces in spacetime, not points.

There is also a growing consensus that the ADM and Bondi-Sachs masses are
simply not enough.  We need some definition of energy that is `more
local' than these, i.e., a quasilocal definition that does not rely on the
existence of an asymptotically flat region \cite{BCM}.  For example,
recent proponents of this movement are Ashtekar {\it et al} \cite{ACK,ABF}, 
who have introduced the quasilocal idea of an `isolated horizon' to
describe a black hole.  They articulate several reasons for this need, and
it is useful to paraphrase here at least part of their argument:  Let us
accept that a black hole is a thermodynamic object, and so obeys the first
law: $\delta E=T\,\delta S+\cdots$.  Now suppose that the universe is
asymptotically flat in spatial directions, and contains a single black
hole.  Then $E$ in the first law is the ADM mass.  But if there is
anything else in the universe then $E$ is not the ADM mass, and the
question arises, What expression is to be used for $E$ in the first
law?  In other words, we expect that we can put something else in the
universe, say a galaxy somewhere, such that the black hole we started
with, considered `by itself', will still behave as more or less the same
thermodynamic object, with the same mass, radiating at the same
temperature as before, and with the same entropy equal to one quarter its
area.  This expectation requires the ability to compute the energy of a
given system contained within a finite closed surface, rather than merely
the total energy of all such systems comprising the whole universe.

Thus quasilocal energy lies between the notions of local energy density 
and total energy of an isolated system, in the sense that it is expected
to give the energy contained in any volume, no matter how small or
large.  Although the equivalence principle precludes the existence of
a local gravitational energy density, it does not prevent us from
evaluating the (quasilocal) gravitational energy in an arbitrarily small
but nonvanishing volume $\Sigma$.  This is because no matter how small
$\Sigma$ is, $S=\partial\Sigma$ is not a point, but rather the boundary of
some neighborhood of a point, and so we are always inherently making a
`tidal force measurement'.  In this sense quasilocal energy is distinct
from attempts to define a local gravitational energy density based on
certain symmetries of the action, and the concept of a N\"{o}ther
charge.\footnote{A recent discussion of the connection between
pseudotensor methods and the quasilocal idea can be found in
Ref.~\cite{CNC}.}  At the other extreme is the Komar mass (or the closely
related ADM mass), i.e., the total energy associated with the time
translation symmetry of an isolated system.  As emphasized in
Ref.~\cite{BD}, this gravitational conserved charge is intimately
connected with a lapse function, whereas quasilocal energy need not make
any reference to a lapse function.  The point is, the two are conceptually
distinct~\cite{BCM,BD}, even though in some circumstances one might expect
their numerical values to coincide.

Currently there are several contenders for a good definition of
quasilocal energy (see Refs.~\cite{BY,Hayw,ChenNester} and the references
therein).  The two that interest us at the moment are
Brown and York's `canonical quasilocal energy' (or CQE)~\cite{BY}, and the
various definitions based on the integral over $S$ of the Witten-Nester
two-form (the two-form used in Witten's proof of the positive energy 
theorem~\cite{Nest,Witt}).  The latter approach uses spinorial 
methods, and the different definitions are distinguished by the choice of
supplementary equation the $S$-spinors are supposed to satisfy, for
example the Sen-Witten equation~\cite{Witt,Sen}, the Dougan-Mason
equation~\cite{DM}, or the Ludvigsen-Vickers equation~\cite{LV}.  The
Brown-York definition of quasilocal energy has the form\footnote{We use a 
sign convention for extrinsic curvatures opposite to that of Brown and
York, hence the negative sign in front of this integral.} 
\beq
{\rm CQE}=-{1\over{8\pi}}\,\int_S\,dS\,k - {\rm CQE}^{\rm ref}\;,
\label{BY_QE}
\eeq
in geometrized units, with $G=c=1$.  The CQE is supposed to be the energy
of the gravitational and matter fields contained in a finite spatial 
volume $\Sigma$, whose boundary two-surface is $S=\partial\Sigma$.  $dS$
is the induced integration measure on $S$, and $k$ is the trace of the
extrinsic curvature of $S$ as embedded in $\Sigma$.  Thus, $-k/(8\pi)$
is the Brown-York quasilocal energy surface density.  When $\Sigma$ is
asymptotically flat the integral in Eq.~(\ref{BY_QE}) (the `unreferenced'
CQE) diverges as $S$ is taken to infinity, and a reference term, denoted
${\rm CQE}^{\rm ref}$, is required to regulate the energy.  Brown and
York's prescription is to choose
\begin{equation}
{\rm CQE}^{\rm ref}=-{1\over{8\pi}}\,\int_S\,dS\,k^{\rm ref},
\label{BY_REF}
\end{equation}
where $k^{\rm ref}$ is the trace of the extrinsic curvature of an
isometric embedding of $S$ into some reference space, usually taken to be 
flat $\IR^3$.  With this choice the resulting CQE reduces to the ADM mass
when $S$ is taken to infinity~\cite{BY}.  While the CQE has a host of
desirable properties, neatly summarized in Ref.~\cite{Lau1}, the embedding
prescription needed to evaluate ${\rm CQE}^{\rm ref}$ is not entirely
satisfactory, because not all two-surfaces that arise in practice can be
embedded into flat $\IR^3$.  A ready example is the horizon of the Kerr
black hole, which fails to be embeddable in flat $\IR^3$ when
the angular momentum exceeds the irreducible mass (but is not yet an
extremal black hole), and the two-sphere develops regions with negative
scalar curvature~\cite{Smar}.  While this is but one example, it is
noteworthy that the breakdown of embeddability is in this case associated
with angular momentum and negative scalar curvature.  It is precisely such
issues: embeddability, angular momentum, and negative scalar curvature,
that will figure prominently in this paper, and will be seen to be
subtly intertwined.

Although a relationship between the Brown-York quasilocal energy and
the spinorial definitions based on the Witten-Nester integral is not
immediately obvious, Lau~\cite{Lau1} has shown that spinors may always be
chosen so that the resulting spinorial definition is equal to the
unreferenced Brown-York quasilocal energy in Eq.~(\ref{BY_QE}).  Moreover,
he shows that the role of the Sen-Witten equation is to provide a
definite reference point for the energy, which is not in general the same
as ${\rm CQE}^{\rm ref}$ in Eq.~(\ref{BY_REF}).  The point being made here
is two-fold: (i) the unreferenced Brown-York quasilocal energy seems to be
robust, and (ii) all of the problems lie in choosing a suitable reference
energy.  The various prescriptions are either not generally well defined,
or they do not agree with each other.

I will now present a brief review of the Brown-York approach in a form
that will be useful to us later.  The classical energy-momentum tensor of
matter is a local concept, associated with a spacetime point.  It is
defined for any field theory residing on a nondynamical background
spacetime $(M,g)$ via the functional derivative of the (first order) matter 
action with respect to the metric, as follows:
\begin{equation}
2\,\delta_{g}I^{\rm mat}[\varphi,g]=\,\int_{M}\,d^{4}x\,\sqrt{-g}\,T_{\rm
mat}^{ab}\,\delta g_{ab}.
\label{matter_action}
\end{equation}
Here $\varphi$ denotes the matter field(s) in question, and the factor of 
two on the left is a convention.  Usually, as we will assume here, there
is no boundary term arising from this variation (for `minimally coupled
matter'), but in case there is it does not change the essence of the
following argument, it just adds an interesting dimension to it.
$T_{\rm mat}^{ab}$ so defined is covariantly conserved, as follows from
the matter Euler-Lagrange equations.  This prescription for learning about
matter energy-momentum gives reasonable answers for all field theories,
and so it is natural to try the same thing for gravity.  In this case one
finds, for the usual first order action~\cite{York},
\begin{equation}
2\,\delta_{g}I^{\rm
grav}[g]=\int_{M}\,d^{4}x\,\sqrt{-g}\,\left(-{1\over{8\pi}}G^{ab}\right)
\,\delta g_{ab}+\int_{\cal B}\,d^{3}x\,\sqrt{-\gamma}\,
\left(-{1\over{8\pi}}\Pi^{ab}\right)\,\delta\gamma_{ab}.
\label{gravity_action}
\end{equation}
Inspecting the bulk term one is thus tempted to define $T_{\rm
grav}^{ab}:=-G^{ab}/(8\pi)$ as the local energy-momentum tensor of the
gravitational field, where $G^{ab}$ is the Einstein tensor.  And this is
perfectly reasonable: it is covariantly conserved---in this case
identically so, via the contracted Bianchi identity.  Moreover, its
on-shell value is zero, in full accord with the equivalence 
principle, i.e., there is no nontrivial local energy-momentum tensor for
the gravitational field.

In fact this absence of a nontrivial local energy-momentum tensor is true
not only for the gravitational field, but also for any system comprised
of both matter {\it and} gravity.  To see this we need only make the
spacetime metric dynamical, in which case the matter action in
Eq.~(\ref{matter_action}) must be augmented by the gravity action.  Adding
Eqs.~(\ref{matter_action}) and (\ref{gravity_action}) one finds for the
total action
\begin{equation}
2\,\delta_{g}I^{\rm
tot}[\varphi,g]=\int_{M}\,d^{4}x\,\sqrt{-g}\,\left(T_{\rm 
mat}^{ab}-{1\over{8\pi}}G^{ab}\right)
\,\delta g_{ab}+\int_{\cal B}\,d^{3}x\,\sqrt{-\gamma}\,
\left(-{1\over{8\pi}}\Pi^{ab}\right)\,\delta\gamma_{ab}.
\label{total_action}
\end{equation}
Thus one is led to identify $T_{\rm tot}^{ab} := T_{\rm mat}^{ab}-
G^{ab}/(8\pi)$ as the total local energy-momentum tensor for matter plus
gravity.  It has the desirable property of being covariantly conserved,
but turns out to be just zero by the Einstein equations.  If this argument
is taken seriously we learn that, as soon as we add gravity to any matter
system, the notion of a nontrivial local energy-momentum tensor disappears.
Furthermore, one might interpret the Einstein equations, written in the
form $T_{\rm mat}^{ab}+T_{\rm grav}^{ab}=0$, as a `micro-balancing' of
local stress-energy-momentum at each spacetime point: wherever a component
of matter stress-energy-momentum is positive, the corresponding component
of gravitational stress-energy-momentum is negative, and vice versa, such
that the total is always zero.  The idea that $-G^{ab}/(8\pi)$ is the
local energy-momentum tensor of gravity is, of course, a very old idea,
first put forward by Lorentz and Levi-Civita.  It was rejected by Einstein, 
since it implies that the total energy of a closed system would always be 
zero, which is obviously problematic.\footnote{See the historical
discussion given on pages 176-7 in Ref.~\cite{Paul}.  I thank L. de
Menezes for bringing this reference to my attention.}  It is only with
hindsight that we now realize why the problem was not resolved much
sooner.  People then did not think about boundary terms as much as they do
today.  Thanks to Brown and York we now know that what comes to the rescue
is the boundary term in Eq.~(\ref{total_action}).

In this equation the spacetime is assumed to be the topological product of
a three-space $\Sigma$ and a real line interval.  The boundary component
$\cal B$ is a timelike `tube', topologically the product of $S=\partial\Sigma$ 
and the real line interval.  (The two spacelike `end-cap' boundary components 
of $\partial M$ have been omitted, as they play no role in this discussion.)  
The quantity $-\sqrt{-\gamma}\,\Pi^{ab}/(16\pi)$, constructed in the usual
way out of the extrinsic curvature of $\cal B$, is the gravitational
momentum conjugate to the three-metric $\gamma_{ab}$ induced on $\cal
B$.  Now, in the spirit of identifying the energy-momentum tensor as the
functional derivative of the action with respect to the metric, one reads
off from Eq.~(\ref{total_action}) the energy-momentum tensor $T_{\cal
B}^{ab} := -\Pi^{ab}/(8\pi)$, which is inherently associated with the
boundary $\cal B$, rather than the bulk spacetime.  Like any acceptable
energy-momentum tensor, $T_{\cal B}^{ab}$ is covariantly conserved (as a
tensor residing in $\cal B$)---this follows from the analogue of the
diffeomorphism constraint of general relativity for the three-surface
$\cal B$ (rather than $\Sigma$).  Physically, $\cal B$ is to be thought of
as the congruence of world lines of a two-parameter family of observers
with four-velocity $u^a$, hypersurface orthogonal to a one-parameter
foliation of $\cal B$ by spacelike two-surfaces with the topology of
$S$.  Within their three-dimensional spacetime $({\cal B},\gamma)$, the
observers measure a spatial energy density $T_{\cal B}^{ab}u_{a}u_{b}$,
which is precisely $-k/(8\pi)$, and thus one is led to
Eq.~(\ref{BY_QE}).  Finally, observe that one can add to the action any
covariant functional of the boundary three-metric $\gamma_{ab}$ without
affecting the previous argument.  This is the source of the reference
point ambiguity ${\rm CQE}^{\rm ref}$ in Eq.~(\ref{BY_QE}).  This
summarizes the central idea of the Brown-York approach~\cite{BY}.

Now if energy is really quasilocal, and calculated via a surface integral
involving $T_{\cal B}^{ab}$, one comes to the conclusion that {\it a
priori} neither $T_{\rm mat}^{ab}$ nor $T_{\rm grav}^{ab}$ has anything
to do with energy!  While this might be unsettling at first, it is
reassuring to know that a satisfactory notion of local {\it matter} energy
density can be recovered from the small sphere limit of quasilocal
energy.  For example, in Ref.~\cite{BLY} it is shown that, for a certain
choice of reference term ${\rm CQE}^{\rm ref}$, the Brown-York quasilocal
energy contained in an infinitesimal sphere of proper radius $r$ is the
volume of the sphere ($4\pi r^{3}/3$) times the local matter energy
density $T_{\rm mat}^{ab}u_{a}u_{b}$ (evaluated at the center of the
sphere) that would be measured by an observer with four-velocity $u^a$.  
Moreover, this is a well established property of most quasilocal energy
definitions~\cite{BLY,HS,Doug,Berg1,Berg2,Szab1}, so the result is quite
robust.  And at higher order in $r$, gravitational energy begins to
appear, as will be discussed in detail later.  The point is there is no
contradiction between (i) the local energy-momentum tensor $T_{\rm
mat}^{ab}+T_{\rm grav}^{ab}$ being zero, and (ii) there being nonzero
stress-energy-momentum in a finite spatial volume.  This is because
$T_{\rm mat}^{ab}+T_{\rm grav}^{ab}$ is {\it not} a local energy-momentum
tensor---indeed, if we accept the previous argument, there is no such
thing.  There is only $T_{\cal B}^{ab}$, associated with the fact that
energy is fundamentally quasilocal.

The main purpose of this Introduction is to emphasize, firstly, that
energy is fundamentally quasilocal, i.e., associated with closed spacelike
two-surfaces---not points---in spacetime; and secondly, there are strong
reasons to believe that $-k/(8\pi)$ is the correct quasilocal energy surface 
density.  The major unresolved problem is how to choose the right well
defined energy reference term, ${\rm CQE}^{\rm ref}$.  Rather than address
this problem per se, I will begin with $-k/(8\pi)$ as an energy surface
density and construct a new definition of quasilocal energy based on
analogy with the special relativity formula: $E^{2}-\vec{p}^{\;2}=m^{2}$.
The new definition is both physically and geometrically natural, and lies
somewhere between the Brown-York CQE and the Hawking~\cite{Hawk} or
Hayward~\cite{Hayw} definitions.  A reference subtraction procedure is
still required, that involves a reference embedding, but this is a
codimension-two embedding that is not subject to the problem that afflicts
the Brown-York embedding prescription.  Moreover, there is a shift in the
physics: the reference embedding is not associated with determining a
reference energy so much as a `reference angular momentum', so to
speak.  Why angular momentum?  Because angular momentum contributes to
energy, and the new definition can be seen as a precise formulation of
this fact at the quasilocal level.

The paper is organized as follows.  In Sec.~2 we introduce the geometrical
quantities we will use later.  Sec.~3 contains the physical and
geometrical motivations behind the new definition of quasilocal energy (as
well as the definition itself).  The reference subtraction term is
discussed in Sec.~4.  In Sec.~5 both the spatial and future null infinity
limits of the energy are examined; the small sphere limit is
considered in Sec.~6.  Finally, in Sec.~7 we examine the new energy in the
context of asymptotically anti-de Sitter spacetimes.  Since this is a
lengthy paper I have provided a summary of its results at the end, which
also includes some additional discussion.

\section{The geometry of two-dimensional spacelike submanifolds}
\noindent

Let $(M,g)$ be a four-dimensional Lorentzian geometry with signature $+2$,
and $S$ be a closed two-dimensional spacelike submanifold.  Let $u^a$ and
$n^a$ be timelike and spacelike unit normals to $S$ that are orthogonal
to each other: $u^a u_a = -1$, $n^a n_a = 1$, and $u^a n_a = 0$.  We will
assume that $S$ is orientable, and an open neighborhood of $S$ in $M$ is
space and time orientable, so that $u^a$ and $n^a$ are globally well
defined~\cite{Szab2}.  $u^a$ and $n^a$ are fixed up to an arbitrary
local boost transformation:
\begin{eqnarray}
u^{\prime a} & = & u^a \cosh\lambda + n^a \sinh\lambda \nonumber \\
n^{\prime a} & = & u^a \sinh\lambda + n^a \cosh\lambda .
\label{boost}
\end{eqnarray}
The physical picture to keep in mind is that of a finite spatial volume
$\Sigma$, i.e., a three-dimensional spacelike submanifold, whose boundary
is $S$.  Although $S$ need not be connected, nor simply connected, we will
often think of $\Sigma$ as having the topology of a three-ball, and $S$
that of a two-sphere, and thus will sometimes refer to the direction of
$n^a$ (assumed outward directed) as the `radial' direction.  Given such a
three-surface $\Sigma$ spanning $S$ it is natural to choose $u^a$ to be
orthogonal to $\Sigma$ (and future directed), in which case $n^a$ is
tangential to $\Sigma$.  Physically, $u^a$ is the instantaneous four-velocity 
of a two-parameter family of observers
on $S$.  With $u^a$ thus tied to the spanning surface $\Sigma$, a
deformation of $\Sigma$ (preserving $S$) will in general effect a radial
boost, Eqs.~(\ref{boost}).

The remainder of this section is a summary of some standard facts about
the geometry of the submanifold $S$, as can be found, e.g., in
Ref.~\cite{Chen}, except here we follow a notation similar to that used in
Ref.~\cite{Szab2}.  The surface projection operator, ${\cal P}^{a}_{b}$,
is a tensor
defined on $S$ by
\beq
{\cal P}^{a}_{b}:=\delta^{a}_{b} + u^a u_b - n^a n_b .
\eeq
(All raising and lowering of indices $a,b,c,\ldots$ will be effected with
the metric $g_{ab}$, or its inverse, $g^{ab}$.)  A `surface tensor' is
defined as a tensor on $S$ that is left invariant under projection of all
its indices with the surface projection operator.  Obviously one such
tensor is the spatial two-metric
\beq
\sigma_{ab} := {\cal P}^{c}_{a}{\cal P}^{d}_{b}g_{cd} = 
g_{ab} + u_a u_b - n_a n_b
\label{spatialmetric}
\eeq
induced on $S$.  Another is the corresponding volume form on $S$, given by
\beq
\epsilon_{ab} := \epsilon_{abcd} u^c n^d ,
\label{epsilon}
\eeq
where $\epsilon_{abcd}$ is the volume form on $M$.  The symbol $dS$ will
be used in place of $\epsilon_{ab}$ as the integration measure for
$(S,\sigma)$.

If $\nabla_a$ denotes the Levi-Civita connection of $(M,g)$, then ${\cal
D}_{a}$, the Levi-Civita connection induced on $(S,\sigma)$, is defined by
\beq
{\cal D}_a T^{b\ldots}_{c\ldots} = 
{\cal P}_{a}^{d}{\cal P}^{b}_{e}{\cal P}_{c}^{f}\cdots
\nabla_{d}T^{e\ldots}_{f\ldots} ,
\eeq
where $T^{b\ldots}_{c\ldots}$ is any surface tensor.  Then for any two 
surface vector fields $X^a$ and $Y^a$, the Gauss formula reads
\beq
X^a \nabla_a Y^c = X^a {\cal D}_a Y^c + h^c_{\;\;ab} X^a Y^b ,
\label{GaussFormula}
\eeq
where $h^c_{\;\;ab}$ is the second fundamental form.  Its first index is
normal to $S$, i.e., ${\cal P}^{d}_{c} h^{c}_{\;\;ab}=0$, whereas the
remaining two are surface tensor indices, that are symmetric under
interchange (as can be easily seen by interchanging $X$ and $Y$ in
Eq.~(\ref{GaussFormula}) and subtracting the two equations).  Thus the
second fundamental form can be decomposed into components along the two
unit normals:
\beq
h^{c}_{\;\;ab} =  u^c l_{ab} - n^c k_{ab},
\eeq
where the two extrinsic curvatures are (symmetric) surface tensors given by
\begin{eqnarray}
l_{ab} &=& - u_c h^{c}_{\;\;ab} = {\cal P}_{a}^{c}{\cal P}_{b}^{d}
\nabla_{c}u_{d} 
\nonumber\\
k_{ab}  &=& - n_c h^{c}_{\;\;ab} = {\cal P}_{a}^{c}{\cal P}_{b}^{d}
\nabla_{c}n_{d}.
\label{extrinsic_curvatures}
\end{eqnarray}
It is useful to decompose the extrinsic curvatures into trace and
trace-free parts:
\begin{eqnarray}
l_{ab}  &=&  {1\over 2}l\sigma_{ab}+\tilde{l}_{ab} 
\nonumber\\
k_{ab}  &=&  {1\over 2}k\sigma_{ab}+\tilde{k}_{ab},
\label{trace-trace-free}
\end{eqnarray}
where $l=\sigma^{ab}l_{ab}$ and $k=\sigma^{ab}k_{ab}$, and a tilde
appearing over any quantity in this paper will always mean `trace-free
part of'.  The mean curvature vector is then
\beq
H^c := {1\over 2}\sigma^{ab} h^{c}_{\;\;ab} = {1\over 2}(lu^c - kn^c ),
\label{mean_curvature_vector}
\eeq
and $H\cdot H  = (k^2 - l^2 )/4$ is the square of the mean curvature.

Extrinsic curvature is a measure of how a unit normal vector rotates as it
is parallelly propagated tangent to $S$ in the ambient space $(M,g)$.  Two
normal vectors means two extrinsic curvatures.  However, from
Eqs.~(\ref{extrinsic_curvatures}) we see that $l_{ab}$ and $k_{ab}$
measure only the components of this rotation tangent to $S$.  There is
also a normal component, i.e., the component of the rotation of one normal
vector along the other.  Thus a complete characterization of the 
extrinsic geometry of $S$ requires also the surface one-form
\beq
A_a := {\cal P}_{a}^{b}n^{c}\nabla_{b}u_{c} .
\label{A}
\eeq
This is an $SO(1,1)$ connection in the normal bundle of $S$, and its
associated curvature two-form is
\beq
{\cal F}_{ab} := {\cal D}_{a}A_b - {\cal D}_{b}A_a .
\eeq
We will see later that the curvature of the normal bundle of $S$ plays a
key role with regard to angular momentum.

While the second fundamental form (including $H^c$ and $H\cdot H$) and
the curvature of the normal bundle are invariant under local radial
boosts, the extrinsic curvatures and the connection on the normal bundle
are not.  They transform as
\begin{eqnarray}
l^{\prime}_{ab} & = & l_{ab} \cosh\lambda + k_{ab} \sinh\lambda 
\nonumber \\
k^{\prime}_{ab} & = & l_{ab} \sinh\lambda + k_{ab} \cosh\lambda 
\nonumber \\
A^{\prime}_a & = & A_a + {\cal D}_a \lambda .
\label{extrinsic_boost}
\end{eqnarray}
Observe that $A_a$ is different from the other extrinsic curvatures in
that its transformation law, which is a gauge transformation, depends on
the derivative of $\lambda$.

Our sign conventions are such that the Riemann tensor of $(M,g)$ is
defined by $(\nabla_a \nabla_b - \nabla_b \nabla_a )X_c =
R_{abc}^{\;\;\;\;\;\;d}X_d$, 
and similarly that of $(S,\sigma)$ by 
$({\cal D}_a {\cal D}_b - {\cal D}_b {\cal D}_a )X_c =
{\cal R}_{abc}^{\;\;\;\;\;\;d}X_d$ ($X_c$ is a surface one-form in the
latter case).  Appropriate projections of the Riemann tensor of $(M,g)$
yield the Gauss equation:
\beq
{\cal P}_{a}^{e}{\cal P}_{b}^{f}{\cal P}_{c}^{g}{\cal P}_{d}^{h}
R_{efgh} = {\cal R}_{abcd} + (l_{ac}l_{bd}-l_{bc}l_{ad})
- (k_{ac}k_{bd}-k_{bc}k_{ad}),
\label{Gauss}
\eeq
the Codazzi equations:
\begin{eqnarray}
{\cal P}_{a}^{e}{\cal P}_{b}^{f}{\cal P}_{c}^{g}u^{h}R_{efgh} &=&
({\cal D}_{a}l_{bc}-{\cal D}_{b}l_{ac})-(A_{a}k_{bc}-A_{b}k_{ac}),
\nonumber\\
{\cal P}_{a}^{e}{\cal P}_{b}^{f}{\cal P}_{c}^{g}n^{h}R_{efgh} &=&
({\cal D}_{a}k_{bc}-{\cal D}_{b}k_{ac})-(A_{a}l_{bc}-A_{b}l_{ac}),
\label{Codazzi}
\end{eqnarray}
and the Ricci equation:
\beq
{\cal P}_{a}^{e}{\cal P}_{b}^{f}u^{g}n^{h}R_{efgh}=
-{\cal F}_{ab}+(k_{a}^{\;\;c}l_{bc}-l_{a}^{\;\;c}k_{bc}).
\label{Ricci}
\eeq
These are the integrability conditions for the isometric embedding of
$(S,\sigma)$ into $(M,g)$, and so by definition of $S$ are necessarily
satisfied.

\section{The invariant quasilocal energy}

A physical interpretation of the various geometrical quantities
introduced in the previous section can be given as follows.  The
expansion $k$ measures the fractional expansion of the area of a small
element of $S$ when each point in the element is projected a unit distance
radially outward.  It will have a certain positive value if, for
example, $S$ is a round sphere enclosing a volume of flat $\IR^3$.  (For
our present purposes, imagine `flat $\IR^3$' as a $t={\rm constant}$
surface in Minkowski space.)  Now if $S$ is a round sphere of the same
area enclosing some matter, then, according to the Einstein equations, the
matter curves the space inside $S$ in such a way that its volume is
greater than one would infer by measuring just the area of the sphere and
using Euclidean geometry.  Thus the expansion measured at $S$ must be
smaller, i.e., the areas of spherical shells at larger radii will not
increase as rapidly as expected.   So we see that the unreferenced
Brown-York quasilocal energy in Eq.~(\ref{BY_QE}) is greater (less
negative) when $S$ contains matter, than when it does not.  This is an
intuitive reason why $k$ is a measure of the energy inside $S$.  (It also
explains the need to subtract off a reference energy of the form given in
Eq.~(\ref{BY_REF}): $k^{\rm ref}$ is the nonzero value of $k$ when $S$
merely encloses a volume of flat $\IR^{3}$, i.e., no energy.)

$l$ is similar to $k$, except that it measures the expansion of $S$ in
time, i.e., in the direction of the observer's four-velocity $u^a$.
Intuitively, if the observers tend to be moving radially outward then the
area of the two-surface they are on will be expanding, i.e., $l>0$.
Conversely, a radially inward motion corresponds to $l<0$.  Thus $l$
(more precisely, $l/(8\pi)$) can be interpreted as a `radial momentum
surface density'~\cite{Lau2}.  In the case that $(k^{2}-l^{2})$ is
positive, the observers can always make appropriate local radial boosts 
such that $l=0$ at each point of $S$, a situation corresponding to a
`quasilocal rest frame'.  I will comment on this notion more precisely
at the end of this section.

The trace-free quantities $\tilde{k}_{ab}$ and $\tilde{l}_{ab}$ measure
the shear of $S$, and are intimately connected with angular momentum (or 
at least $\tilde{l}_{ab}$ is).  For example, consider a set of locally
nonrotating observers who at coordinate time $t$ are on a constant $r,t$
sphere of the Kerr black hole in Boyer-Lindquist coordinates.  Their
four-velocity is given by
\beq
u^a = {1\over N}\left( {\partial\over{\partial t}}
+ \omega {\partial\over{\partial\phi}}\right)^a ,
\eeq
where $N$ is the lapse function, and $\omega(r,\theta)=
-g_{t\phi}/g_{\phi\phi}$ is an observer's angular velocity as measured
from infinity~\cite{MTW}.  Starting at Eqs.~(\ref{extrinsic_curvatures}) it
is not difficult to show that in this case $l=0$, so here is an example
of observers in a `quasilocal rest frame' as defined above.  Furthermore,
one can show that the nonvanishing components of the shear in the time
direction are given by
\beq
\tilde{l}_{\theta\phi}=\tilde{l}_{\phi\theta}=
{{g_{\phi\phi}}\over{2N}}{{\partial\omega}\over{\partial\theta}} .
\label{frame_dragging}
\eeq
Physically, a nonzero $\omega$ reflects the frame dragging caused by the
rotating black hole.  The fact that the degree of frame dragging depends
on $\theta$ is what makes the observers at different latitudes of the
sphere rotate at different rates relative to `the distant stars', and more
to the point, relative to each other.  This causes a shear effect between
observers at neighboring latitudes, which obviously disappears when
the angular momentum is zero.

Furthermore, let the locally nonrotating observers label themselves with
coordinates $(\theta^{\prime},\phi^{\prime})$, which at $t=0$ coincide
with the Boyer-Lindquist $(\theta,\phi)$ coordinates on $S$.  Then
although the observers always measure the same two-geometry of $S$ as time
$t$ goes on, the components of the two-metric $\sigma_{ab}$ in their 
$(\theta^{\prime},\phi^{\prime})$ coordinates will differ from those in
the $(\theta,\phi)$ coordinates by a $t$- and $\theta$-dependent
diffeomorphism along the $\phi$-direction.  So although one usually
associates shear with a geometrical deformation, for instance a round
sphere evolving into an ellipsoid, one can also have a physically
meaningful shear associated with a continuous parameter family of
isometric surfaces.  This fact plays an important role in understanding
certain embedding equations we will encounter later, and will be discussed
in detail elsewhere~\cite{me}.

The last geometrical quantity to interpret is $A_a$, the connection in
the normal bundle.  In the Brown-York analysis the quantity $-A_a/(8\pi)$
is called the `momentum surface density', and is denoted as
$j_a$~\cite{BY}.  The momentum vector $j^a$ is tangential to $S$,
corresponding to a `rotating two-surface', and thus should be associated
with angular momentum.  Indeed, this is correct:  Let $\cal B$ denote the
timelike three-surface that is the congruence of world lines belonging to
the two-parameter family of observers on a two-sphere $S$.  If $\cal B$
admits a Killing vector field $\phi^a$, whose orbits lie in $S$, then one
can define the angular momentum charge
\beq
J := \int_{S}\,dS\,\phi^a j_a,
\eeq
which can be shown to coincide with the ADM angular momentum at infinity
for asymptotically flat spacetimes~\cite{BY}.  Thus we expect both the
shear and the connection in the normal bundle to play a role in angular
momentum at the quasilocal level, and indeed we will see that this turns
out to be the case.

Now the first goal of this paper is to provide a physical motivation for 
the general relativistic analogue of the special relativity formula: $E^2
-\vec{p}^{\;2} =m^2$.  First of all, this formula applies strictly to
point particles (as opposed to extended objects).  One imagines
determining, say, the instantaneous three-velocity of such a particle by
measuring its location in space at two closely separated points in time, 
in some inertial reference frame.  In the spirit of the quasilocal idea,
the analogue of this in general relativity would be to first replace
measurements at a point with measurements on a closed spacelike
two-surface $S$.  But measurements of what?  It would seem that
`measurements of the location of the point particle' `in some inertial
frame' is to be replaced with `measurements of the two-geometry of $S$'
`in a generic spacetime'.  These measurements are to be repeated at two
closely separated points in time.  In the point particle case this yields
the three-velocity (or the three-momentum $\vec{p}$ if one also knows
$m$); in the two-surface case it yields $l_{ab}$, the time
component of the extrinsic curvature of $S$.  Now I pointed out above
that the trace of $l_{ab}$---more precisely $l/(8\pi)$---indeed has the
interpretation of a momentum: it is the normal (or radial) momentum
surface density~\cite{Lau2}.  So it seems reasonable to replace $\vec{p}$
with $l/(8\pi)$.  What about the trace-free part of $l_{ab}$?  It was
argued above that $\tilde{l}_{ab}$ is associated with angular
momentum.  Insofar as angular momentum is qualitatively distinct from
linear momentum, its role at least at this point of the argument is not
clear, and we will simply drop it for now (however, its role will become
clear later).  Notice that dropping $\tilde{l}_{ab}$ is at least roughly
consistent with being interested only in the two-geometry of $S$, i.e.,
the two-metric $\sigma_{ab}$ modulo diffeomorphisms, since, as indicated
above, $\tilde{l}_{ab}$ is in some cases associated with just 
diffeomorphisms of $S$.

Thus, in the expression $E^2-\vec{p}^{\;2}$ we will replace $\vec{p}$ with
$l/(8\pi)$.  What should replace $E$?  Given the previous discussion, the
obvious answer is the Brown-York energy surface density,
$-k/(8\pi)$.  Clearly $l/(8\pi)$ and $-k/(8\pi)$ are on exactly the same
geometrical footing, being proportional to the timelike and spacelike
components of the mean curvature vector $H^c$ (see 
Eq.~(\ref{mean_curvature_vector})).  Thus we arrive at the generalization
\beq
E^2-\vec{p}^{\;2} \longrightarrow {1\over {(8\pi)^2}}(k^2 -l^2 ).
\label{generalization}
\eeq
Now before we accept this generalization, let us observe that there is
something unexpected about it.  The four-momentum $(E,\vec{p}\;)$ has
become a two-momentum, $(-k,l)/(8\pi)$.  What happened to the other two
components of spatial momentum?  $l/(8\pi)$ is just the radial
component; shouldn't the Brown-York momentum surface density $j^a$
(in our notation, $-A^{a}/(8\pi)$), which is tangent to $S$, be the
analogue of the two missing components of $\vec{p}\;$?  If so, then
instead of Eq.~(\ref{generalization}) we should have
\beq
E^2 -\vec{p}^{\;2} \stackrel{?}{\longrightarrow} {1\over {(8\pi)^2}}(k^2
-l^2 -A^a A_a ).
\label{trial}
\eeq

At first sight this expression is appealing because it manifestly includes
a contribution from angular momentum, and it is known that in general
relativity angular momentum contributes to mass.  A simple example that
illustrates this phenomenon is the Kerr black hole, where the ADM mass in
excess of the irreducible mass is due to rotational energy.  The precise
relationship is~\cite{MTW}
\beq
M_{\rm ir}^2 = M^2 - \left({{J}\over{2M_{\rm ir}}}\right)^2 ,
\label{Kerr}
\eeq
where $M$ is the ADM mass, $M_{\rm ir}$ the irreducible mass, and $J$ the
angular momentum of the black hole.  Comparing the right hand sides of
the previous two equations suggests we conceptually identify
$|A|/(8\pi)$ with the angular momentum term, $J/(2M_{\rm ir})$,
which seems reasonable.  This leaves $\sqrt{k^2 -l^2}/(8\pi)$ to be
interpreted as an object like $M$, viz., a total mass, `total' in the
sense that it includes the contribution from angular momentum.  But here
then is the
point: $\sqrt{k^2 -l^2}/(8\pi)$ somehow implicitly already includes the
angular momentum contribution to mass.  Precisely how will become clear
later, but to see immediately that this is at least plausible, consider
the case $l=0$.  Then $-\sqrt{k^2 -l^2}/(8\pi)$ reduces to the Brown-York
energy surface density, at least when $k$ is nonnegative, and it is known
that the (referenced) Brown-York quasilocal energy yields the ADM mass
at spatial infinity, which includes the correct angular momentum
contribution to mass.  So we do not need the $A^a A_a$ term in
Eq.~(\ref{trial}).  Besides, putting it in is counter to our goal of
seeing if general relativity admits an analogue of the {\it invariant}
mass, $m$:  While the combination $k^2 -l^2$ is invariant under radial
boosts,\footnote{This was first noted in Ref.~\cite{Lau2}.  A further
discussion of boosted observers in the Brown-York framework appears in
Ref.~\cite{Ivan}.} $A^a A_a$ is not---see  Eqs.~(\ref{extrinsic_boost}).  
So from this point of view the right hand side of Eq.~(\ref{trial}) is
defective, not to mention the generally unsavory fact that it {\it mixes}
objects with different transformation 
properties.\footnote{Hayward's~\cite{Hayw} definition of quasilocal energy
includes an angular momentum contribution of the form $-\omega^a
\omega_a$, analogous to the $-A^a A_a$ term in Eq.~(\ref{trial}).  
Hayward's $\omega_a$ is a suitably normalized anholonomicity, or `twist',
of the pair of null normals to $S$, and encodes essentially the same
information as $A_a$.  The important distinction is that, unlike the
connection $A_a$, the object $\omega_a$ is boost invariant, and so
representing angular momentum with a term proportional to $-\omega^a
\omega_a$, as Hayward does, is perfectly acceptable.  (The relationship
between $A_a$ and $\omega_a$ is discussed in Appendix~B of
Ref.~\cite{Hayw2}.)  However, there is no need, or even natural way for
$\omega_a$ to enter our work here.  For instance, the $A^a A_a$ term in
Eq.~(\ref{trial}) cannot simply be replaced with $\omega^a \omega_a$,
since the (tentative) inclusion of this term is suggested by the physical
interpretation of $A_a$ as a momentum surface density.  This
interpretation arises from Brown and York's Hamilton-Jacobi analysis of
the gravitational action~\cite{BY}, and it is not clear that a similar
interpretation can be given to $\omega_a$.}  The question of `missing
momentum components' can also be thought about as follows.  A point
particle has three components of spatial momentum.  Likewise, each point
on a two-surface $S$ also has three components of spatial momentum (more 
properly, momentum surface density): one normal, and two tangential to
$S$.  But being tangential, the latter two are associated with a `rotating
surface', and hence with angular momentum.  In going from a point to a
two-surface, two components of the linear momentum have become angular
momenta.  So they do not (directly at least) contribute to the expression
for an invariant mass given on the right hand side of
Eq.~(\ref{generalization}) because, as claimed, this expression already
inherently includes the contribution from angular momentum.

Thus we are led to propose the following definition of an `invariant
quasilocal energy' (or IQE):
\beq
{\rm IQE} = -{1\over{8\pi}}\,\int_{S}\,dS\,\sqrt{k^2 -l^2} - {\rm
IQE}^{\rm
ref} ,
\label{QIM}
\eeq
where ${\rm IQE}^{\rm ref}$ is a reference subtraction term that will be
defined later.  The word `invariant' in `IQE' refers to the fact just
mentioned, that $k^{2}-l^{2}$ is invariant under local radial boosts of
the observers on $S$.  And the word `energy' is used instead of
`mass'---despite our analogy between $\sqrt{k^{2}-l^{2}}/(8\pi)$ and the
mass $m$---because, as we will see in Sec.~7, the IQE behaves more like an
energy than a mass.  So the IQE can be thought of as the amount of `rest
energy' contained in $S$, a quantity independent of the motion of the
observers measuring it.  Notice that the unreferenced IQE is
negative.  Nominally the reference energy ${\rm IQE}^{\rm ref}$ is {\it
more} negative, so that the referenced IQE is positive.

It is useful to express the integrand of the unreferenced
IQE in two other equivalent forms.  Define the pair of null normals
$\xi_{\pm}^{a} := u^a \pm n^a$ on $S$, and the corresponding null
expansions
\beq
\theta_{\pm} := \sigma^{ab}\nabla_{a}\xi_{\pm \,b} = l \pm k ,
\label{null_expansions}
\eeq
cf. Eqs.~(\ref{extrinsic_curvatures}) and (\ref{trace-trace-free}).  Then
we have the following three equivalent expressions:
\beq
{1\over{8\pi}}\sqrt{k^2 -l^2}=
{1\over{8\pi}}\sqrt{-\theta_{+}\theta_{-}}=
{1\over{4\pi}}\sqrt{H\cdot H} .
\label{equiv_forms}
\eeq
For the last expression recall the definition of the mean curvature given
after Eq.~(\ref{mean_curvature_vector}).  Thus the unreferenced IQE in
Eq.~(\ref{QIM}) has a very simple geometrical interpretation: up to a
proportionality constant, it is just the mean curvature of $S$, averaged
over $S$.  Because of the square root it is defined only when $k^2 -
l^2 \geq 0$, i.e., at each point of $S$ the mean curvature vector $H^c$ in
Eq.~(\ref{mean_curvature_vector}) must be either spacelike or
null, but never timelike.  Roughly speaking, this
means that the area of $S$ changes more rapidly in a radial direction,
than in time.  For example, this condition is satisfied for the constant
$r,t$ two-spheres outside the horizon of a Schwarzschild black hole, but
not for those inside; on the horizon the unreferenced IQE is zero.

In terms of the null expansions, recall that a future (past) trapped
surface is one for which both ingoing and outgoing null expansions,
$\theta_{-}$ and $\theta_{+}$, are everywhere negative (positive) on
$S$~\cite{Wald}.  Thus, the unreferenced IQE is imaginary when $S$ is a
future or past trapped surface.  It is real only when no point on $S$ is
`trapped'.  Now a future trapped surface does not quite characterize a
black hole, and more subtle characterizations have been proposed for a
local definition of a black hole horizon~\cite{ACK,ABF,Hayw2}.
For example, Hayward~\cite{Hayw2} has introduced the notion of a `future
outer trapping horizon', $H$, characterized by: (i) $\theta_{-}|_H <0$
(in-going light rays converging), (ii) $\theta_{+}|_H =0$ (outgoing light
rays instantaneously parallel on the horizon), (iii) $\theta_{+}|_{H^+}>0$
(outgoing light rays diverging just outside the horizon), and (iv)
$\theta_{+}|_{H^-}<0$ (outgoing light rays converging just inside the
horizon).  According to this general definition of a black hole, the
unreferenced IQE is nonzero just outside the horizon, zero on the horizon,
and undefined (or imaginary) just inside the horizon.  In this connection
see also Ref.~\cite{Hayw3}. 

Furthermore, observe that the condition for the integrand of the
unreferenced IQE to be real and nonzero, namely $k^2 -l^2 >0$, is
precisely the same condition that ensures that the observers can always,
by appropriate local boosts, go to a `quasilocal rest frame' in which
$l=0$ at each point of $S$.  Such a two-surface is analogous to a massive
particle.  The case $k^2 -l^2 =0$ everywhere on $S$, for instance when $S$
is a future outer trapping horizon, is analogous to a massless particle,
for which no quasilocal rest frame exists.  And finally, the case $k^2
-l^2 <0$, say `inside' a future outer trapping horizon, corresponds to a
superluminal particle.\footnote{\label{string}As noted in the text, the
spirit of the quasilocal idea is to replace measurements at a point (of
certain aspects of a point particle, say) with measurements on a closed
spacelike two-surface.  If one takes seriously that Eq.~(\ref{generalization}) 
is the generalization of point particle rest mass, then one is quickly led
to speculate that a closed spacelike two-surface is the generalization of
a point particle.  This is curiously reminiscent of string theory, except
that the one-dimensional string is replaced by a two-dimensional surface.}

The situation is actually more subtle than indicated in the previous two
paragraphs.  The conditions for $\sqrt{-\theta_{+}\theta_{-}}$ to be real
are reminiscent of the condition $\theta_{-}\leq 0$ required for the
holomorphic case of the Dougan-Mason quasilocal energy to be nonnegative.  
The conditions $\theta_{+}\geq 0$ and $\theta_{-}\leq 0$ essentially imply
that the two-surface $S$ is `suitably' convex~\cite{DM}.  To emphasize
that `` `suitably' convex'' is not a serious restriction, in particular it
does not mean that $S$ cannot be concave, consider a two-parameter family
of observers at rest in an inertial frame in flat spacetime.  Suppose that
at $t=0$ they lie on a two-sphere $S$ that is round except for a small
indentation.  Then $l=0$ at each point of $S$, and $k$ is positive
everywhere except in a small region near the center of the indentation,
where it is negative.  Thus there will be a circle of points $C$ at which
$k=0$.  So at each point of $S$ we have $k^{2}-l^{2}\geq 0$, equality
holding on $C$.  One might worry that a radial boost at a point on $C$
will make $l^{2}>0$, and hence  $k^{2}-l^{2}< 0$.  But of course this will
not happen:  If we consider a second set of observers, boosted relative to
the first, then $k=l=0$ on $C$ implies $k^{\prime}=l^{\prime}=0$ on
$C$.  So we can consider the second set of observers to be boosted
radially outward in the region of the indentation, such that the indentation, 
and its attendant set of fixed points $C$, smoothly disappear as the
sphere evolves in time.  $k$ switches from negative to positive by passing
through the origin of a $k$--$l$ diagram.  Thus we can imagine a wide
class of two-surfaces, including ones with indentations, and dynamically
changing in time, for which $k^{2}-l^{2}\geq 0$ everywhere on $S$.  Moreover, 
bear in mind that the observers are allowed to accelerate, so there is a
great deal of freedom for them to maintain a `physically reasonable'
$S$.  Nevertheless, what is needed here is a careful analysis based on
Raychaudhuri-like equations for a two-parameter family of accelerated
timelike curves (as opposed to the more usual case of timelike geodesics).  
Such a detailed analysis is outside the scope set for this introductory
paper.

\section{The reference invariant quasilocal energy}

As in the Brown-York case, the unreferenced IQE diverges in an 
asymptotically flat spacetime as the two-surface $S$ is taken to
(spatial or null) infinity, and so must be regulated with a reference
term, ${\rm IQE}^{\rm ref}$, as already anticipated in Eq.~(\ref{QIM}).
To better understand the nature of our definition of the invariant 
quasilocal energy, and to help suggest a natural choice for ${\rm
IQE}^{\rm ref}$, we now make use of the Gauss embedding equation given in
Eq.~(\ref{Gauss}).  This equation has only one independent 
component.  Transvecting both sides with $\sigma^{ac}\sigma^{bd}$ reduces
it to the scalar equation
\beq
\sigma\sigma R = {\cal R} - {1\over 2}(k^2 -l^2 ) + (\tilde{k}^2
-\tilde{l}^2 ) ,
\label{scalar_Gauss}
\eeq
where $\sigma\sigma R$ is shorthand for $\sigma^{ac}\sigma^{bd} R_{abcd}$,
$(\tilde{k}^2 -\tilde{l}^2 )$ is shorthand for
$(\tilde{k}^{ab}\tilde{k}_{ab}-\tilde{l}^{ab}\tilde{l}_{ab})$,
and ${\cal R}$ is the scalar curvature of $(S,\sigma)$.  Using this
equation we can express the IQE given in Eq.~(\ref{QIM}) in the equivalent
form
\beq
{\rm IQE} = -{1\over{8\pi}}\,\int_{S}\,dS\,\sqrt{2\left[{\cal R} - 
\sigma\sigma R + (\tilde{k}^2 -\tilde{l}^2 )\right]} - {\rm IQE}^{\rm
ref}.
\label{QIM1}
\eeq
We remark here that $\sigma\sigma R$ is a natural geometrical object
called the sectional curvature of $(S,\sigma)$ as embedded in
$(M,g)$~\cite{Chen}.  It will play an important role in what follows.

Now the definition of the unreferenced IQE is rooted in the extrinsic 
geometry of the submanifold $(S,\sigma)$, thought of as a two-surface
isometrically embedded in the spacetime $(M,g)$.  It is then natural to
define the reference IQE to be of the same form as the unreferenced IQE in
Eq.~(\ref{QIM1}), i.e., to be the same geometrical object, except with
$(S,\sigma)$ now isometrically embedded in a different spacetime---some
reference spacetime, $(M^{\rm ref},g^{\rm ref})$.  Thus ${\rm IQE}^{\rm
ref}$ will be the integral in Eq.~(\ref{QIM1}) (or (\ref{QIM})), except
with all quantities referred to the reference spacetime, which we indicate
with a superscript `ref'.  Note that although the extrinsic geometry of
$S$ will be different in $(M^{\rm ref},g^{\rm ref})$, its intrinsic geometry, 
by assumption, will not.  So in the ${\rm IQE}^{\rm ref}$ integral we are
constructing we can set $dS^{\rm ref} = dS$, ${\cal R}^{\rm ref} = {\cal
R}$, and $(\sigma\sigma R)^{\rm ref}=\sigma\sigma R^{\rm ref}$.  Also note
that, in general, $M^{\rm ref}\not = M$ (topologically).  For example, $S$
may be a two-sphere embedded in a black hole spacetime, with $M=R^2 \times
S^2$, whereas the reference spacetime might be Minkowski space, with
$M^{\rm ref}=R^4$.  With this understanding, we define
\beq
{\rm IQE}^{\rm ref} = -{1\over{8\pi}}\,\int_{S}\,dS\,\sqrt{(k^{2} -
l^{2})^{\rm ref}} = -{1\over{8\pi}}\,\int_{S}\,dS\,\sqrt{2\left[{\cal R}
- \sigma\sigma R^{\rm ref} + (\tilde{k}^2 -\tilde{l}^2 )^{\rm
ref}\right]}.
\label{reference_QIM}
\eeq
The term $\sigma\sigma R^{\rm ref}$ is shorthand for
$\sigma^{ac}\sigma^{bd}R_{abcd}^{\rm ref}$, where $R_{abcd}^{\rm ref}$ is
the Riemann tensor of the reference spacetime.  

Typically one is motivated to choose a reference spacetime of constant
curvature, the geometrical reason being that then the Gauss, Codazzi, and
Ricci embedding equations make no reference to `where' $(S,\sigma)$ is
embedded in $(M^{\rm ref},g^{\rm ref})$.  In other words, the conditions
placed on $k^{\rm ref}_{ab}$, $l^{\rm ref}_{ab}$, and $A^{\rm ref}_{a}$ by
the `reference version' of Eqs.~(\ref{Gauss}--\ref{Ricci})---which are
just integrability conditions for the reference embedding---do not depend
on knowing the embedding itself~\cite{Spiv}.  This is a pleasing
criterion because it keeps the reference spacetime `abstract', rather than
`concrete'.  For a four-dimensional space of constant curvature we have 
\beq
R^{\rm ref}_{abcd}={C\over{12}}(g^{\rm ref}_{ac}g^{\rm ref}_{bd}-
g^{\rm ref}_{bc}g^{\rm ref}_{ad}),
\eeq
where $C$ is the constant value of its scalar curvature.  For this choice
of reference spacetime one gets
\beq
\sigma\sigma R^{\rm ref}={C\over 6}.
\label{ssRref}
\eeq
For example, for Minkowski space we have $C=0$, and for anti-de~Sitter
space we have $C=-12/\ell^2$, where $\ell$ is the radius of curvature of
the anti-de~Sitter space, and is related to the (negative) cosmological
constant $\Lambda$ by $\Lambda = -3/\ell^2$.  We will return to these two
examples later.

The idea of embedding $(S,\sigma)$ into some reference space(time) is in
the same spirit as the Brown-York approach, but an important difference
that arises out of using the invariant quantity $\sqrt{k^2 -l^2 }$,
rather than $k$, deserves further comment.  In the Brown-York approach 
$k$ is the trace of the extrinsic curvature of $(S,\sigma)$ as embedded in
a three-geometry, $(\Sigma,h)$, where $\Sigma$ is a spacelike three-surface 
spanning $S$, with induced metric $h_{ab}$.  Thus it is natural to take
$k^{\rm ref}$ as the trace of the extrinsic curvature of $(S,\sigma)$ as
embedded in some three-dimensional reference space, $(\Sigma^{\rm
ref},h^{\rm ref})$.  So the embeddings of $S$, for both the unreferenced
and reference CQE, inherently have a {\it three}-dimensional target
space.  On the other hand, $\sqrt{k^2 -l^2 }$ is proportional to a
geometrical invariant of $S$, namely its mean curvature, and makes no
essential reference to a spanning three-surface $\Sigma$ (making the IQE
`truly quasilocal' in the sense that it depends on $S$
alone\footnote{The CQE can also be made `truly quasilocal', in a
slightly different sense: by relaxing the restriction that the
foliation of the spacetime (i.e., $\Sigma$) be orthogonal to the
boundary $\cal B$~\cite{BY}, it is shown in Ref.~\cite{Ivan} that
the resulting CQE no longer depends on $\Sigma$, but instead just
depends on the foliation of $\cal B$.}).  As a consequence, the
embeddings inherently have a {\it four}-dimensional target space(time).  

The advantage of a three-dimensional target reference space, say flat
$\IR^3$, is that when the embedding exists, it is unique (up to 
translations and rotations), and so the Brown-York ${\rm CQE}^{\rm ref}$
is unique.  The disadvantage, as is well known, is that such embeddings do
not exist for all $(S,\sigma)$ of interest, and this problem
is not limited to just a few isolated exceptional cases.  

For a four-dimensional target reference spacetime, say Minkowski
space, the situation is reversed:  an embedding always exists, but it is
not unique.  Regarding the first half of this statement,
Brinkmann~\cite{Brin} has shown, by a simple explicit construction, that
any $n$-dimensional conformally flat Riemann space can be considered as a
particular cut of a light cone in $(n+2)$-dimensional Minkowski
space.  And conversely, any cut of such a light cone gives an
$n$-dimensional conformally flat Riemann space.  Now any $n=2$ space is,
of course,  conformally flat, and thus any $(S,\sigma)$ can always be so
embedded, even if $S$ has regions with negative scalar curvature.  In the
Introduction I mentioned the example of the horizon of the Kerr black
hole, which cannot be globally embedded into flat $\IR^3$ when the angular
momentum exceeds the irreducible mass, which coincides with the two-sphere
developing regions with negative scalar curvature~\cite{Smar}.  However,
it is a simple exercise to apply Brinkmann's construction and thus
globally embed the horizon into a light cone of four-dimensional Minkowski
space.  I will omit the details of this calculation, and just note that
the embedding is valid for all angular momentum $J$ (up to and including
the extremal black hole case), and changes smoothly with $J$, including at
the critical point when $J$ equals the irreducible mass.

On the other hand, in a codimension-two (versus -one) embedding there is
more `elbow room', and consequently the embedding is not unique---there is
a function worth of freedom (which will be discussed in detail in 
Ref.~\cite{me}).   This results in an ambiguity in the reference energy,
${\rm IQE}^{\rm ref}$, which enters via the reference shear term
$(\tilde{k}^2 -\tilde{l}^2)^{\rm ref}$---see Eq.~(\ref{reference_QIM}).  
This is the only term in ${\rm IQE}^{\rm ref}$  not yet determined, and
the only one for which we require a reference embedding.  Observe that in
the Brown-York approach the undetermined quantity is $k^{\rm ref}$, an
expansion.  Here it is $(\tilde{k}^2 -\tilde{l}^2 )^{\rm ref}$, a
shear.  I will now argue that it is precisely this term that plays the key
role in properly incorporating angular momentum into the IQE.  The basic
idea is simple, but first we will introduce some notation.

In the previous section we introduced the null normals
$\xi_{\pm}^{a}=u^{a}\pm n^{a}$, and corresponding null expansions
$\theta_{\pm}=l\pm k$ in Eq.~(\ref{null_expansions}).  Similarly, the
(trace-free) shears in the two null directions are defined by $s_{\pm \,
ab} :=\tilde{l}_{ab}\pm\tilde{k}_{ab}$.  The curvature of the normal
bundle has only one independent component, and can be written as
${\cal F}_{ab}=({\cal F}/2)\epsilon_{ab}$ for some scalar field $\cal
F$, where $\epsilon_{ab}$ is the volume form on $S$ defined earlier in
Eq.~(\ref{epsilon}).  With this notation, and assuming that
the reference spacetime is one of constant curvature, i.e.,
Eq.~(\ref{ssRref}) holds, the Gauss, Codazzi, and Ricci embedding 
equations given at the end of Sec.~2 take the form
\begin{eqnarray}
{C\over 6} & = & {\cal R} + {1\over 2}\theta_{+}^{\rm ref}
\theta_{-}^{\rm ref} - 
s_{+\;\;\;\;b}^{{\rm ref}\;a}
s_{-\;\;\;\;a}^{{\rm ref}\;b},
\label{ref_Gauss}
\\
0 & = & {1\over 2}({\cal D}_{a}\mp A_{a}^{\rm ref})\theta_{\pm}^{\rm ref}
-({\cal D}_{b}\mp A_{b}^{\rm ref})
s_{\pm\;\;\;\;a}^{{\rm ref}\;b},
\label{ref_Codazzi}
\\
0 & = & {\cal F}^{\rm ref}+{1\over 2}\epsilon^{a}_{\;\;b} 
[s_{+}^{\rm ref},s_{-}^{\rm ref}]^{b}_{\;\;a}.
\label{ref_Ricci}
\end{eqnarray}
In the Ricci equation, $[s_{+}^{\rm ref},s_{-}^{\rm ref}
]^{b}_{\;\;a}$ denotes the commutator of the shears:
$s_{+\;\;\;\;c}^{{\rm ref}\;b}
 s_{-\;\;\;\;a}^{{\rm ref}\;c}- 
 s_{-\;\;\;\;c}^{{\rm ref}\;b}
 s_{+\;\;\;\;a}^{{\rm ref}\;c}$.
Notice that by using null directions, rather than $u^a$ and $n^a$, the
Codazzi equations have decoupled into a `$+$' and a `$-$' set.

Our task is thus: Given $\sigma_{ab}$, and hence $\cal R$, ${\cal
D}_a$, and $\epsilon_{ab}$, solve these embedding equations for the
unknown quantities $\theta_{\pm}^{\rm ref}$, $s_{\pm \, ab}^{\rm ref}$,
and $A_{a}^{\rm ref}$.  (Of course ${\cal F}^{\rm ref}=2\epsilon^{ab}
{\cal D}_{a}A_{b}^{\rm ref}$ is not an independent quantity.)  In
particular, we are interested in the solution for the boost
invariant reference shear term
\beq
(\tilde{k}^2 -\tilde{l}^2 )^{\rm ref} \equiv -
s_{+\;\;\;\;b}^{{\rm ref}\;a}
s_{-\;\;\;\;a}^{{\rm ref}\;b},
\label{RefShearTerm}
\eeq
appearing in the Gauss equation, which is to then be substituted into the
second integral of Eq.~(\ref{reference_QIM}).  Or equivalently, solve for
$(k^2 -l^2 )^{\rm ref}\equiv -\theta_{+}^{\rm ref}\theta_{-}^{\rm ref}$
and substitute the answer into the first integral of
Eq.~(\ref{reference_QIM}).  This is how ${\rm IQE}^{\rm ref}$ is
determined.

However, as already noted, any solution we obtain is not unique.  We can
see this immediately by counting functional degrees of freedom.  
$\theta_{\pm}^{\rm ref}$ are two functions, $s_{\pm\, ab}^{\rm ref}$ are
four (the two shears are symmetric and trace-free), and $A_{a}^{\rm
ref}$ are two.  These eight functions are subject to six equations: Gauss
is one, Codazzi are four, and Ricci is one.  This leaves two arbitrary
functions in the solution.  But owing to the invariance of the embedding
equations under a local boost transformation (see 
Eqs.~(\ref{extrinsic_boost})), one of these functions is just the boost
parameter, $\lambda$, leaving one nontrivial arbitrary function in the
solution.  

The question then arises, Is there a natural way to impose one
additional functional condition on the unknowns so that the embedding,
subject to this additional condition, is unique, and hence ${\rm IQE}^{\rm
ref}$ is unique?  One of the central ideas in this paper is to
impose the additional condition
\beq
{\cal F}^{\rm ref}={\cal F} ,
\label{F=F_ref}
\eeq
i.e., the curvature of the normal bundle of $S$ as embedded in the
reference spacetime $(M^{\rm ref},g^{\rm ref})$ should equal that of $S$
as embedded in the original physical spacetime, $(M,g)$.  

There are several reasons why it is geometrically natural to demand ${\cal
F}^{\rm ref}={\cal F}$.  First, the two-surface $S$ has two connections: 
one is an $SO(2)$ connection on the tangent bundle of $S$, associated
with the curvature ${\cal R}_{abcd}$ (which has only one independent
component, namely ${\cal R}$), and the other is an $SO(1,1)$
connection on the normal bundle of $S$, associated with the curvature
${\cal F}_{ab}$ (which also has only one independent component,
namely ${\cal F}$).  In fact both of these connections are metric
connections, associated with the metrics in the tangent and normal
bundles to $S$, respectively~\cite{Chen}.  Furthermore,  Szabados~\cite{Szab2} 
has considered the two-dimensional version of the Sen connection for
spinors and tensors on a submanifold such as $S$, and has found that the
two-surface spinor curvature has, essentially, imaginary part equal to
${\cal R}$, and real part equal to ${\cal F}$.  
Finally, although $A_{a}$ is a measure of extrinsic geometry, as pointed
out earlier it is not really on the same footing as the extrinsic
curvatures $k_{ab}$ and $l_{ab}$, since its transformation law under local
radial boosts is qualitatively different---see Eqs.(\ref{extrinsic_boost}).  
It transforms like the connection that it is, and gives rise to a
curvature, and so arguably has more in common with ${\cal R}$ than with
$k_{ab}$ and $l_{ab}$.  The point is, $\cal F$ is really on the same
geometrical footing as $\cal R$.  We have already demanded that ${\cal
R}^{\rm ref}={\cal R}$, as a necessary condition for the embedding of
$(S,\sigma)$ into $(M^{\rm ref},g^{\rm ref})$ to be isometric.  So
demanding also that ${\cal F}^{\rm ref}={\cal F}$ is thus seen to be quite
natural.

Unfortunately, implementing Eq.~(\ref{F=F_ref}) seems like an intractable
task.  Embedding equations involving curvature of the normal bundle, i.e., 
codimension-two (and higher) embeddings, have, of course, been studied
for a long time.  With regard to solutions, although one expects to be
able to express $(\tilde{k}^2 -\tilde{l}^2 )^{\rm ref}$ in terms of $\cal
F$, $\cal R$, and their derivatives, I am not aware of any such general
results in the literature.  In fact, much of the literature on such
embeddings considers the case ${\cal F}=0$, which is not the case we are
particularly interested in here (a notable exception is Ref.~\cite{Chen}).  
One possible way to proceed is as follows.  Given ${\cal F}^{\rm ref}$
($={\cal F}$ by Eq.~(\ref{F=F_ref})), choose $A_{a}^{\rm ref}$ such that
${\cal F}^{\rm ref}=2\epsilon^{ab}{\cal D}_{a}A_{b}^{\rm ref}$.  There may
be a convenient gauge choice, such as ${\cal D}\cdot A^{\rm ref}=0$, or
$l^{\rm ref}=0$.  Then view the Codazzi equations (\ref{ref_Codazzi}) as a
set of four linear partial differential equations for the four independent
degrees of freedom in the two shears.  These equations are of
second order if one makes use of the fact that any trace-free symmetric
tensor $s_{ab}$ on a two-surface $(S,\sigma)$ with two-sphere topology can
be expressed as $s_{ab}={\cal D}_{a}v_{b}+{\cal D}_{b}v_{a}-\sigma_{ab}{\cal
D}\cdot v$ for some vector field $v^{a}$.  Thus solve for $s_{\pm\,
ab}^{\rm ref}$ in terms of the expansions, $\theta_{\pm}^{\rm ref}$, and
their derivatives.  The expressions one obtains at this stage are, in
general, nonlocal.  Then substitute these into the Gauss and Ricci
equations, (\ref{ref_Gauss}) and (\ref{ref_Ricci}), which are really just
nonlinear algebraic constraints.  But because the shears involve nonlocal
operators acting on the expansions, one ends up with two nonlocal and
nonlinear partial differential equations for the two expansions.  Remarkably, 
it is almost possible to solve these equations, but in the end one
encounters a certain combination of nonlocality and nonlinearity that
makes the final step to a solution seem impossible.  Nevertheless,  it
appears that the solution for $(\tilde{k}^2 -\tilde{l}^2 )^{\rm ref}$, if
it can be found, almost certainly depends in a simple way on both $\cal
R$ {\it and} $\cal F$, and derivatives (of a finite or possibly infinite
order) of these two curvatures.  I am suggesting that it is through this
subtle presence of ${\cal F}$ in $(\tilde{k}^2 -\tilde{l}^2 )^{\rm ref}$
that angular momentum is properly incorporated into the IQE.  

So although a direct attack on the embedding equations has not yet
yielded a solution, fortunately one can make some progress of a general
nature by calculating the first and second order variations of ${\cal
F}^{\rm ref}$ and $(\tilde{k}^2 -\tilde{l}^2 )^{\rm ref}$ under isometric
deformations of a given embedding.  The idea is to see how both of these
quantities change under such a deformation, and thereby infer how 
$(\tilde{k}^2 -\tilde{l}^2 )^{\rm ref}$ depends on ${\cal F}^{\rm
ref}$, and hence angular momentum.  The results are somewhat involved, and
will be given elsewhere~\cite{me}.  For now let us start by making some
simple observations regarding the enigmatic object $(\tilde{k}^2
-\tilde{l}^2 )^{\rm ref}$.

To begin with, one might object to our argument thus far because it
implies that $(\tilde{k}^2 -\tilde{l}^2 )^{\rm ref}$, and hence ${\rm
IQE}^{\rm ref}$, depends on the extrinsic geometry of $S$ as embedding in
the physical spacetime.  In particular, through Eq.~(\ref{F=F_ref}) and
the reference embedding equations, $(\tilde{k}^2 -\tilde{l}^2 )^{\rm ref}$
depends on $\cal F$.  On the other hand, it is often stated that a
reference subtraction term should be a functional of only the intrinsic
geometry of $(S,\sigma)$.  However, notice that there is no dependence on
the extrinsic curvatures proper, i.e. $l_{ab}$ and $k_{ab}$, only a
dependence on $\cal F$, a quantity which I argued above is really on the
same geometrical footing as the intrinsic quantity $\cal
R$.  Moreover, as discussed in the Introduction (refer to
Eq.~(\ref{total_action})), in the Brown-York approach one is free to add
to the action any functional of the boundary {\it three}-metric,
$\gamma_{ab}$, which contains information about the two-metric
$\sigma_{ab}$, as well as information about how $(S,\sigma)$ is
embedded in the three-boundary $\cal B$.  For instance, one could add to
the action a boundary integral of the scalar curvature of $\cal B$, whose
variation would add to $\Pi^{ab}$ in Eq.~(\ref{total_action}) a term
proportional to the Einstein tensor of $\gamma_{ab}$, as is done in 
Ref.~\cite{BK}.  Such a term obviously depends on some extrinsic
geometry of $(S,\sigma)$.  In Brown and York's work this fact is of course
recognized, but being in a Hamiltonian framework, they restrict the form
of the arbitrary boundary functional such that the energy surface density
($-k/(8\pi)$) and momentum surface density ($-A_{a}/(8\pi)$) of $S$ in a
particular spacelike hypersurface $\Sigma$ depend only on the canonical
data on $\Sigma$.  This effectively means that their reference subtraction
term can depend only on $\sigma_{ab}$~\cite{BY}.  But as I emphasized
earlier, our approach is based on the invariant object $\sqrt{k^2 -l^2}$, 
and makes no essential reference to a three-surface $\Sigma$ spanning
$S$.  The invariant quasilocal energy constructed here does not
come out of a canonical analysis, so there is no reason that our subtraction 
term cannot depend on $\cal F$.

So the shear term $(\tilde{k}^2 -\tilde{l}^2 )^{\rm ref}$ is allowed, but
is it really necessary?  Perhaps it is just an unsavory term resulting
from a poor definition of the IQE.  For instance, looking at the Gauss
embedding equation (\ref{scalar_Gauss}) one might be tempted to write,
instead of Eq.~(\ref{generalization}), 
\beq
E^2-\vec{p}^{\;2} \stackrel{?}{\longrightarrow} {1\over {(8\pi)^2}}[(k^2
-l^2 ) -2(\tilde{k}^2 -\tilde{l}^2 )] ,
\label{trial2}
\eeq
where the additional shear term on the right hand side is perhaps the 
proper way to include angular momentum, somewhat like the $A^{a}A_{a}$
term we attempted in Eq.~(\ref{trial}).  This would have the advantage of
changing ${\rm IQE}^{\rm ref}$ in Eq.~(\ref{reference_QIM}) to
\beq
{\rm IQE}^{\rm ref} \stackrel{?}{=}
-{1\over{8\pi}}\,\int_{S}\,dS\,\sqrt{2\left[{\cal R}
- \sigma\sigma R^{\rm ref} \right]},
\label{trial_reference_QIM}
\eeq
which is clearly unique, and moreover, requires no reference embedding.  
Eq.~(\ref{trial_reference_QIM}) is a more general case of the zero point
energy suggested by Lau~\cite{Lau3} (except that his derivation of it
requires a reference embedding---we will return to this point later).  But
unfortunately it cannot be correct.  For example, when the reference
spacetime is Minkowski space, Eq.~(\ref{ssRref}) tells us that
$\sigma\sigma R^{\rm ref}=0$ and thus the radical reduces to $\sqrt{2{\cal
R}}$, which is not defined for negative $\cal R$.  Nor is this problem
properly solved by taking $C\not=0$ in Eq.~(\ref{ssRref}), since this
would put an ad hoc fixed lower bound on $\cal R$.

So the shear term $(\tilde{k}^2 -\tilde{l}^2 )^{\rm ref}$ is not only
allowed, it is necessary (or at least its absence leads to an unsatisfactory
result).  In fact its role seems to be to keep nonnegative what is under
the square root in Eq.~(\ref{reference_QIM}).  To see this more clearly,
consider the special case that $(S,\sigma)$ is embeddable in flat $\IR^3$.
If our reference spacetime is Minkowski space, we can then choose to embed
$(S,\sigma)$ in a $t={\rm constant}$ slice and, within this slice, the
embedding is essentially unique.  In this case it is easy to see
that we will have $l_{ab}^{\rm ref}=0$.  And by assumption, $\sigma\sigma 
R^{\rm ref}=0$, so Eq.~(\ref{reference_QIM}) reduces to
\beq
{\rm IQE}^{\rm ref}|_{l_{ab}^{\rm ref}=0} =
-{1\over{8\pi}}\,\int_{S}\,dS\,| k^{\rm ref} | =
-{1\over{8\pi}}\,\int_{S}\,dS\,\sqrt{2\left[{\cal R}
+ (\tilde{k}^{\rm ref})^{2}\right]}.
\label{reference_QIM_l=0}
\eeq
The uniqueness of the embedding means that $k^{\rm ref}$ and
$\tilde{k}_{ab}^{\rm ref}$ are unique.  Now clearly, no matter what the
surface is, the spatial shear $\tilde{k}_{ab}^{\rm ref}$ must be such that
what is under the square root in Eq.~(\ref{reference_QIM_l=0}) is
nonnegative, because $|k^{\rm ref}|$ is real.  (More properly, one should
look at the `reference version' of Eq.~(\ref{scalar_Gauss}), with $(l^{\rm
ref})^2 = (\tilde{l}^{\rm ref})^2 =\sigma\sigma R^{\rm ref} = 0$.)
For example, consider a dumb-bell-shaped surface of revolution in flat
$\IR^3$.  In a region near the throat of this surface $\cal R$ is
negative, nevertheless at every point of the surface we have ${\cal
R}+(\tilde{k}^{\rm ref})^{2}\geq 0$.  

I emphasize that, even when $(S,\sigma)$ can be embedded in flat $\IR^3$,
its embedding in Minkowski space need not be chosen to be in a $t={\rm
constant}$ slice, as was done in the previous paragraph.  One may also
embed it in a light cone, or in a host of other ways---remember that there
is a function-worth of freedom in our choice.  I argued in the context of
Eq.~(\ref{F=F_ref}) that this freedom has to do with angular momentum, or
more precisely, the curvature of the normal bundle.  For the $t={\rm
constant}$ embedding, $\tilde{l}_{ab}^{\rm ref}=0$ implies $s_{\pm\,
ab}^{\rm ref}=\pm\tilde{k}_{ab}^{\rm ref}$, and so inspection of
Eq.~(\ref{ref_Ricci}) reveals that in this case ${\cal F}^{\rm ref}=0$.
However, it is not hard to see that starting with such a $t={\rm
constant}$ embedding one can perform an infinitesimal isometric
deformation of the embedding out of the $t={\rm constant}$ plane, i.e., in
a direction $\varphi\partial/\partial t$, where $\varphi$ is an arbitrary
function.  Furthermore, I show in Ref.~\cite{me} that after such an
infinitesimal deformation ${\cal F}^{\rm ref}$ is no longer zero, and can
in fact be made to be
essentially any infinitesimal function we like by a suitable choice of
$\varphi$.  It is not hard to imagine (just hard to do!) that by
integrating such isometric deformations one may be able to achieve a
two-geometry $(S,\sigma)$ isometrically embedded with any desired
curvature of the normal bundle.  With this in mind, it would seem
unnatural to reference-embed, e.g., a constant $r,t$ two-sphere of the
Kerr geometry (which is easily shown to have ${\cal F} \not= 0$) as a
two-sphere in a Minkowski reference spacetime with ${\cal F}^{\rm ref}=0$
(say, in a $t={\rm constant}$ 
slice), when it seems possible to instead embed it with
${\cal F}^{\rm ref}={\cal F}$.  Note, however, that as the embedding is 
deformed out of the $t={\rm constant}$ surface, $\tilde{l}_{ab}^{\rm ref}$
will also cease to be zero, and so will introduce a negative contribution
to the quantity under the square root in Eq.~(\ref{reference_QIM}).  This
jeopardizes the nonnegativity of this quantity.  But at
the same time we clearly cannot simply throw away the $-(\tilde{l}^{\rm
ref})^2$ term, since this would violate a key property of the IQE,
namely its invariance under local boosts.  Short of solving the embedding
equations for a generic ${\cal F}^{\rm ref}$ and explicitly checking, I do
not know of any guarantee of nonnegativity.  Equivalently, in question
here is the nonnegativity of the quantity $(k^{2}-l^{2})^{\rm ref}$, which
is the same type of question as the nonnegativity of $(k^{2}-l^{2})$
discussed at the end of Sec.~3.  I do not at present have a complete
answer to either of these difficult questions.

Finally, one might guess that it is possible to avoid the embedding
problem entirely by simply setting $(\tilde{k}^2 -\tilde{l}^2 )^{\rm ref}
= (\tilde{k}^2 -\tilde{l}^2 )$, which is in the same spirit as
Eq.~(\ref{F=F_ref}) in that, like ${\cal F}$, the shear term has something
to do with angular momentum.  But this does not work.  For example, it is
not hard to show that, although $(\tilde{k}^2 -\tilde{l}^2 )$ is in
general nonvanishing on constant $r,t$ spheres of the Kerr geometry, it
happens to vanish on the horizon.  So if we set $(\tilde{k}^2 -\tilde{l}^2
)^{\rm ref} = (\tilde{k}^2 -\tilde{l}^2 )$, then in calculating ${\rm
IQE}^{\rm ref}$ for the Kerr horizon example we would run into the same
problem with negative $\cal R$ as we did in Eq.~(\ref{trial_reference_QIM}).  
So such a prescription must not be valid, and we cannot avoid the
embedding problem this way.

Let us conclude this section by addressing the question, What is the
relationship between the IQE defined
here and the Brown-York CQE?  We begin by supposing that the following
four conditions are satisfied: (i) $(S,\sigma)$ is a two-surface in the
physical spacetime such that $k^2 -l^2 > 0$; (ii) $k > 0$; (iii) 
$(S,\sigma)$ is such that it can be embedded in flat $\IR^3$; and
(iv) for the embedding in (iii), $k^{\rm ref} \geq 0$.  As discussed
earlier, condition (i) ensures that the unreferenced IQE is well
defined---roughly speaking, $S$ is strictly outside of a black hole.  Then
it is always possible to go to a `quasilocal rest frame' where $l=0$ on
$S$, and the integrand in Eq.~(\ref{QIM}) is just $|k|$.  Given condition
(ii), the unreferenced IQE thus reduces to the unreferenced CQE in
Eq.~(\ref{BY_QE}), provided the observers in the Brown-York case are in a
`quasilocal rest frame'.  Condition (iii) ensures that the Brown-York
prescription is well-defined, and allows us to choose a $t={\rm constant}$
embedding in Minkowski space, as above, and get 
Eq.~(\ref{reference_QIM_l=0}).  The first integral in this equation,
together with condition (iv), shows that our ${\rm IQE}^{\rm ref}$ reduces
to the Brown-York ${\rm CQE}^{\rm ref}$ in Eq.~(\ref{BY_REF}).  So if
these conditions hold, and we choose to use a $t={\rm constant}$ embedding
to calculate ${\rm IQE}^{\rm ref}$, then our invariant quasilocal energy is
the same as the Brown-York `rest energy'.  In most applications considered
in the literature these conditions are satisfied, and the IQE will then
share all of the desirable properties of the CQE.  For example, it will be
the thermodynamic energy that appears in the first law of black hole
thermodynamics for Schwarzschild black holes, as considered in
Ref.~\cite{BY}.

On the other hand, I emphasize that conditions (ii) and (iv) are easily
violated.  One need only think of a round sphere with a small indentation,
an example discussed at the end of Sec.~3 (except here the spacetime is
generic).  So in general, the IQE defined here is not simply the
Brown-York `rest energy'.  Furthermore, we need not choose a $t={\rm
constant}$ embedding to calculate ${\rm IQE}^{\rm ref}$.  Indeed, as I
have argued, such a choice is unnatural when ${\cal F}\not=0$.  In short,
the invariant quasilocal energy defined here is not quite the same object
as the Brown-York quasilocal energy.   Note that the aforementioned
thermodynamic nature of the CQE is derived in Ref.~\cite{BY} assuming that
$S$ is a round sphere.  It would be interesting to extend this analysis to
indented spheres, for instance, and determine which, if either of the CQE
or IQE, is the `correct' thermodynamic energy.

\section{The large sphere limit of the IQE}

Let us now assume that the spacetime $(M,g)$ is asymptotically flat, and
evaluate the limit of the invariant quasilocal energy as $(S,\sigma)$
tends to a large sphere at infinity.  At spatial and null infinity we
might expect these limits to be the ADM and Bondi-Sachs masses,
respectively.  Let us see if this is so.

Let $\tau$ be a `time' function on $M$ such that $\tau=\tau_{*}$ defines a
spacelike (respectively, null) hypersurface ${\cal H}_{\tau_{*}}$ of
topology $R\times S^{2}$ extending to spatial (respectively, null)
infinity.  Letting the parameter $\tau_{*}$ vary over some range gives a
foliation of a part of $M$.  Let $r$ be a function on $M$ such that
$r=r_{*}$ defines a hypersurface that intersects each leaf ${\cal 
H}_{\tau_{*}}$ (over the allowed range of $\tau_{*}$) in a spacelike
two-sphere, $S_{\tau_{*},r_{*}}$.  The parameter $r_{*}$ ranges to
infinity, and over its range the surfaces $S_{\tau_{*},r_{*}}$ provide a
foliation of ${\cal H}_{\tau_{*}}$.  We are interested in the limit
$r_{*}\rightarrow\infty$, with $\tau_{*}$ arbitrary but fixed. 
In a rather benign abuse of notation we will refer to $S_{\tau_{*},r_{*}}$
as simply $S$, and `take the limit as $r\rightarrow\infty$ with $\tau$
fixed'.  The metric induced on $S$ will, as usual, be denoted as
$\sigma_{ab}$ (in abstract index notation).

Now assume that the functions $\tau$ and $r$ have been chosen such that 
$(S,\sigma)$ tends to a round sphere at infinity.  Thus the components
of its metric in spherical coordinates $x^{i}=(\theta,\phi)$ have an
asymptotic expansion of the form
\beq
\sigma_{ij}=
r^2\left(
\begin{array}{cc}
   1 & 0 \\
   0 & \sin^{2}\theta
\end{array}
\right) +2r\left(
\begin{array}{cc}
   X & Y\sin\theta \\
   Y\sin\theta & Z\sin^{2}\theta
\end{array}
\right)+O_{<}(r).
\label{two-sphere}
\eeq
In this expansion, $X$, $Y$, and $Z$ are each arbitrary functions of
$\tau$, $\theta$, and $\phi$.  The symbol
$O_{<}(r^{-n})$ denotes a term that falls off {\it faster} (or grows more
slowly, depending on the sign of $n$) than $r^{-n}$, but not necessarily
according to a power of $r$.  For example, rather than $O(1)$, the
remainder term $O_{<}(r)$ might grow as
$\ln r$.  The motivation for this increased generality will
be explained below when we consider the large sphere limit at null
infinity.  Furthermore, we can choose (the function) $r$ to be an areal
radius, in which case we may take $\sqrt{\sigma}=r^{2}\sin\theta$, where
$\sigma=\det\sigma_{ij}$.  It is easy to see
that this requires $Z=-X$ in Eq.~(\ref{two-sphere}).  

The scalar curvature of a round sphere of areal radius $r$ is $2/r^2$.  
Since the metric in Eq.~(\ref{two-sphere}) differs from that of a round
sphere by a term one power lower in $r$, we immediately have that its
scalar curvature ${\cal R}$ is given by
\beq
{\cal R}={2\over{r^2}}+{{\Delta_{\cal R}}\over{r^3}},
\label{R}
\eeq
where the remainder term $\Delta_{\cal R}$ is of order one.\footnote{
In Ref.~\cite{BLY2} it is shown that $\Delta_{\cal R}={\cal D}\cdot
v+O_{<}(1)$ for some vector field $v$ in $S$.  In other words, to leading
order $\Delta_{\cal R}$ is a divergence.  This plays a crucial role
in some of the results in Ref.~\cite{BLY2}.  However, we will not need to
use this fact, except to provide some insight into our discussion of the
`solution' of the Ricci embedding equation in Sec.~5.2.}  In our
asymptotically flat spacetime the
components of the Riemann tensor fall off as $1/r^3$, and so the same will
be true of $\sigma\sigma R$, the sectional curvature of $(S,\sigma)$.  In
the present context the appropriate reference spacetime is Minkowski
space, and so $\sigma\sigma R^{\rm ref}=0$.  The only other terms to
consider in Eqs.~(\ref{QIM1}) and (\ref{reference_QIM}) are the shear
terms, $(\tilde{k}^{2}-\tilde{l}^{2})$ and 
$(\tilde{k}^{2}-\tilde{l}^{2})^{\rm ref}$.  We will see below that
these, too, fall off at least as fast as $1/r^3$ in both the spatial and
null infinity limits.  In the large sphere limit the unreferenced IQE thus
behaves as
\begin{eqnarray}
{\rm IQE}^{\rm unref}&=&
-{1\over{8\pi}}\,\int_{S}\,dS\,\sqrt{2\left[{2\over{r^2}}+
{{\Delta_{\cal R}}\over{r^3}} - \sigma\sigma R + (\tilde{k}^2
-\tilde{l}^2 )\right]}\nonumber\\
&=&
-{1\over{8\pi}}\,\int_{S}\,dS\,{2\over r}\left\{1+{{r^{2}}\over 4}
\left[{{\Delta_{\cal R}}\over{r^3}} - \sigma\sigma R + (\tilde{k}^2
-\tilde{l}^2 )\right]+O(r^{-2})\right\}.
\end{eqnarray}
The reference IQE behaves similarly, except we have $\sigma\sigma R^{\rm
ref}=0$.  Thus
\beq
{\rm IQE}^{\rm ref}=
-{1\over{8\pi}}\,\int_{S}\,dS\,{2\over r}\left\{1+{{r^{2}}\over 4}
\left[{{\Delta_{\cal R}}\over{r^3}} - 0 + (\tilde{k}^2
-\tilde{l}^2 )^{\rm ref}\right]+O(r^{-2})\right\}.
\eeq
In forming the difference of the previous two expressions it is important
to observe that not only do the divergent terms coming from the $2/r^2$
piece of $\cal R$ cancel, but also the (finite) remainder terms
$\Delta_{\cal R}$ are the same in both, and thus also cancel, independent
of what $\Delta_{\cal R}$ is.
Thus we find that the large sphere behavior of the (referenced) IQE is
given by
\beq
{\rm IQE} = {1\over{16\pi}}\int_{S}\,dS\,r\,[\sigma\sigma R -
(\tilde{k}^{2}-\tilde{l}^{2}) + (\tilde{k}^{2}-\tilde{l}^{2})^{\rm ref} +
O(r^{-4})].
\label{largeSQIM}
\eeq
Of course if the shear terms fall off as $1/r^4$ or faster they get
absorbed into the $O(r^{-4})$ remainder term.  It is worth
emphasizing that in the large sphere limit the square root in the IQE is
eliminated by the fact that $\cal R$ dominates over the other terms.  The
areal radius factor $r$ outside the brackets in Eq.~(\ref{largeSQIM}) is
really 
$\sqrt{2/{\cal R}}$.\footnote{\label{mechanism}This mechanism works even
when the sphere is not asymptotically round.  In this case the shear terms
contribute at a higher order, viz. $1/r^2$, in an effort to keep what is
under the square root sign positive, as discussed earlier.  In other
words, the factor $\sqrt{2/{\cal R}}$ is modified in such a way that
negative $\cal R$ is likely not a problem.  We will not consider this
more complicated case here.}  In Sec.~6 we will see a similar mechanism at
work in the small sphere limit.  But in the intermediate regime the IQE
is, in general, an integral of the difference of two radicals.

As a quick check of Eq.~(\ref{largeSQIM}) let us evaluate the right hand
side for the Schwarzschild geometry.  In the usual Schwarzschild
coordinates $r$ is an areal radius, and it is a simple exercise to compute
the sectional curvature of a $t,r={\rm constant}$ two-sphere.  The result
is: $\sigma\sigma R=4M/r^3$.  The shear terms obviously vanish, and with
$dS=r^{2}d\Omega$ ($d\Omega$ the measure on the unit round sphere) one
immediately gets ${\rm IQE}=M$, the ADM mass of the black hole.  Now
the main task is to investigate in detail the shear terms in 
Eq.~(\ref{largeSQIM}), which we will do separately for the spatial and
null infinity limits, respectively.

\subsection{The spatial infinity limit}

Rather than proceed with complete generality, it is more instructive to
consider an asymptotically flat metric that exhibits angular momentum
explicitly, and then see how this angular momentum works its way into the
shear terms.  (We will be completely general in the more interesting null 
infinity limit case.)  The spacetime far from any isolated
stationary (nonradiating) rotating source is described asymptotically by
the Kerr metric (see Secs.~19.3 and 33.3 of Ref.~\cite{MTW}), so let us
take $(M,g)$ to be the Kerr spacetime.  We choose the following basis of
orthonormal one-forms:
\beq
e^{0}=N\,dt,\;\;e^{1}={{\rho}\over{\sqrt{\Delta}}}\,dr,\;\;
e^{2}=\rho\,d\theta,\;\;e^{3}=\sqrt{g_{\phi\phi}}(d\phi-\omega\,dt),
\label{KerrOneForms}
\eeq
which are associated with locally nonrotating observers.  The corresponding 
basis of orthonormal vector fields is
\beq
e_{0}={1\over N}(\partial_{t}+\omega\partial_{\phi}),\;\;
e_{1}={\sqrt{\Delta}\over\rho}\partial_{r},\;\;
e_{2}={1\over\rho}\partial_{\theta},\;\;
e_{3}={1\over\sqrt{g_{\phi\phi}}}\partial_{\phi}.
\label{KerrVectors}
\eeq
The notation used is standard: $x^{a}=(t,r,\theta,\phi)$ are
Boyer-Lindquist
coordinates, $\omega(r,\theta)=-g_{t\phi}/g_{\phi\phi}$ is an observer's
angular velocity as measured from infinity, $N=\sqrt{\omega^2
g_{\phi\phi}-g_{tt}}$ is the lapse function, etc. (see, e.g.,
Sec.~33.4 of Ref.~\cite{MTW}).  Let $A,B,\ldots$ be indices labeling the 
basis vectors and one-forms, ranging from 0 to 3, and $I,J,\ldots$ denote
the subset of these taking values 2 and 3.  These indices are raised and
lowered with the flat Lorentz metric $\eta_{AB}=\eta^{AB}={\rm
diagonal}(-1,1,1,1)$.   

The vector fields $e_{I}^{\;\;a}$ are tangent to
any $r,t={\rm constant}$ two-sphere $S$, and so $u^{a}:=e_{0}^{\;\;a}$ and
$n^{a}:=e_{1}^{\;\;a}$ are, respectively, timelike and spacelike unit
vectors orthogonal to $S$.  From Eqs.~(\ref{extrinsic_curvatures}) we see
that the orthonormal basis components of the extrinsic curvatures $l_{ab}$
and $k_{ab}$ are given by
\begin{eqnarray}
l_{IJ}&=&e_{I}^{\;\;a}e_{J}^{\;\;b}\nabla_{a}u_{b}=-\omega_{0JI},
\\
k_{IJ}&=&e_{I}^{\;\;a}e_{J}^{\;\;b}\nabla_{a}n_{b}=-\omega_{1JI}.
\end{eqnarray}
Here $\omega_{CBA}=-e_{A}^{\;\;a}e_{B}^{\;\;b}\nabla_{a}e_{Cb}$
are Ricci rotation coefficients.  Working out these 
coefficients\footnote{\label{omega_calc}The easiest way to do this is to
recognize that, with $Z=0$ or 1, $\omega_{ZJI}=\alpha_{(IJ)Z}$, where
$\alpha^{C}_{\;\;AB}=i_{e_B}i_{e_A}\,de^{C}$.  Thus we need only compute
the exterior derivative of $e^{I}$.} I find that the trace-free parts of
$l_{IJ}$ and $k_{IJ}$ are given by 
\begin{eqnarray}
\tilde{l}_{IJ}&=&\alpha\left(
\begin{array}{cc}
   0 & 1 \\
   1 & 0
\end{array}\right), \;\;\;\;{\rm where}\;\;
\alpha={{\sqrt{g_{\phi\phi}}}\over{2N\rho}}\partial_{\theta}\omega,
\label{alpha}
\\
\tilde{k}_{IJ}&=&\beta\left(
\begin{array}{cc}
   1 & 0 \\
   0 & -1
\end{array}\right), \;\;\;\;{\rm where}\;\;
\beta={{\sqrt{\Delta}}\over
{2\rho}}\partial_{r}\ln{{\rho}\over{\sqrt{g_{\phi\phi}}}}.
\end{eqnarray}
Geometrically, the coefficient $\alpha$ is just
$e_{1}\cdot\Omega^{\rm (precess)}$, i.e., the radial component of
the angular velocity vector $\Omega^{\rm (precess)}$ that measures the
precession of a gyroscope carried by a locally nonrotating observer,
relative to the observer's orthonormal frame (see Eq.~(33.24) of
Ref.~\cite{MTW}; compare also with Eq.(\ref{frame_dragging}) above, and
the discussion following it).  Thus, the unreferenced shear term in
Eq.~(\ref{largeSQIM}) is given by
\beq
(\tilde{k}^2-\tilde{l}^2)=2(\beta^2-\alpha^2)
={{a^4}\over{2r^6}}{\sin}^{4}\theta+O\left({1\over{r^7}}\right),
\label{ShearTermSpatial}
\eeq
where the last expression on the right hand side is the large $r$
asymptotic expansion.  So clearly, being of order $1/r^{6}$, 
$(\tilde{k}^2-\tilde{l}^2)$ does not contribute to the large sphere limit
of the IQE at spatial infinity.

What about the reference term $(\tilde{k}^2-\tilde{l}^2 )^{\rm ref}$?  It
is plausible that the reference term is of the same order in $1/r$ as the
unreferenced term, viz. $1/r^6$, or less, and so also does not contribute.
However, to be certain one needs to solve the embedding equations
(\ref{ref_Gauss}-\ref{ref_Ricci}), subject to the condition
${\cal F}^{\rm ref}={\cal F}$, as argued in Sec.~4.  We will not attempt
to do so here, but it is instructive to at least work out what $\cal F$ is
for the Kerr geometry.  From Eq.~(\ref{A}) we see that the orthonormal
basis components of the connection in the normal bundle are given by
\beq
A_{I}=e_{I}^{\;\;b}n^{c}\nabla_{b}u_{c}=-\omega_{01I}.
\label{spatialAI}
\eeq
Evaluating these Ricci rotation coefficients reveals that the
one-form $A=A_{I}e^{I}$ (pulled back to the two-sphere $S$) is given by
\beq
A=\gamma\,d\phi,\;\;\;\;{\rm where}\;\;
\gamma={{g_{\phi\phi}\sqrt{\Delta}}\over{2N\rho}}\partial_{r}\omega
=-{{3aM}\over{r^2}}{\sin}^{2}\theta+O\left({1\over r^4}\right).
\label{AforKerr}
\eeq
Recall that $\omega=\omega(r,\theta)$ is a measure of the frame dragging
produced by the rotating geometry.  While the shear in the time direction 
measures the $\theta$ dependence of $\omega$ (see $\alpha$ in 
Eq.~(\ref{alpha})), Eq.~(\ref{AforKerr}) shows that the connection in the
normal bundle measures its $r$ dependence.  Both are measures of angular
momentum.

Calculating the exterior derivative of $A$ leads to the curvature in
the normal bundle:
\beq
{\cal F}={{1}\over{\sqrt{g_{\phi\phi}}\rho}}\partial_{\theta}\left(
{{g_{\phi\phi}\sqrt{\Delta}}\over{2N\rho}}\partial_{r}\omega\right)
=-{{6aM}\over{r^4}}\cos\theta+O\left({{1}\over{r^6}}\right),
\label{KerrF}
\eeq
which is of order $1/r^4$.  Inspection of the Ricci equation (\ref{ref_Ricci}),
or Eq.~(\ref{Ricci}) with the left hand side set to zero, reveals that a
solution to the reference embedding equations, subject to
Eq.~(\ref{F=F_ref}), requires that $\tilde{l}^{\rm ref}_{IJ}$ and
$\tilde{k}^{\rm ref}_{IJ}$ be two matrices whose commutator is of order
$1/r^4$.  On the other hand, one might guess that the trace of the
difference of their squares, $(\tilde{k}^{2}-\tilde{l}^{2})^{\rm ref}$,
might be of order $1/r^6$, as suggested above.  It is not difficult to
convince oneself that these two conditions are not incompatible, so the
reference embedding equations at least do not obviously forbid the
reference shear term $(\tilde{k}^{2}-\tilde{l}^{2})^{\rm ref}$ from being
of the same order of magnitude as $(\tilde{k}^{2}-\tilde{l}^{2})$, such
that neither contributes to the IQE.

In any case, assuming just that $(\tilde{k}^{2}-\tilde{l}^{2})^{\rm ref}$ 
is at most $O_{<}(r^{-3})$, which is almost certainly true, we find that
the large sphere limit of the IQE at spatial infinity is given by
\beq
\lim_{r\rightarrow\infty}{\rm IQE} =
\lim_{r\rightarrow\infty}{1\over{16\pi}}\int_{S}\,dS\,r\,\sigma\sigma R.
\label{largeSQIMspatial}
\eeq
This limit of the IQE thus has a simple geometrical interpretation: apart
from a factor proportional to the areal radius $r$, it is just the
average over $S$ of the sectional curvature of $(S,\sigma)$ as embedded
in the physical spacetime $(M,g)$.  

Now let us assume that $(M,g)$ is vacuum ($R_{ab}=0$) near spatial
infinity, so that there the Riemann tensor reduces to the Weyl tensor,
$C_{abcd}$.  From the definition of the two-surface metric given in
Eq.~(\ref{spatialmetric}), and the fact that the Weyl tensor is traceless,
one immediately gets
\beq
\sigma\sigma R = 2E_{ab}n^{a}n^{b}.
\label{coulomb}
\eeq
Thus the sectional curvature of $(S,\sigma)$ is just (twice) the
radial-radial (Coulomb) component of the electric part of the Weyl
tensor, $E_{ab}:= -C_{acbd}u^{c}u^{d}$.  Inserting this result into
Eq.~(\ref{largeSQIMspatial})
we see that in this limit the IQE is precisely the coordinate-independent
expression of the ADM mass given by Ashtekar and 
Hansen~\cite{AH}.\footnote{In Ref.~\cite{AH} a different
definition of $E_{ab}$ is used, namely $E_{ab}=C_{acbd}n^{c}n^{d}$, but
accounting for this difference in notation the two results agree.}

One more remark is in order here: Hayward's work on quasilocal
energy~\cite{Hayw} resembles what is done here in the sense that the Gauss
embedding equation plays a central role, and that the analysis is boost
invariant in spirit, i.e., no reference is made to a spacelike
three-surface spanning $S$, with its attendant preferred time direction on
$S$, and so on.  However, Hayward's quasilocal energy is distinct
from the IQE here: it does not involve a square root.  Basically,
Hayward starts with an integral over $S$ of the `2+2' Hamiltonian density,
which yields a dimensionless quantity, and then multiplies this quantity 
by the areal radius of $S$ to correct this `defect', i.e., give it the
dimensions of energy.  This is in the same spirit as the areal radius
factor appearing in the Hawking mass~\cite{Hawk}.   For the large sphere
limit at spatial infinity Hayward
arrives at the same result given in Eq.~(\ref{largeSQIMspatial}), except 
with $r$ `outside the integral', so to speak.  His quasilocal energy has
the very appealing feature of not requiring a reference subtraction term,
at least when the sectional curvature $\sigma\sigma R$ falls off as
$1/r^3$ (however it diverges if, e.g.,  the spacetime is asymptotically
anti-de Sitter space---recall Eq.~(\ref{ssRref})).  In our case,
the square root in Eq.~(\ref{QIM1}) ensures that the IQE has the
dimensions of energy, but the price paid is that a reference subtraction
term is needed.  (Without the square root the large sphere limit of the
unreferenced IQE would just be (negative) the Euler number of $S$, which
is finite, but carries no information about energy.)  The areal radius
factor $r$ in Eq.~(\ref{largeSQIMspatial}) appears `inside the
integral': as mentioned before, it arises from the dominant scalar
curvature term
$\cal R$, and is really $\sqrt{2/{\cal R}}$.  I emphasize that this is a
geometrically natural mechanism---$r$ is not put in by hand.  Finally,
while one might feel that there is something unattractively ad hoc about a
reference subtraction term, the flexibility it affords makes it possible
to deal with the wide range of boundary conditions possible in general
relativity.  In Sec.~7 we will consider an interesting example
of this.

\subsection{The null infinity limit}

We now suppose that the physical spacetime $(M,g)$ is asymptotically flat
at future null infinity.  As in the previous subsection we begin with the
generic large sphere form of the IQE given in Eq.~(\ref{largeSQIM}), 
except now we take the large $S$ limit in the future null direction.  More
precisely, on $M$ we introduce the Bondi coordinates~\cite{Bond}
$x^{a}=(w,r,\theta,\phi)$, and as before, denote the subset
$(\theta,\phi)$ of spherical coordinates by $x^i$.
The retarded time $w$ labels a one-parameter family of
outgoing null hypersurfaces, and $r$ is an areal radius (luminosity
parameter) along the outgoing null geodesic generators of these
hypersurfaces.  The $w,r={\rm constant}$ surfaces are topologically
two-spheres, any one of which we denote as $S$.  This setup is the same as
discussed at the beginning of Sec.~5, where $w$ here is what we there 
called the `time' coordinate, $\tau$.  We are interested in the
one-parameter family of two-spheres $S$ in the limit as
$r\rightarrow\infty$, with $w$ arbitrary but fixed.

In the Bondi coordinates our asymptotically flat metric takes the
standard form~\cite{Sach,CMS}
\beq
g_{ab}\,dx^{a}\,dx^{b}=
-UV\,dw^{2}-2U\,dw\,dr+\sigma_{ij}(dx^{i}+W^{i}\,dw)(dx^{j}+W^{j}\,dw).
\label{BondiMetric}
\eeq
We assume the following expansions for the various terms in
this metric:\footnote{We are following closely the notation used in
Ref.~\cite{BLY2}, as well as the spirit of the discussion in their
footnote 2.  The meaning of the notation $O_{<}(r^{-n})$ was described
following Eq.~(\ref{two-sphere}).  The motivation for this level of
generality is that Chru\'{s}ciel {\it et al}~\cite{CMS} have recently
shown that one can allow `polyhomogeneous' terms of the form
$r^{-n}{\ln}^{m}r$ in these expansions and still have a consistent
framework for solving the Bondi-Sachs-type characteristic initial value
problem.  Allowing only expansions in powers of inverse $r$ is tantamount
to Sachs' `outgoing radiation condition'~\cite{Sach}, which they argue is
overly restrictive.  However, besides making the calculations `tighter' as
regards remainder terms, and slightly more general, we would get the same
results had we assumed Sachs' outgoing radiation condition.} 
\begin{eqnarray}
V&=&1-2mr^{-1}+O_{<}(r^{-1}), \label{V}
\\
U&=&1-{1\over 2}(X^{2}+Y^{2})r^{-2}+O_{<}(r^{-2}),
\label{U}
\\
W^{\theta}&=&
(2X\cot\theta+\partial_{\theta}X+\csc\theta\partial_{\phi}Y)r^{-2}
+O_{<}(r^{-2}), \nonumber 
\\
W^{\phi}&=&
\csc\theta\,(2Y\cot\theta+\partial_{\theta}Y-\csc\theta\partial_{\phi}X)
r^{-2}+O_{<}(r^{-2}),
\label{W}
\\
\sigma_{ij}&=&
r^{2}\left(
\begin{array}{cc}
1 & 0 \\
0 & {\sin}^{2}\theta
\end{array}\right)
+2r\left(
\begin{array}{cc}
X & Y\sin\theta \\
Y\sin\theta & -X{\sin}^{2}\theta
\end{array}\right)
+O_{<}(r).
\label{sigma}
\end{eqnarray}
The function $V$ contains the mass aspect, 
$m(w,\theta,\phi)$.  Observe that the metric on $S$ is of the same form
given in Eq.~(\ref{two-sphere}) (with $Z=-X$ because $r$ is an areal
radius here), except now $X(w,\theta,\phi)$ and $Y(w,\theta,\phi)$
have a significant physical interpretation: they are the real and
imaginary parts of Sachs' complex asymptotic shear $c=X+iY$~\cite{Sach}.
Thus the scalar curvature of $(S,\sigma)$ will be given by
Eq.~(\ref{R}), and we can begin our discussion of the IQE at
Eq.~(\ref{largeSQIM}).  Our first task is to compute the unreferenced
shear term, $(\tilde{k}^{2}-\tilde{l}^{2})$.

Inspecting the metric in Eq.~(\ref{BondiMetric}), we choose the following
basis of one-forms:
\beq
e^{-}={1\over2}U\,dw,\;\;
e^{+}=dr+{1\over2}V\,dw,\;\;
e^{I}=\gamma^{I}_{\;\;i}(dx^{i}+W^{i}\,dw),
\label{BondiOne-Forms}
\eeq
where indices $I,J,\ldots$ take the values 2 and 3, and
$\gamma^{I}_{\;\;i}$
is defined by demanding that 
$\sigma_{ij}=\delta_{IJ}\gamma^{I}_{\;\;i}\gamma^{J}_{\;\;j}$.  
A suitable choice for $\gamma^{I}_{\;\;i}$ is given by
\beq
\gamma^{I}_{\;\;i}=\left(
\begin{array}{cc}
r+X & 0 \\
2Y & (r-X)\sin\theta
\end{array}\right)+O_{<}(1).
\eeq
In this matrix expression, $I$ ($i$) is a row (column) index.  Let the
indices $A,B,\ldots$ take values in the set $\{-,+,I\}$.  Then the metric
is given by $g_{ab}=\eta_{AB}e^{A}_{\;\;a}e^{B}_{\;\;b}$, where 
\beq
\eta_{AB}=\left(
\begin{array}{cccc}
 0 & -2 & 0 & 0 \\
-2 &  0 & 0 & 0 \\
 0 &  0 & 1 & 0 \\
 0 &  0 & 0 & 1 
\end{array}\right).
\eeq
This matrix, and its inverse, $\eta^{AB}$, are used to raise and lower the
the basis indices.  The vector fields dual to the one-forms in 
Eq.~(\ref{BondiOne-Forms}) are given by $e_{A}^{\;\;\;a}=\eta_{AB}g^{ab}
e^{B}_{\;\;b}$, or explicitly:
\beq
e_{-}={2\over
U}\left(\partial_{w}-{1\over2}V\partial_{r}-W^{i}\partial{i}\right),\;\;
e_{+}=\partial_{r},\;\;
e_{I}=\gamma_{I}^{\;\;i}\partial_{i},
\label{BondiVectors}
\eeq
where $\gamma_{I}^{\;\;i}$ is defined by
$\gamma_{I}^{\;\;i}=\delta_{IJ}\sigma^{ij}\gamma^{J}_{\;\;j}$,
$\sigma^{ij}$ being the inverse of the matrix $\sigma_{ij}$.

Inspection of $e_{I}$ in Eq.~(\ref{BondiVectors}) shows that these vectors
are tangent to $S$, and so $e_{\pm}$ are two null normals to $S$.  Since
their normalization is such that $e_{+}\cdot e_{-}=-2$, we can set
$\xi_{\pm}^{a}=e_{\pm}^{\;\;a}$, where $\xi_{\pm}^{a}$ was previously
defined by $\xi_{\pm}^{a}:=u^{a}\pm n^{a}$ (see Eq.~(\ref{null_expansions})).
Thus, from Eqs.~(\ref{extrinsic_curvatures}) we have the following result, 
in basis components:
\beq
l_{IJ}\pm k_{IJ}=e_{I}^{\;\;a}e_{J}^{\;\;b}\nabla_{a}e_{\pm b}=
-\omega_{\pm JI},
\label{lpmk}
\eeq
Working out the required Ricci rotation coefficients\footnote{See the
footnote on page~\pageref{omega_calc}, with $Z=\pm$.} I find
\beq
l_{IJ}+k_{IJ}=B_{IJ},\;\;
l_{IJ}-k_{IJ}={2\over U}\left(A_{IJ}-{V\over 2}B_{IJ}-{\cal D}_{(I}W_{J)}
\right),
\label{lpmkdetail}
\eeq
where
\begin{eqnarray}
A_{IJ}&=&\gamma_{(I}^{\;\;\;\;i}\dot{\gamma}_{J)i}=
{1\over r}\left(
\begin{array}{cc}
\dot{X} &  \dot{Y} \\
\dot{Y} & -\dot{X}
\end{array}\right)+O_{<}(r^{-1}),
\label{AIJ}\\
B_{IJ}&=&\gamma_{(I}^{\;\;\;\;i}\gamma^{\prime}_{J)i}=
{1\over r}\left(
\begin{array}{cc}
1 & 0 \\
0 & 1
\end{array}\right)-
{1\over {r^2}}\left(
\begin{array}{cc}
X &  Y \\
Y & -X
\end{array}\right)+O_{<}(r^{-2}).
\label{BIJ}
\end{eqnarray}
Here we use an overdot (prime) to denote differentiation with respect to
$w$ ($r$).  Observe that ${\cal D}_{(I}W_{J)}$ in Eq.~(\ref{lpmkdetail}) 
is of order $1/r^2$.  Taking the trace-free part of
Eqs.~(\ref{lpmkdetail}) gives us
the basis components of the shears in the two null directions: $s_{\pm\,IJ}=
\tilde{l}_{IJ}\pm\tilde{k}_{IJ}$.  Explicitly:
\beq
s_{+\,IJ}=-{1\over r^2}\left(
\begin{array}{cc}
X &  Y \\
Y & -X
\end{array}\right)+O_{<}(r^{-2}),\;\;
s_{-\,IJ}={2\over r}\left(
\begin{array}{cc}
\dot{X} &  \dot{Y} \\
\dot{Y} & -\dot{X}
\end{array}\right)+O_{<}(r^{-1}).
\label{BondiShears}
\eeq
Thus the unreferenced shear term in Eq.~(\ref{largeSQIM}) is given by
\beq
(\tilde{k}^{2}-\tilde{l}^{2})\equiv -s_{+\,IJ}s_{-}^{\;\;IJ}
={4\over{r^3}}(X\dot{X}+Y\dot{Y})+O_{<}(r^{-3})
\label{BondiShearTerm}
\eeq
(cf. Eq~(\ref{RefShearTerm})).  

We thus learn that, in contrast to the spatial infinity limit (see
Eq.~(\ref{ShearTermSpatial})), in the null infinity limit the unreferenced
shear term is of order $1/{r^3}$, and so {\it does} contribute to the
IQE.  We will argue below that the reference shear term
$(\tilde{k}^{2}-\tilde{l}^{2})^{\rm ref}$ is also of order $1/r^3$, but
that it is a total derivative and therefore does not contribute.  So as
not to interrupt the flow or our discussion, for the moment let us assume
this is true, in which case Eq.~(\ref{largeSQIM}) becomes
\beq
\lim_{r\rightarrow\infty}{\rm IQE} =
\lim_{r\rightarrow\infty}{1\over{16\pi}}\int_{S}\,dS\,r\,
\left[\sigma\sigma R-{4\over{r^3}}(X\dot{X}+Y\dot{Y})\right].
\label{largeSQIMnull}
\eeq
Because of the $c\dot{\bar{c}}+\dot{c}\bar{c}=2(X\dot{X}+Y\dot{Y})$ term,
this result looks like it could very well be the Bondi-Sachs
mass~\cite{Sach}.  To see that in fact it is, a straightforward
calculation of the Riemann tensor of $g_{ab}$ projected into $S$ gives the
following sectional curvature of $(S,\sigma)$:
\beq
\sigma\sigma R = {4\over{r^3}}(m+X\dot{X}+Y\dot{Y})+O{<}(r^{-3}).
\label{ssRBondi}
\eeq
Inserting this result into Eq.~(\ref{largeSQIMnull}) we see that the
shear terms cancel, leaving only the mass aspect, $m$:
\beq
\lim_{r\rightarrow\infty}{\rm IQE} =
{1\over{4\pi}}\int_{S}\,d\Omega\,m(w,\theta,\phi).
\label{BondiMass}
\eeq
In obtaining this expression, recall that because $r$ is an areal radius
we can (and did) take $\sqrt{\sigma}=r^{2}\sin\theta$, and so
$dS=r^{2}\,d\Omega$, where $d\Omega=\sin\theta\,d\theta\,d\phi$ is the
measure on the unit round sphere.\footnote{On this note, it might be
helpful to point out an important detail in the calculation of the
sectional curvature given in Eq.~(\ref{ssRBondi}).  One of the the terms
that arises in the calculation is the trace of $A_{IJ}$ in 
Eq.~(\ref{AIJ}), which appears to be of order $O_{<}(r^{-1})$.  If this 
were so it would be problematic.  But in fact it is zero, because
$A_{I}^{\;\;I}=\gamma_{I}^{\;\;i}\dot{\gamma}^{I}_{\;\;i}=
(1/2)\dot{\sigma}/\sigma=0$, since $\sigma=r^{4}{\sin}^{2}\theta$ 
does not depend on $w$.}  Thus the future null infinity limit of the IQE
is the Bondi-Sachs mass~\cite{Sach}.

Now there is an important lesson to be learned from this result.  The
unreferenced shear term $(\tilde{k}^{2}-\tilde{l}^{2})$ is solely
responsible for producing the all-important $c\dot{\bar{c}}+\dot{c}\bar{c}$ 
term that accounts for the mass loss due to gravitational
radiation.  Hence this term is {\it necessary} under the square root in
Eq.~(\ref{QIM1}), and so there is no natural way to avoid 
$(\tilde{k}^{2}-\tilde{l}^{2})^{\rm ref}$ in ${\rm IQE}^{\rm ref}$, and
its attendant embedding problem.  Moreover, we learn that these shear
terms are not only associated with angular momentum, as I have been
stressing, but also encode information about gravitational
radiation.  We will see precisely the same phenomenon emerge in the small
sphere limit in Sec.~6.  Furthermore, it is emphasized in Ref.~\cite{Gold}
that it is easy to construct, {\it ab initio}, an integral expression
involving the Riemann tensor (e.g., an integral of $\sigma\sigma R$ over
$S$) that is conserved under certain circumstances.  One is thus tempted
to interpret such a conserved quantity as an energy.  However, such
attempts fail to produce, in the null infinity limit, the crucial
null-surface-dependent shear terms seen in Eq.~(\ref{largeSQIMnull}), and it 
is difficult to see how to modify them in a {\it covariant} way to produce
these terms~\cite{Gold}.  The shear term $(\tilde{k}^{2}-\tilde{l}^{2})$
is precisely such a covariant modification.  Moreover, it arises
naturally from simply replacing the Brown-York $k$ with the boost
invariant quantity $\sqrt{k^2 -l^2 }$.  (Of course a similar
observation can be made concerning, say, the Hawking mass~\cite{Hawk},
which has the same large sphere limit as in Eq.~(\ref{largeSQIM}), except
without the {\it reference} shear term.)

These clean results rely on our assumption that the reference shear term
$(\tilde{k}^{2}-\tilde{l}^{2})^{\rm ref}$ does not contribute to the null
infinity limit of the IQE.  I claimed above that this is so because it is
a total derivative.  To prove this would require solving the embedding
equations (\ref{ref_Gauss}-\ref{ref_Ricci}), which we know is a very
difficult task.  However, I will now present a heuristic solution of 
the Ricci equation that leads to a substantiation of this
claim.  Moreover, we will see how demanding ${\cal F}^{\rm ref}={\cal F}$
plays a crucial role in achieving this result, which provides our first
bit of indirect but concrete evidence that this condition is required to
properly account for angular momentum (and as we now see, also
gravitational radiation).

To begin we need to calculate the connection in the normal bundle,
$A_{a}$, and then its corresponding curvature, $\cal F$.  Proceeding as we
did in the spatial infinity case (see Eq.~(\ref{spatialAI})) we find that
the basis components of $A_{a}$ are given by
\beq
A_{I}={1\over2}\omega_{+-I}={1\over{rU}}W_{I}+{1\over2}e_{I}(\ln U),
\label{nullAI}
\eeq
where $e_{I}(\ln U)$ denotes the derivative of $\ln U$ along the
vector field $e_{I}$.  This term is pure gauge.  As for the other term,
since we only know $W_{I}$ to leading order we can put $U=1$ here---see
Eqs.~(\ref{U}) and (\ref{W}).  Thus, up to a gauge transformation, the
connection one-form $A$ is just 
($1/r$ times) the one-form $W_{i}\,dx^{i}$.  And the curvature
is thus proportional to the curl of $W$:
\beq
{\cal F}={2\over r}\epsilon^{IJ}{\cal D}_{I}W_{J}=
{2\over r}({\cal D}_{2}W_{3}-{\cal D}_{3}W_{2}).
\label{BondiF}
\eeq
Keep in mind that the numerical indices here refer to basis components,
not coordinate components.  It is easy to see that $\cal F$ is of order
$1/r^3$.  It is interesting to compare $\cal F$ with the scalar curvature
$\cal R$, whose form was given in Eq.~(\ref{R}).  Using the metric in
Eq.~(\ref{sigma}) it is not difficult to evaluate the remainder term,
$\Delta_{\cal R}$.  The net result is~\cite{BLY2}:
\beq
{\cal R}={{2}\over{r^{2}}}+{{2}\over{r}}{\cal D}\cdot W+O_{<}(r^{-3}),
\label{BondiR}
\eeq
Note from Eq.~(\ref{W}) that $W^i$ is of order $1/r^2$, so the term above
involving $W$ is, indeed, of order $1/r^3$, as it should be.
Thus we see that $\cal R$ is associated with the
divergence of $W$, and $\cal F$ with its curl.  This is an explicit 
example of a point made earlier, namely that both curvatures are on the
same geometrical footing:  To capture the two pieces of information in
$W$---its divergence and its curl---requires precisely both $\cal R$ and
$\cal F$.  

Let us now turn to the null shears of $S$ as they appear in the
physical spacetime, Eq.~(\ref{BondiShears}).  The form of these shears
suggests we make the following ansatz for the null shears of $S$ in the
reference (Minkowski) spacetime:
\beq
s_{+\,IJ}^{\rm ref}={1\over r^2}\left(
\begin{array}{cc}
\alpha &  \beta \\
\beta & -\alpha
\end{array}\right)+O_{<}(r^{-2}),\;\;
s_{-\,IJ}^{\rm ref}={1\over r}\left(
\begin{array}{cc}
\gamma &  \delta \\
\delta & -\gamma
\end{array}\right)+O_{<}(r^{-1}).
\label{refBondiShears}
\eeq
Observe that we might expect the pair $(\alpha,\beta)$ to play a role
distinct from the pair $(\gamma,\delta)$.  Comparing
Eqs.~(\ref{refBondiShears}) and (\ref{BondiShears}) suggests that $\alpha$
and $\beta$ will be like $X$ and $Y$ in that they have something to do
with the {\it intrinsic} geometry of $(S,\sigma)$.  The `more important'
terms will be $\gamma$ and $\delta$, because they occur at the
dominant power of inverse $r$.  Also, we expect them to be
related to the {\it extrinsic} geometry of $S$, since their counterparts,
$\dot{X}$ and $\dot{Y}$, measure how $\sigma_{ij}$ changes as a
function of the retarded time $w$---they are the two `news' 
functions~\cite{Sach}.

With these observations in mind, we will now `solve' the Ricci embedding
equation, (\ref{ref_Ricci}).  We first compute (in basis components)
\beq
-{1\over2}\epsilon^{I}_{\;\;J}[s_{+}^{\rm ref},s_{-}^{\rm
ref}]^{J}_{\;\;I}
={2\over{r^3}}(\alpha\delta-\beta\gamma)={2\over{r^3}}\det\left(
\begin{array}{cc}
\alpha & \beta \\
\gamma & \delta
\end{array}\right).
\label{Ricci1}
\eeq
Next we impose the condition ${\cal F}^{\rm ref}={\cal F}$, and observe
that ${\cal F}$ in Eq.~(\ref{BondiF}) can also be expressed as a
`determinant', i.e.,
\beq
{\cal F}^{\rm ref}={\cal F}={2\over r}\det\left(
\begin{array}{cc}
{\cal D}_{2} & {\cal D}_{3} \\
       W_{2} &        W_{3}
\end{array}\right)
\eeq
The Ricci embedding equation instructs us to equate the two previous 
determinant expressions.  One solution is to make the
identifications $\alpha\leftrightarrow r{\cal D}_{2}$ and
$\beta\leftrightarrow r{\cal D}_{3}$ (which are consistent with our
expectation that $\alpha$ and $\beta$ be associated with intrinsic
geometry), together with $\gamma\leftrightarrow rW_{2}$ and
$\delta\leftrightarrow rW_{3}$ (which are consistent with $\gamma$ and
$\delta$ being associated with extrinsic geometry, since $W$ is
proportional to the connection in the normal bundle---a measure of
extrinsic geometry).  Notice that this means $s_{+\,IJ}^{\rm ref}$
is a derivative operator!  To make this more palatable one may
go to a `Fourier transform space', where the derivative operators
${\cal D}_{I}$ become momenta, ${\cal K}_{I}$.
Accepting these heuristic identifications, and recalling 
Eq.~(\ref{RefShearTerm}), the key point now is to observe that
\beq
(\tilde{k}^2 -\tilde{l}^2 )^{\rm ref} \equiv -
s_{+\;\;\;\;J}^{{\rm ref}\;I}
s_{-\;\;\;\;I}^{{\rm ref}\;J} = -{2\over{r^3}}(\alpha\gamma+\beta\delta)
=-{2\over{r}}({\cal D}_{2}W_{2}+{\cal D}_{3}W_{3})=
-{2\over{r}}{\cal D}\cdot W,
\label{RefShearSoln}
\eeq
the result we desired: being a total derivative, 
$(\tilde{k}^2 -\tilde{l}^2 )^{\rm ref}$ does not contribute to the IQE.

Now recall that the purpose of solving the embedding equations is to
establish a relationship between ${\cal F}^{\rm ref}$ and
$(\tilde{k}^{2}-\tilde{l}^{2})^{\rm ref}$.  It is via this relationship
that we envision the angular momentum information in ${\cal F}^{\rm ref}$
($={\cal F}$) to enter ${\rm IQE}^{\rm ref}$, which sees only
$(\tilde{k}^{2}-\tilde{l}^{2})^{\rm ref}$.  The previous result implies
the following relationship: ${\cal F}^{\rm ref}$ is the curl of $W$, and 
$(\tilde{k}^{2}-\tilde{l}^{2})^{\rm ref}$ is the divergence of
$W$.  Hence, no relationship!  Is this a problem?  Certainly not.  We
require the relationship in question only when different isometric
embeddings of the same $(S,\sigma)$ result in a different 
$(\tilde{k}^{2}-\tilde{l}^{2})^{\rm ref}$, making the IQE ambiguous.  Such
a relationship can be used to resolve this ambiguity by setting ${\cal
F}^{\rm ref}={\cal F}$.  But accepting the argument that 
$(\tilde{k}^{2}-\tilde{l}^{2})^{\rm ref}$ is always a total derivative,
any such ambiguity would be of no consequence here.  Furthermore, in this
simple case there is no need for $(\tilde{k}^{2}-\tilde{l}^{2})^{\rm ref}$
to carry any information about angular momentum (or gravitational
radiation) coming from $\cal F$, because all of the relevant information
is already carried in the {\it unreferenced} shear term, 
$(\tilde{k}^{2}-\tilde{l}^{2})$.  This is not to say that the condition
${\cal F}^{\rm ref}={\cal F}$ is not important here.  According to our
heuristic argument, it is precisely this condition which ensures that 
$(\tilde{k}^{2}-\tilde{l}^{2})^{\rm ref}$ is a total derivative, and so
carries no information.

The null infinity limit is a simple case which only minimally exercises
the consequences of the condition ${\cal F}^{\rm ref}={\cal F}$.  For
$(S,\sigma)$ a finite two-surface the situation is much richer: it is
highly nonlinear, since the square roots do not disappear, and almost
certainly requires effectively a one-to-one relationship between ${\cal
F}^{\rm ref}$ and $(\tilde{k}^{2}-\tilde{l}^{2})^{\rm ref}$ to remove the
embedding ambiguity inherent in the IQE.

\section{The small sphere limit of the IQE}

Having considered the large sphere case, we now turn our attention to
evaluating the IQE when $(S,\sigma)$ is a small sphere.  The large and
small sphere limits are similar in that in both cases $S$ approaches an
asymptotically flat region of $(M,g)$.  In the latter case, the 
asymptotically flat region is the infinitesimal neighborhood of a generic
spacetime point $p\in M$, which is the `center' of our shrinking 
sphere.  For simplicity we will suppose that $(S,\sigma)$ is
asymptotically round.  Another feature in common with the large sphere
limit is that in this codimension-two setting, the limit can be approached
from different directions, either spatial or null.  More precisely, fix
a set of Riemann normal coordinates $(t,x^{i})$ about the point $p$, set
$r^{2}:=\delta_{ij}x^{i}x^{j}$, and define $S_{*}$ by the
condition $(r,t)=(r_{*},\alpha r_{*})$, where $\alpha$ is a direction
parameter.  Then consider the limit of $S_{*}$ as $r_{*}\rightarrow 0$. 
As before, we will henceforth omit the subscript `$*$'.  The case
$\alpha=0$ is a spatial limit, since then $S$ always lies entirely in the
$t=0$ spacelike three-surface containing $p$.   $\alpha=\pm 1$ is the null
limit, in which $S$ lies in the future/past light cone of the point
$p$.  The latter case was considered by Horowitz and Schmidt~\cite{HS} in
their classic work on the small sphere limit of the Hawking mass.  Brown,
Lau, and York~\cite{BLY} also consider this same limit of the Brown-York
quasilocal energy.  We will be borrowing some results from these
two references.

Explicitly, for a given value of the parameter $r$, $S$ is defined as a
submanifold of $(M,g)$ by embedding a topological two-sphere with
coordinates $\theta$ and $\phi$ into the Riemann normal coordinate system,
as follows: $t=\alpha r$, $x^{1}=r\sin\theta\cos\phi$,
$x^{2}=r\sin\theta\sin\phi$, $x^{3}=r\cos\theta$.  Since (with $t=\alpha
r$) the deviation from the flat metric in Riemann normal coordinates is
$O(r^{2})$, the
induced metric $\sigma_{ab}$ on $S$ will differ from that of the round
sphere to this same order, and so the scalar curvature of $(S,\sigma)$
will have an expansion in $r$ of the form
\beq
{\cal R}={2\over{r^2}}+{\cal R}^{(0)}+r{\cal R}^{(1)}
+r^{2}{\cal R}^{(2)}+O(r^{3}),
\label{Rexpansion}
\eeq
where each of the coefficients ${\cal R}^{(n)}$ is a function of
$\theta$, $\phi$, and the parameter $\alpha$.  To
evaluate the IQE we will also need similar expansions for the other
quantities appearing in Eqs.~(\ref{QIM1}) and (\ref{reference_QIM}).  
These are written as follows:
\begin{eqnarray}
\sigma\sigma R&=&\sigma\sigma R^{(0)}+r\,\sigma\sigma
R^{(1)}+r^{2}\sigma\sigma R^{(2)}+O(r^{3}),
\label{ssRexpansion} \\
(\tilde{k}^{2}-\tilde{l}^{2})&=&r^{2}(\tilde{k}^{2}-\tilde{l}^{2})^{(2)}
+O(r^{3}),
\label{Sshear}\\
(\tilde{k}^{2}-\tilde{l}^{2})&=&r^{2}(\tilde{k}^{2}-\tilde{l}^{2})^{(2)
\,{\rm ref}}+O(r^{3}),
\label{refSshear}
\end{eqnarray}
where each of the coefficients on the right hand side is similarly a
function of $\theta$, $\phi$, and $\alpha$.  Since the appropriate
reference spacetime in this case is
Minkowski space, we have $\sigma\sigma R^{\rm ref}=0$.   Substituting
these expansions into Eq.~(\ref{QIM1}) we find that in the small sphere
limit the unreferenced IQE behaves as
\begin{eqnarray}
{\rm IQE}^{\rm unref}=
-{1\over{8\pi}}\,\int_{S}\,&&dS\,{2\over r}\left\{1+{{r^{2}}\over 4}
\left[ ({\cal R}^{(0)}-\sigma\sigma R^{(0)})+r({\cal
R}^{(1)}-\sigma\sigma R^{(1)})
\right.\right.
\label{SS1}
\\
&&\left.\left.
+r^{2}\left({\cal R}^{(2)}-\sigma\sigma R^{(2)}+
(\tilde{k}^{2}-\tilde{l}^{2})^{(2)}
-{1\over 8}({\cal R}^{(0)}-\sigma\sigma R^{(0)})^{2}\right)
\right]+O(r^{5})\right\}.
\nonumber
\end{eqnarray}
Similarly, the reference IQE behaves as
\begin{eqnarray}
{\rm IQE}^{\rm ref}=
-{1\over{8\pi}}\,\int_{S}\,dS\,{2\over r}&&\left\{1+{{r^{2}}\over
4}\left[
{\cal R}^{(0)}+r{\cal R}^{(1)}
\right.\right.
\nonumber\\
&&\left.\left.
+r^{2}\left({\cal R}^{(2)}+
(\tilde{k}^{2}-\tilde{l}^{2})^{(2)\,{\rm ref}}
-{1\over 8}({\cal R}^{(0)})^{2}\right)
\right]+O(r^{5})\right\}.
\label{SS2}
\end{eqnarray}
Notice that, unlike in the large sphere limit, neither the unreferenced
nor reference energies diverge as $r\rightarrow 0$.  Nevertheless, the
reference subtraction procedure is still necessary to eliminate the
leading $O(r)$ term in ${\rm IQE}^{\rm unref}$, which has nothing to do
with energy.  Thus, forming the difference of the previous two expressions
we find that the small sphere behavior of the (referenced) IQE is given
by
\beq
{\rm IQE} = {1\over{16\pi}}\int_{S}\,dS\,r\,\left[\sigma\sigma R -
(\tilde{k}^{2}-\tilde{l}^{2}) + (\tilde{k}^{2}-\tilde{l}^{2})^{\rm ref} +
{1\over 8}r^{2}\sigma\sigma R^{(0)}(\sigma\sigma
R^{(0)}-2{\cal R}^{(0)})+
O(r^{3})\right].
\label{smallSQIM}
\eeq
The sectional curvature and shear terms have been resummed according to
Eqs.~(\ref{ssRexpansion}-\ref{refSshear}), and the resulting expression is
valid to the order indicated.  Notice that, as in the large sphere limit,
the scalar curvature $\cal R$ dominates the other terms under the square
root, allowing us to expand the radical about
$\sqrt{4/r^{2}}=2/r$.\footnote{As remarked in the footnote on
page~\pageref{mechanism}, if the sphere is not asymptotically round the
shear terms will contribute at order $1/r^{2}$, and the $2/r$ term outside
the braces in Eqs.~(\ref{SS1}) and (\ref{SS2}) will be modified
accordingly.  We will not consider this more complicated case here.}  And
after the reference subtraction is performed what remains again is the
sectional curvature of $(S,\sigma)$ as the dominant term contributing to
the energy.  Comparing Eqs.~(\ref{smallSQIM}) and (\ref{largeSQIM}) we
observe that both the small and large sphere limits of the IQE are
very nearly formally identical.

We will split the IQE into three pieces, each to be discussed separately:
${\rm IQE} = {\rm IQE}_{1} + {\rm IQE}_{2} + {\rm
IQE}_{3}+O(r^{6})$, where
\begin{eqnarray}
{\rm IQE}_{1} &=& {1\over{16\pi}}\int_{S}\,dS\,r\,\left[\sigma\sigma R -
(\tilde{k}^{2}-\tilde{l}^{2})
\right],
\label{(1)}\\
{\rm IQE}_{2} &=& {1\over{128\pi}}\int_{S}\,dS\,r^{3}\,
\left[\sigma\sigma R^{(0)}(\sigma\sigma R^{(0)}-2{\cal R}^{(0)})\right],
\label{(2)}\\
{\rm IQE}_{3} &=& {1\over{16\pi}}\int_{S}\,dS\,r\,\left[
(\tilde{k}^{2}-\tilde{l}^{2})^{\rm ref}\right].
\label{(3)}
\end{eqnarray}
We begin with ${\rm IQE}_{1}$, and show that this piece is essentially the
Hawking mass~\cite{Hawk}.  To see this, we combine
Eqs.~(\ref{scalar_Gauss}) and (\ref{mean_curvature_vector}) to get
\beq
\sigma\sigma R - (\tilde{k}^{2}-\tilde{l}^{2})=
{\cal R}-{1\over 2}(k^{2}-l^{2})=
{\cal R}-2H\cdot H.
\eeq
Replacing the integrand of ${\rm IQE}_{1}$ with the last expression, and
using the Gauss-Bonnet theorem to integrate the $\cal R$ term, we find
\beq
{\rm IQE}_{1}={1\over{4\pi}}\sqrt{{{A}\over{4\pi}}}\left[
2\pi-{1\over 2}\int_{S}\,dS\,H\cdot H\,\right],
\label{Hawking}
\eeq
where we pulled $r$ outside the integral and replaced it with
$\sqrt{A/(4\pi)}$, where $A$ is the area of $(S,\sigma)$.  This form of 
${\rm IQE}_{1}$ is precisely the expression of the Hawking mass given in
Ref.~\cite{HS}.  (Our $H^{c}$ in Eq.~(\ref{mean_curvature_vector}) is
their $N^{c}/2$, and the sign of their metric signature is opposite to
ours.)  Comparing Eq.~(\ref{Hawking}) with Eqs.~(\ref{QIM}) and
(\ref{equiv_forms}) we observe that, while the unreferenced IQE involves
the mean curvature itself, $\sqrt{H\cdot H}$, the Hawking mass is
constructed from the square of the mean curvature.  As mentioned
above, the square root in $\sqrt{H\cdot H}$ effectively disappears in the
small (and large) sphere limits due to the presence of the dominant scalar
curvature term, and consequently the leading order contribution to the
IQE reduces to essentially the Hawking mass.

There are two subtleties worth mentioning: (i) Replacing $r$ with
$\sqrt{A/(4\pi)}$ is in general not valid because it requires that $r$ be
an areal radius which, in general, it is not.  However, it certainly
{\it is} to lowest order in $r$, which will be sufficient for our purposes
here.  But to higher order, ${\rm IQE}_{1}$ and the Hawking mass will in
general give different results.  (ii) It is well known that the Hawking
mass runs into difficulties when $(S,\sigma)$ is not a round
sphere~\cite{Hayw,HS}, a problem that was addressed by Hayward in
Ref.~\cite{Hayw}.  It might be that this problem is a result of having to
insert by hand the factor $\sqrt{A/(4\pi)}$ outside the integral, versus
having $r$ `inside the integral' generated automatically by $\sqrt{2/{\cal
R}}$.  A related issue was discussed at the end of Sec.~5.1 in connection
with Hayward's definition of quasilocal energy.

The connection between ${\rm IQE}_{1}$ and the Hawking mass allows us to
borrow some results from Ref.~\cite{HS}, which are calculated for
the null limit case ($\alpha=1$).  When matter is present Horowitz 
and Schmidt find that, to lowest order in $r$, the Hawking mass is
\beq
{\rm IQE}_{1}=\left({4\over 3}\pi r^{3}\right)\,T_{ab}^{\rm mat}\,
u^{a}u^{b}|_{p}+O(r^{4}).
\label{Hawking3}
\eeq
Here $T_{ab}^{\rm mat}$ is the energy-momentum tensor of matter, and the
expression is to be evaluated at the point $p$, where the unit timelike
vector $u^a$ is just $(\partial/\partial t)^a$ in our Riemann normal
coordinates.  This is a standard result in the literature on quasilocal
energy~\cite{BLY,HS,Doug,Berg1,Berg2,Szab1}, and a very significant
one.  As emphasized in the Introduction, the quasilocal idea asserts that
the time-time component of the energy-momentum tensor of matter {\it a
priori} has nothing to do with energy.  It is only from the small sphere
limit of the quasilocal energy that we learn this quantity is an energy
volume density, i.e., multiplying it by the volume factor $4\pi
r^{3}/3$ gives the energy in an infinitesimal sphere of proper
radius $r$.  However, integrating this energy volume density over a
finite volume to determine the total energy inside is not, in general, 
valid unless one wishes to ignore gravitational effects which, as we
will see in moment, come at higher order in $r$.\footnote{A well
established example of this phenomenon is the Tolman density, which 
integrates to the Komar mass, and is defined in the special case that
the spacetime is stationary and asymptotically flat~\cite{Wald}.  
It is noteworthy that it is not $T_{ab}$ that appears
in the Tolman density, but rather the combination $T_{ab}-(1/2)Tg_{ab}$.  
The extra term involving the trace of $T_{ab}$ is associated
with gravitational effects---see Ref.~\cite{Tolm}, and problems~4 and 5
in Chapter 11 of Ref.~\cite{Wald}.} It is in this sense that the
quasilocal idea implies that even energy due to matter is not localizable
in the context of general relativity.

Now let us assume the spacetime is vacuum in the neighborhood of
$p$.  Then the leading order contribution to the Hawking mass is~\cite{HS}
\beq
{\rm IQE}_{1}={1\over{90}}r^{5}\,
T_{abcd}u^{a}u^{b}u^{c}u^{d}|_{p}+O(r^{6}),
\label{BR}
\eeq
where $T_{abcd}$ is the Bel-Robinson tensor~\cite{Dese}.  Thus
gravitational
energy begins to appear at $O(r^{5})$.  This same result is obtained for
the Brown-York quasilocal energy for a suitable choice of reference
embedding~\cite{BLY}.  However, this is not a universal result in the
literature~\cite{Doug,Berg1,Berg2,Szab1}.  For example, Hayward's
quasilocal mass gives a similar result
as above, but with the numerical factor $1/90$ replaced
with $-2/45$~\cite{Berg1}.  Given that gravitational energy is such a
difficult problem it is not surprising that a consensus has not yet been
reached.

We now turn our attention to the second contribution to the IQE, namely
${\rm IQE}_{2}$ given in Eq.~(\ref{(2)}).  This quantity represents a
deviation from the Hawking mass due to the fact that the IQE is roughly
the square root of the former.  Actually, it is clearer to compare with
the Brown-York quasilocal energy, since in constructing the IQE we simply
replaced the Brown-York $k$ with $\sqrt{k^{2}-l^{2}}$.  In the context
of our generalization given in Eq.~(\ref{generalization}), we therefore
have the heuristic comparison:
\beq
m=\sqrt{E^{2}-\vec{p}^{\;2}}=E-{{\vec{p}^{\;2}}\over{2E}}-\cdots
\longrightarrow
{1\over{8\pi}}\sqrt{k^{2}-l^{2}}=
{1\over{8\pi}}\left(k-{{l^{2}}\over{2k}}-\cdots\right).
\label{comparison}
\eeq
So ${\rm IQE}_{2}$ might be thought of as the analogue of the term
$-l^{2}/(2k)$, and as such would be expected to reduce the magnitude of
the IQE from the result given in Eq.~(\ref{BR}).  

In order to calculate ${\rm IQE}_{2}$ we need to evaluate the quantities
${\cal R}^{(0)}$ and $\sigma\sigma R^{(0)}$ in Eq.~(\ref{(2)}).  To
do so we appeal to the Gauss equation, (\ref{scalar_Gauss}).  Up to
zeroth order in $r$, this equation reads
\beq
\sigma\sigma R^{(0)}={2\over{r^{2}}}+{\cal R}^{(0)}-
{1\over 2}(k^{2}-l^{2})+O(r),
\eeq
where we made use of Eq.~(\ref{Rexpansion}).  We will show later that
\beq
(k^{2}-l^{2})={4\over{r^{2}}}+{4\over3}(1+2\alpha^{2})E_{ab}n^{a}n^{b}+O(r),
\label{k2-l2}
\eeq
where $\alpha$ is the direction parameter introduced
at the beginning of this section.  $E_{ab}n^{a}n^{b}$ is the
radial-radial component of the electric part of the Weyl tensor, which we
saw earlier in Eq.~(\ref{coulomb}).  This quantity ($E_{ab}n^{a}n^{b}$) 
is to be evaluated at
the point $p$, where in our Riemann normal coordinates the radial unit
vector $n^{a}$ has only spatial components, given by
$n^{i}=x^{i}/r$.  We also need $dS$ to lowest order, which is just
$r^{2}\,d\Omega$, $d\Omega$ being the measure on the unit sphere.  Putting
these results together we have
\beq
{\rm IQE}_{2}=-{1\over{96\pi}}(5+4\alpha^{2})r^{5}E_{ij}E_{kl}
\int d\Omega\,n^{i}n^{j}n^{k}n^{l}.
\eeq
By symmetry, the integral over the product of radial vectors must be
proportional to $\delta^{(ij}\delta^{kl)}$.  Transvecting both this term
and the integral in question with $\delta_{ij}\delta_{kl}$, we easily
obtain that the proportionality constant is $4\pi/5$.  Using the fact that
$E_{ab}$ is symmetric, trace-free, and orthogonal to $u^{a}$, we find
that
\beq
{\rm IQE}_{2}=-{1\over{180}}(5+4\alpha^{2})r^{5}E_{ab}E^{ab}.
\label{(2)calc}
\eeq
So ${\rm IQE}_{2}$ is negative, as expected, and this negative
contribution is to be added to ${\rm IQE}_{1}$ in Eq.~(\ref{BR}).  Of
course we can only consider the $\alpha=1$ case, since this is the case
assumed in Eq.~(\ref{BR}).  Recall that the time component of the
Bel-Robinson tensor can be expressed in terms of the electric and magnetic
parts of the Weyl tensor~\cite{Dese}: $T_{abcd}u^{a}u^{b}u^{c}u^{d}=
E_{ab}E^{ab}+B_{ab}B^{ab}$, so ${\rm IQE}_{1}$ is nonnegative.  Inspection
of Eq.~(\ref{(2)calc}) shows that adding to ${\rm IQE}_{1}$ the
$\alpha=1$ value of ${\rm IQE}_{2}$ makes the energy have indefinite
sign.  It is positive (negative) if the magnetic (electric) part
dominates.  This seems like a strange result, but it is only an
intermediate result.  We  have not yet considered the last contribution,
${\rm IQE}_{3}$, involving the reference shear term.  But unfortunately at
present I do not know how to solve the embedding equations to determine
this term.

Now one can construct a heuristic argument much like the one given at the
end of Sec.~5.2, which suggests that $(\tilde{k}^{2}-\tilde{l}^{2})^{\rm
ref}$ is a total derivative, and so does not contribute.  However, the
argument is much less believable in this case.  In contrast to
Eq.~(\ref{refBondiShears}) it turns out that, because $\cal F$ is $O(1)$
in $r$ (as we shall see later), we must expand the reference shears
$s_{\pm}^{\rm ref}$ over {\it three} orders of magnitude in $r$, from
$O(1)$ to $O(r^{2})$.  One might trust a heuristic argument working to
leading order, but believing the higher order corrections is less
palatable.  In short, I do not know what energy prediction the IQE gives
at $O(r^{5})$, and until a solution to the embedding equations is found
there is no sense in speculating.

However, before leaving this section I will provide an intriguing
interpretation of how a definition of quasilocal energy such as the
Hawking mass (or the IQE) provides a measure of the gravitational energy
contained inside a small sphere.

In the $\alpha=1$ null limit case, the lowest order contribution to the
gravitational energy, namely $(1/90)r^{5}\,
T_{abcd}u^{a}u^{b}u^{c}u^{d}|_{p}$ in Eq.~(\ref{BR}), originates in the
$O(r^{2})$ terms inside the brackets of the integrand of ${\rm IQE}_{1}$
in Eq.~(\ref{(1)}).  In the terminology of Eqs.~(\ref{ssRexpansion}) and
(\ref{Sshear}), this means we are interested in the coefficients
$\sigma\sigma R^{(2)}$ and
$(\tilde{k}^{2}-\tilde{l}^{2})^{(2)}$.  Inspecting the Appendix of
Ref.~\cite{BLY} reveals that these two coefficients differ only by a
numerical factor.  They are both proportional to
$\Psi_{0}\bar{\Psi}_{0}|_{p}$ (in Newman-Penrose notation), and the two
numerical factors conspire to produce the $1/90$ factor in the final
result.  Thus, to understand how the integrand of ${\rm IQE}_{1}$ encodes
information about gravitational energy it suffices to study the shear
term, $(\tilde{k}^{2}-\tilde{l}^{2})^{(2)}$.  We will now compute this
term for arbitrary $\alpha\in [-1,1]$ to see how it behaves in both the
spatial and null limit cases of the small sphere.

Denoting our Riemann normal coordinates $(t,x^{i})$ collectively as
$x^{a}$, the metric in these coordinates take the form
\beq
g_{ab}=\eta_{ab}-{1\over 3}J_{abcd}x^{c}x^{d}+O(x^{3}),
\label{RNcoords}
\eeq
where $J_{abcd}=(R_{acbd}+R_{adbc})/2$ is the Jacobi curvature
tensor~\cite{MTW}.  We first construct a pair of mutually orthogonal unit
normal vector fields $u^a$ and $n^a$ on $S$, with $u^a$ normal to the
$t={\rm constant}$ surface passing through $S$.  These are given by
\begin{eqnarray}
u^{0}&=&{1\over N},\;\; u^{i}=-{1\over 3}r^{2}\beta^{i0}+O(r^{3}),
\;\; u_{0}=-N,\;\; u_{i}=0,
\label{u}\\
n^{0}&=&0,\;\; n^{i}=\rho\left[{{x^{i}}\over{r}}+{1\over
3}r^{2}\beta^{ij}{{x_{j}}\over{r}}+O(r^{3})\right],\;\;
n_{0}=-{1\over 3}r^{2}\beta_{0j}{{x^{j}}\over{r}}
+O(r^{3}),\;\; n_{i}=\rho{{x_{i}}\over{r}},
\label{n}
\end{eqnarray}
where
\begin{eqnarray}
N &=& 1+{1\over 6}r^{2}\beta_{00}+O(r^{3}),
\label{N}\\
\rho &=& 1-{1\over 6}r^{2}\beta_{ij}{{x^{i}x^{j}}\over{r^{2}}}+O(r^{3}),
\label{rho}\\
\beta_{ab} &=& \alpha^{2}J_{ab00}+2\alpha J_{ab0j}{{x^{j}}\over{r}}
+J_{abij}{{x^{i}x^{j}}\over{r^{2}}}.
\label{beta}
\end{eqnarray}
Since the Jacobi tensor in Eq~(\ref{RNcoords}) is evaluated at the
coordinate origin $p$, its indices, and thus those of $\beta_{ab}$, are
raised and lowered with the flat spacetime metric $\eta_{ab}=\eta^{ab}=
{\rm diagonal}(-1,1,1,1)$.  Similarly, $x_{i}:=\delta_{ij}x^{j}$.

Now define on $S$ a pair of mutually orthogonal unit tangent vector fields
$e_{I}^{\;\;a}$, where indices $I,J,\ldots$ take the values 2 and 3.  The
set $\{e_{0}^{\;\;a}:=u^{a},e_{1}^{\;\;a}:=n^{a},e_{I}^{\;\;a}\}$ thus
comprises an orthonormal basis adapted to $S$.  Let basis indices
$A,B,\ldots$ run from 0 to 3, and $\alpha,\beta,\ldots$ from 1 to
3.  Beginning with this setup it is straightforward to compute the basis
components of the extrinsic curvatures defined in
Eqs.~(\ref{extrinsic_curvatures}).  I find:
\begin{eqnarray}
l_{IJ}&=&e_{I}^{\;\;a}e_{J}^{\;\;b}\nabla_{a}u_{b}=
-{2\over 3}r[\alpha J_{00IJ}+J_{01IJ}]+O(r^{2}),
\label{lIJSmall}\\
k_{IJ}&=&e_{I}^{\;\;a}e_{J}^{\;\;b}\nabla_{a}n_{b}=
{1\over r}\delta_{IJ}
-{1\over 3}r\left[J_{11IJ}-\alpha^{2}\left(J_{00IJ}-
{1\over2}J_{0011}\delta_{IJ}\right)\right]+O(r^{2}).
\label{kIJSmall}
\end{eqnarray}
In these equations a quantity such as $J_{01IJ}$ means $[e_{0}^{\;\;a}
e_{1}^{\;\;b}e_{I}^{\;\;c}e_{J}^{\;\;d}J_{abcd}]|_{p}$, which is a
function of only the angles $\theta$ and $\phi$ on $S$.  

Since we are interested in purely gravitational energy we shall restrict
ourselves to the vacuum case.  In our basis components the electric and
magnetic parts of the Weyl tensor are defined by~\cite{DeFe}:
\beq
E_{\alpha\beta}=-C_{0\alpha 0\beta}\;\;\;{\rm and}\;\;\;
B_{\alpha\beta}=-{*}C_{0\alpha 0\beta},
\label{defEB}
\eeq
where $*C_{ABCD}=(1/2)\epsilon_{AB}^{\;\;\;\;\;\;EF}C_{EFCD}$.  These are
symmetric trace-free three-dimensional tensors associated with $t={\rm
constant}$ spacelike hypersurfaces.  As such, each has five independent
components, which together comprise the ten independent components of the
Weyl tensor.  In terms of these fields, the components of the Jacobi
curvature tensor relevant to Eqs.~(\ref{lIJSmall}-\ref{kIJSmall}) read:
\beq
J_{0011}=-E_{11},\;\;
J_{00IJ}=-E_{IJ},\;\;
J_{01IJ}=-{*}\tilde{B}_{IJ},\;\;
J_{11IJ}=-E_{IJ}-E_{11}\delta_{IJ}.
\eeq
Here $\tilde{B}_{IJ}$ is the trace-free part of $B_{IJ}$, and
$*\tilde{B}_{IJ}=\epsilon_{I}^{\;\;K}\tilde{B}_{KJ}$ is its dual in
$(S,\sigma)$, which is also trace-free.  The trace of $E_{IJ}$ is
$\delta^{IJ}E_{IJ}=\delta^{\alpha\beta}E_{\alpha\beta}-E_{11}=-E_{11}$,
since $E_{\alpha\beta}$ is trace-free.  Thus, the trace of the extrinsic
curvatures in Eqs.~(\ref{lIJSmall}-\ref{kIJSmall}) is found to be
\begin{eqnarray}
l&=&-{2\over3}r\alpha E_{11}+O(r^{2}),\\
k&=&{2\over r}+{1\over3}r(1+2\alpha^{2})E_{11}+O(r^{2}).
\end{eqnarray}
Squaring these and forming their difference leads to
Eq.~(\ref{k2-l2}) written earlier.  The trace-free parts are
\begin{eqnarray}
\tilde{l}_{IJ}&=&{2\over3}r(\alpha\tilde{E}_{IJ}+\,*\tilde{B}_{IJ})+O(r^{2}),
\label{tflIJ}\\
\tilde{k}_{IJ}&=&{1\over3}r(1-\alpha^{2})\tilde{E}_{IJ}+O(r^{2}).
\label{tfkIJ}
\end{eqnarray}
Now $\tilde{E}_{IJ}$ and $\tilde{B}_{IJ}$ each have two independent
components, and it is useful to introduce the notation
\begin{eqnarray}
\tilde{E}_{IJ}&=&\left(
\begin{array}{cc}
{1\over2}(E_{22}-E_{33}) & E_{23} \\
E_{23} & -{1\over2}(E_{22}-E_{33})
\end{array}\right)=:\left(
\begin{array}{cc}
e^{2} & e^{3} \\
e^{3} & -e^{2}
\end{array}\right),
\label{e}\\
\tilde{B}_{IJ}&=&\left(
\begin{array}{cc}
{1\over2}(B_{22}-B_{33}) & B_{23} \\
B_{23} & -{1\over2}(B_{22}-B_{33})
\end{array}\right)=:-\left(
\begin{array}{cc}
b^{2} & b^{3} \\
b^{3} & -b^{2}
\end{array}\right),
\label{b}
\end{eqnarray}
and thus define a pair of two-vectors $\vec{e}:=e^{I}e_{I}^{\;\;a}$ and 
$\vec{b}:=b^{I}e_{I}^{\;\;a}$ tangent to $S$.  Since $\vec{e}$ and
$\vec{b}$ represent the pullbacks to $S$ of $E_{\alpha\beta}$ and
$B_{\alpha\beta}$, respectively, they are to be thought of as `electric
and magnetic fields' induced on $S$ by the Weyl curvature that $S$ is
embedded in.  It is easy to see that under a rotation of the basis vectors
$e_{I}^{\;\;a}$ through an angle $\gamma$, the components of $\vec{e}$ 
and $\vec{b}$ rotate through an angle $2\gamma$, so $\vec{e}$ and
$\vec{b}$ are not true (spin one) vectors, but rather spin two objects, as
one would expect.

Thus we arrive at the results we are interested in.  To lowest order in
$r$ we have
\begin{eqnarray}
\tilde{l}^{2}&=&{4\over9}r^{2}(\alpha\tilde{E}+{*}\tilde{B})^{2}
={8\over9}r^{2}(\alpha^{2}\vec{e}\cdot\vec{e}+\vec{b}\cdot\vec{b}-
2\alpha\vec{e}\times\vec{b}),
\label{tfl2}\\
\tilde{k}^{2}&=&{1\over9}r^{2}(1-\alpha^{2})^{2}\tilde{E}^{2}
={2\over9}r^{2}(1-\alpha^{2})^{2}\vec{e}\cdot\vec{e}.
\label{tfk2}
\end{eqnarray}
The difference of these two gives the $O(r^{2})$ piece of the unreferenced
shear term appearing in the integrand of ${\rm IQE}_{1}$ in
Eq.~(\ref{(1)}).  Notice that it is the appearance of $*\tilde{B}_{IJ}$
(rather than $\tilde{B}_{IJ}$) that gives rise to the cross product term
$\vec{e}\times\vec{b}=e^{2}b^{3}-e^{3}b^{2}$ in Eq.~(\ref{tfl2}).

We first consider the case $\alpha=1$, in which $S$ lies in the future
light cone of the point $p$.  Then $\tilde{k}^{2}=0$ and so the shear term
$(\tilde{k}^{2}-\tilde{l}^{2})$ is proportional to
$r^{2}(\tilde{E}+{*}\tilde{B})^{2}$.  As a quick check, it is easy to
verify that $(\tilde{E}+{*}\tilde{B})^{2}$, in turn, is proportional to
$\Psi_{0}\bar{\Psi}_{0}|_{p}$, in agreement with
Ref.~\cite{BLY}.  I mentioned above that in this case the $\sigma\sigma R$
term in Eq.~(\ref{(1)}) also contributes a term proportional to
$r^{2}\Psi_{0}\bar{\Psi}_{0}|_{p}$~\cite{BLY}.  Putting in the numerical
factors I find that
\beq
{\rm IQE}_{1}=\int_{S}\,dS\,{{r^{3}}\over{9}}\left[
{{1}\over{8\pi}}(\vec{e}\cdot\vec{e}+\vec{b}\cdot\vec{b})
-{{1}\over{4\pi}}\vec{e}\times\vec{b}\right]+O(r^{6}).
\label{QIMem}
\eeq
Now $(\vec{e}\cdot\vec{e}+\vec{b}\cdot\vec{b})/(8\pi)$ looks like the
energy surface density of the `electromagnetic field', but we must
be careful about its  
dimension.  $(\vec{e}\cdot\vec{e}+\vec{b}\cdot\vec{b})/(8\pi)$
has dimension $L^{-4}$, where $L$ means `length', which is not correct.  
However, the additional factor of $r^{3}/9$ in the integrand suggests that
it is really 
${\cal E}:=r^{3}(\vec{e}\cdot\vec{e}+\vec{b}\cdot\vec{b})/(72\pi)$ that
is the proper energy density.  $\cal E$ has dimension $L^{-1}$, consistent
with it being interpreted as the `electromagnetic' energy per unit area of
$S$.  Besides giving the right dimension, the additional $r^{3}$ factor
forces $\int_{S}\,dS\,{\cal E}$ to go to zero as $r^{5}$, consistent with
the fact that there can be no gravitational energy at order $r^{3}$.  We
interpret $\int_{S}\,dS\,{\cal E}$ as the total `electromagnetic' energy
that was in $S$ at $t=0$, or equivalently, the total gravitational energy
that was in the small volume spanning $S$ at $t=0$.

To further justify this interpretation we now turn to the radiation term
in Eq.~(\ref{QIMem}).  Clearly $\vec{e}\times\vec{b}/(4\pi)$ might be
thought of as the gravitational analogue of the electromagnetic Poynting
flux, directed radially outward from $S$.  But again, its dimension is
wrong.  Of course the factor of $r^{3}/9$ will fix this problem, as
before, but the situation is more interesting this time.  Multiplying by
$r^{2}/9$ we get the proper Poynting flux, ${\cal
P}:=r^{2}\vec{e}\times\vec{b}/(36\pi)$.  $\cal P$ has dimension $L^{-2}$,
consistent with interpreting it as the `electromagnetic' energy per unit
time per unit area.  So $\int_{S}dS\,{\cal P}$ gives the `electromagnetic'
energy per unit time radiating from (or through) the surface $S$.  The
factor of $r^{2}$ indicates that the efficiency of a small volume to
radiate gravitationally grows in proportion to its surface area, in
analogy with an electromagnetic antenna.  But there is one more factor of
$r$, which one might imagine is the $r$ outside the brackets in
Eq.~(\ref{(1)}), i.e., $\sqrt{2/{\cal R}}$.  This distinction between $r$s
is suggested by the close analogy between $\cal P$ and the shear term
responsible for radiation in the null infinity limit---see
Eqs.~(\ref{BondiShearTerm}-\ref{largeSQIMnull}).  This additional factor
of $r$ has the interpretation of a time lapse, i.e., $r\int_{S}dS\,{\cal
P}$ is the amount of `electromagnetic' energy radiated from $S$ between
time $t=0$ and $t=r$.  Thus, the following picture has emerged regarding
Eq.~(\ref{QIMem}).  The `electromagnetic' energy in $S$ (or equivalently,
the gravitational energy in the volume spanning $S$) at time $t=r$ is the
energy at $t=0$ minus the amount of energy radiated during this time
interval.  (Keep in mind that $\vec{e}$ and $\vec{b}$ are evaluated at
$p$, and hence at $t=0$.)

The case $\alpha=-1$ is similar, except now $S$ lies in the past light
cone of the point $p$.  Inspection of Eq.~(\ref{tfl2}) reveals that the
radiation term in Eq.~(\ref{QIMem}) now appears with the opposite
sign.  The fact that this sign change comes out correctly is reassurance
that our picture is correct:  The energy at time $t=-r$ is the energy at
$t=0$ {\it plus} the amount of energy that will be radiated from the
sphere during the interval from $t=-r$ to $t=0$.\footnote{The reader may
have noticed that $\vec{b}$ in Eq.~(\ref{b}) was defined with an awkward
minus sign.  This sign was chosen to give the picture just
described.  Reversing the sign is equivalent to replacing $\alpha$ with
$-\alpha$.  Insofar as $\vec{e}$ and $\vec{b}$ (like $\vec{E}$ and
$\vec{B}$ in electromagnetism) are defined by their physical
interpretation, choosing the sign of $\vec{b}$ to give a
result with the correct interpretation is legitimate.  But this assumes we
know what the correct interpretation is, and it is not certain we do.  For
example, I mentioned above that at $O(r^{5})$ Hayward's quasilocal energy
gives a negative gravitational energy~\cite{Berg1}.  If this is correct,
then we should replace the definition of $\vec{b}$ with $-\vec{b}$.}

Finally, we consider the spatial limit case, $\alpha=0$.  According to our
discussion above we would expect ${\rm IQE}_{1}$ to be the same as in
Eq.~(\ref{QIMem}), except with the radiation term absent.  Inspection of
Eqs.~(\ref{tfl2}-\ref{tfk2}) reveals that this is not the case.  However,
it is only when $\alpha=1$ (and presumably also when $\alpha=-1$) that we
know that $\sigma\sigma R^{(2)}$ in Eq.~(\ref{(1)}) is proportional to
$(\tilde{k}^{2}-\tilde{l}^{2})^{(2)}$, in which case it suffices to
consider only the shear term.  Unfortunately, it is not possible to
compute $\sigma\sigma R$ to $O(r^{2})$ within the framework of our
$O(r^{2})$ Riemann normal coordinates, so we cannot learn if this simple
proportionality between the two persists when $|\alpha|<1$.  One might
guess that it almost certainly does not, but I will leave this question
for future work.  Nevertheless, since we expect $\tilde{l}^{2}$ to play
the key role with regards to radiation, Eq.~(\ref{tfl2}) is still of some
qualitative value when $|\alpha|<1$.  From this equation we see that the
radiation term is zero when $\alpha$ is zero, and turns on in proportion
to $\alpha$, precisely as it should since the time lapse is now $\alpha
r$, instead of $r$.

As satisfying as it might seem, the picture just given is not truly
quasilocal, in the sense that $\vec{e}$ and $\vec{b}$ are evaluated at the
point $p$.  To be truly quasilocal we need `electromagnetic' fields, call
them $\vec{E}$ and $\vec{B}$, evaluated on $S$.  This is where observers
reside, and measurements are made, according to the quasilocal idea.  Such
a quasilocal picture is achieved very naturally as follows.  The basic
idea is: $\vec{e}$ and $\vec{b}$ are certain components of the Weyl tensor
evaluated at $p$ ($r=0$).  But this information is contained in the $O(r)$
piece of certain connection coefficients evaluated on $S$ ($r>0$).  Thus
we expect the desired $\vec{E}$ and $\vec{B}$ fields to be associated with 
connection coefficients.

For simplicity we will restrict ourselves to the case $\alpha=1$, which
also allows us to borrow some results from Ref.~\cite{BLY}.  We first
observe
that
\beq
\Psi_{0}|_{p}=-2[(e^{2}-b^{3})+i(e^{3}+b^{2})].
\eeq
On the left hand side is a component of the Weyl tensor in Newman-Penrose
(NP) notation.  Using Eqs.~(\ref{defEB}) and (\ref{e}-\ref{b}), $\Psi_{0}$
is easily converted to the expression given on the right hand side.  From
Eq.~(B5b) of Ref.~\cite{BLY} we have
\beq
\sigma={r\over3}\Psi_{0}|_{p}+O(r^{2}),
\eeq
where $\sigma$ is one of the NP spin coefficients.  Thus we find that
\beq
{1\over4}\sigma\bar{\sigma}={{r^2}\over{9}}[\vec{e}\cdot\vec{e}+
\vec{b}\cdot\vec{b}-2\vec{e}\times\vec{b}]+O(r^{3}).
\eeq
Comparing this with the integrand in Eq.~(\ref{QIMem}) we are led to
define
\beq
\vec{E}:={r\over3}\vec{e}+O(r^{2}),\;\;\;\vec{B}:={r\over3}\vec{b}
+O(r^{2}),
\label{defnewEB}
\eeq
or in other words,
\beq
\sigma=-2[(E^{2}-B^{3})+i(E^{3}+B^{2})]+O(r^{2}).
\eeq
Thus, to $O(r)$,  $\vec{E}$ and $\vec{B}$ are related to connection
coefficients, as we expected.  

Substituting Eq.~(\ref{defnewEB}) into Eq.~(\ref{QIMem}) we have
\beq
{\rm IQE}_{1}=\int_{S}\,dS\,r\left[
{{1}\over{8\pi}}(\vec{E}\cdot\vec{E}+\vec{B}\cdot\vec{B})
-{{1}\over{4\pi}}\vec{E}\times\vec{B}\right]+O(r^{6}).
\label{QIMemEB}
\eeq
Observe that the mysterious $r^{2}/9$ factor has disappeared, and the
analogy with electromagnetism is improved: $\vec{E}$ and $\vec{B}$ now
have their usual dimension ($L^{-1}$), as do the energy density and
Poynting flux terms.  I emphasize again that, in contrast to $\vec{e}$
and $\vec{b}$ in Eq.~(\ref{QIMem}), $\vec{E}$ and $\vec{B}$ are fields
measured by observers residing in $S$, in the true quasilocal
spirit.  $\vec{E}$ is clearly associated with tidal forces {\it
tangential} to $S$, and $\vec{B}$ is a measure of frame dragging
effects.  Notice that $\vec{E}$ and $\vec{B}$ vanish as $r\rightarrow 0$,
in accord with the equivalence principle.

To conclude this section we consider the connection in the normal bundle
and its associated curvature.  I find that
\beq
A_{J}={2\over3}r\left[\alpha E_{1J}+{1\over2}{*}B_{1J}\right]+O(r^{2}),
\label{smallA}
\eeq
where $*B_{1J}=\epsilon_{J}^{\;\;K}B_{1K}$, and
\beq
{\cal F}=2B_{11}+O(r).
\label{smallF}
\eeq
So to leading order the curvature of the normal bundle is (twice) the
radial-radial component of the magnetic part of the Weyl tensor, and is
thus associated with `gravitational magnetic charge'.  There are both 
local and global dimensions to this result.  Locally, $\cal F$ is
associated with frame dragging, a ready example being $\cal F$ for the
Kerr black hole given in Eq.~(\ref{KerrF}), which is proportional to the
angular momentum.  Globally, it is known that in exact analogy with the
scalar curvature $\cal R$, the integral of ${\cal F}$ over $S$ is
proportional to the Euler number of the normal bundle~\cite{Chen}.  For a
Euclidean-signature spacetime the normal bundle is an $SO(2)$ (rather than
$SO(1,1)$) bundle, and there can be a nontrivial winding number,
corresponding to a gravitational magnetic monopole.  In the case of the
Kerr spacetime there is no monopole present since, as is obvious from
inspection of Eq.~(\ref{KerrF}), the integral of ${\cal F}$ is zero.  It
might be interesting to explore topologically nontrivial cases in the
context of the IQE.

The result in Eq.~(\ref{smallF}) can actually be obtained immediately by
inspection of Eq.~(\ref{Ricci}), assuming that the shear terms are higher
order in $r$ than $\cal F$ is.  To lowest order in $r$ we then see that
${\cal F}=-2R_{0123}$.  But $R_{0123}=C_{0123}$ is identically true, and
thus we are led to Eq.~(\ref{smallF}).  So this equation is true whether
or not matter is present, and is also independent of $\alpha$.  It is
instructive to compare this result with the sectional curvature of $S$ in
vacuo:
\beq
\sigma\sigma R=2E_{11} + O(r).
\label{smallssR}
\eeq
This is essentially the same as the spatial infinity limit result given in
Eq.~(\ref{coulomb}), and is derived similarly.  Comparing the previous
two equations we see a striking
electric/magnetic duality between the sectional curvature of $S$
(electric), and the curvature of its normal bundle (magnetic).  When
matter is present, the right
hand side of the equation above acquires an additional term, and it is
precisely this term that is responsible for the $O(r^{3})$ matter
contribution seen in Eq.~(\ref{Hawking3}).  So the sectional curvature is
the dominant term in the energy that encodes information about the matter
content of the spacetime.  It seems reasonable to expect inertial effects
(frame dragging) produced by this matter to also play a role in the
energy.  But consideration of such effects is subtle, because the
magnetic part of the Weyl tensor has no Newtonian gravity analogue.  I
have argued that the procedure suggested in Sec.~4 is a
geometrically natural way to incorporate such inertial effects into the
energy: one demands ${\cal F}^{\rm ref}={\cal F}$, and then solves the
embedding equations for the reference shear term, present in the
reference energy.  In this way the inertia information contained in $\cal
F$ makes its presence felt in the energy.  Moreover, by inspection of the
purely spatial ($\alpha=0$) case of Eq.~(\ref{smallA}), one observes that
the set of magnetic quantities: $\cal F$, $A$, and $\vec{b}$, precisely
encode the five independent components of the magnetic part of the Weyl
tensor.  It seems likely that the phenomenon of gravitational energy is
subtle enough to be sensitive to this full set.  Out of this set, in this
section we have seen only the role of $\vec{b}$.  To see whether or not
the other components play a role (via $(\tilde{k}^{2}-\tilde{l}^{2})^{\rm
ref}$) will have to wait until a solution to the embedding equations is
found.

\section{Asymptotically anti-de Sitter spacetimes}

In this last section we will explore the significance of the $\sigma\sigma
R^{\rm ref}$ term in Eq.~(\ref{reference_QIM}).  Suppose our physical
spacetime $(M,g)$ is asymptotically anti-de Sitter space.  The
$\sigma\sigma R^{\rm ref}$ term in ${\rm IQE}^{\rm ref}$ gives us the
freedom to specify the Riemann tensor of a reference spacetime, which in
this case is naturally the Riemann tensor of anti-de Sitter space.  Thus,
according to Eq.~(\ref{ssRref}) we have $\sigma\sigma R^{\rm
ref}=-2/\ell^2$, and so
\beq
{\rm IQE}^{\rm ref} = -{1\over{8\pi}}\,\int_{S}\,dS\,\sqrt{2\left[
{{2}\over{\ell^2}}+{\cal R} + (\tilde{k}^2 -\tilde{l}^2 )^{\rm
ref}\right]}.
\label{AdSQIMref}
\eeq
In the large sphere limit it is clear that the cosmological constant term
will dominate, rather than $\cal R$, and the behavior of the IQE is
qualitatively different from that for asymptotically flat spacetimes.

Let us now specialize to the case that $(M,g)$ is the ${\rm
AdS}_{4}$-Schwarzschild spacetime, so that our main argument is not
obscured by consideration of the shear terms, which will obviously be just
zero.  The line element in this case is given by
\beq
ds^2=-N^2\,dt^2+{1\over{f^2}}\,dr^2+r^2\,d\Omega^2,\;\;\;\;{\rm where}\;\;
N(r)=f(r)=\left({{r^2}\over{l^2}}+1-{{2M}\over{r}}\right)^{1\over 2},
\eeq
and $d\Omega^2=d\theta^2+{\sin}^{2}\theta\,d\phi^2$ is the line element on
the unit round sphere.  Let $S$ be a $t,r={\rm constant}$ two-sphere.  
Its scalar curvature is ${\cal R}=2/r^2$, and a simple calculation shows
that its sectional curvature is given by 
\beq
\sigma\sigma R=-{{2}\over{\ell^2}}+{{4M}\over{r^3}},
\label{sectionalAdS4}
\eeq
the dominant term coming from the anti-de Sitter `background'.
Substituting these results into Eqs.~(\ref{QIM1}) and (\ref{AdSQIMref}) we
find
\beq
{\rm IQE} =
-{1\over{8\pi}}\,\int_{S}\,dS\,\sqrt{2\left[{{2}\over{\ell^2}}
+{{2}\over{r^2}}-{{4M}\over{r^3}}\right]}
+{1\over{8\pi}}\,\int_{S}\,dS\,\sqrt{2\left[{{2}\over{\ell^2}}
+{{2}\over{r^2}}\right]}
={{M\ell}\over{r}}+O\left({{1}\over{r^3}}\right).
\label{AdSSchw}
\eeq
The divergent terms due to the cosmological constant cancel, so the limit
of the IQE as $r\rightarrow\infty$ exists,
and this limit is zero.  This would be the expected result if the IQE had
the interpretation of an {\it energy}, which should be red shifted to zero
by the cosmological horizon.  In contrast, we do not expect an invariant
mass to be redshifted.  This is why the terminology `invariant quasilocal
energy' was chosen rather than `invariant quasilocal mass', even though
the IQE is the analogue of the mass $m$ in the formula: 
$m=\sqrt{E^{2}-\vec{p}^{\;2}}$.

However, one can easily modify the definition of the IQE---to give it the
interpretation of mass---by multiplying the right hand side of
Eq.~(\ref{generalization}) by a lapse function.  Thus one replaces
Eq.~(\ref{QIM1}) with
\beq
{\rm IQE}[N_{\cal B}] = -{1\over{8\pi}}\,\int_{S}\,dS\,N_{\cal B}
\sqrt{2\left[{\cal R} - \sigma\sigma R \right]} 
+ {1\over{8\pi}}\,\int_{S}\,dS\,N_{\cal B}^{\rm ref}
\sqrt{2\left[{\cal R} - \sigma\sigma R^{\rm ref} \right]}
\label{AdSQIM1}
\eeq
(ignoring the shear terms).  Here the smearing function, $N_{\cal B}$, is
the lapse function in the timelike three-boundary, $\cal B$, swept out by
the two-parameter family of observers (cf. Eqs.~(11) and (13) in
Ref.~\cite{BK}).  In the ${\rm AdS}_{4}$-Schwarzschild example 
$\cal B$ is an $r={\rm constant}$ surface, and $N_{\cal B}=N$.  The
question arises, What are we to put for $N_{\cal B}^{\rm ref}$?  The
answer that works is $N_{\cal B}^{\rm ref}=N_{\cal B}$, which is
intuitively justified as follows: we are already isometrically embedding
$(S,\sigma)$ into $(M^{\rm ref},g^{\rm ref})$, and the condition $N_{\cal
B}^{\rm ref}=N_{\cal B}$ represents the next would-be step towards an
isometric embedding of $({\cal B},\gamma)$ into $(M^{\rm ref},g^{\rm
ref})$, where $\gamma_{ab}$ is the three-metric in $\cal B$.  (`Would-be'
in the sense that, while the lapse carries some information about
$\gamma_{ab}$, we still only need to embed $(S,\sigma)$ into $(M^{\rm
ref},g^{\rm ref})$, not $({\cal B},\gamma)$ into $(M^{\rm ref},g^{\rm
ref})$.)  By comparing Eq.~(\ref{AdSQIM1}) with (\ref{AdSSchw}), and using
the fact that the lapse function goes as $r/\ell$ for large $r$, it is
easy to see that with $N_{\cal B}^{\rm ref}=N_{\cal B}$ we get
$\lim_{r\rightarrow\infty}{\rm IQE}[N_{\cal B}]=M$ for the ${\rm
AdS}_{4}$-Schwarzschild case.  Thus ${\rm IQE}[N_{\cal B}]$ has the
interpretation of a mass, as claimed.  Unlike the original IQE, it is not
red shifted to zero, and is thus a different physical quantity.  Regarding
the comment at the end of the previous paragraph, since special relativity
does not know about lapse functions, the generalization given in
Eq.~(\ref{generalization}) is open to this ambiguity: one can define both
an invariant quasilocal energy and an `invariant' quasilocal mass.

It is instructive to evaluate ${\rm IQE}[N_{\cal B}]$ also at 
the horizon, $r=r_{+}$ (where $N_{\cal B}(r_{+})=0$), and compare with
what one gets using the unsmeared IQE.  The following results for the
${\rm AdS}_{4}$-Schwarzschild example are easily established:
\begin{eqnarray}
{\rm IQE}&=&\left\{
\begin{array}{ll}
\sqrt{2Mr_{+}} & \mbox{at $r=r_{+}$} \\
0 & \mbox{at $r=\infty$}
\end{array}
\right.
\\
{\rm IQE}[N_{\cal B}]&=&\left\{
\begin{array}{ll}
0 & \mbox{at $r=r_{+}$} \\
M & \mbox{at $r=\infty$}
\end{array}
\right.
\end{eqnarray}
We thus learn that the IQE {\it decreases} with $r$, which can be
interpreted as the result of
negative binding energy---another reason to think of the IQE as an
energy.  On the other hand, ${\rm IQE}[N_{\cal B}]$ {\it increases} with
$r$, which might be interpreted as saying that, for larger $r$,  `more
mass is enclosed'; it starts from zero at the horizon (no mass in the
interior of the black hole) and accumulates to $M$ at infinity.  This is
reminiscent of the old notion that the substance of mass is nothing but
the curvature of spacetime itself.  A similar behavior is observed for
the usual Schwarzschild case:
\begin{eqnarray}
{\rm IQE}&=&\left\{
\begin{array}{ll}
2M & \mbox{at $r=2M$} \\
M & \mbox{at $r=\infty$}
\end{array}
\right.
\\
{\rm IQE}[N_{\cal B}]&=&\left\{
\begin{array}{ll}
0 & \mbox{at $r=2M$} \\
M & \mbox{at $r=\infty$}
\end{array}
\right.
\end{eqnarray}
The only qualitative difference occurs at $r=\infty$, where ${\rm
IQE}[N_{\cal B}]={\rm IQE}$ in the Schwarzschild case because, of course,
the lapse function goes to one in this limit.  There is no cosmological
horizon.

Despite these appealing features of ${\rm IQE}[N_{\cal B}]$, it is
unsatisfactory from the point of view taken here because the presence of
the lapse function means it depends on a choice of three-surface passing
through $S$ (hence the use of inverted commas in the terminology
`invariant' quasilocal mass above).  The situation might be improved by
replacing $k$ with $N_{\cal B}k$ and $l$ with $N_{\Sigma}l$ in
Eq.~(\ref{generalization}), and then proceeding as before.  Here $N_{\cal
B}$ and $N_{\Sigma}$ are time and radial lapse functions, respectively
(equal to $N$ and $1/f$ in the ${\rm AdS}_{4}$-Schwarzschild example).
Admittedly such a procedure is ad hoc, and unless it can be improved upon
we are not particularly interested in ${\rm IQE}[N_{\cal B}]$.  It was
introduced simply to illustrate the distinction between mass and energy,
but for the remainder of this section we will return to the original
definition of the IQE.

The main point of this section is to draw attention to a remarkable
similarity between the reference subtraction term given in 
Eq.~(\ref{AdSQIMref}), and a certain counterterm action recently suggested
in the context of the conjectured AdS/CFT correspondence.  We begin by
observing that when the reference shear term vanishes, 
Eq.~(\ref{AdSQIMref}) reduces to precisely the same reference subtraction
term suggested by Lau~\cite{Lau3}, in the context of the Brown-York 
quasilocal energy (except that Lau's expression has a lapse function
present in the manner discussed above).  However, our derivations of this
expression are different.  Lau employs a `light cone reference' embedding
of $(S,\sigma)$, together with a `rest frame' assumption, $l^{\rm ref}=0$,
to derive an expression for $k^{\rm ref}$, which is then used to construct
his reference subtraction term.  In our case we get the same end result,
but we get it without any recourse to a reference embedding!  This is
because the cosmological constant term in Eq.~(\ref{AdSQIMref}) comes
directly from the explicit dependence of ${\rm IQE}^{\rm ref}$ on the
Riemann tensor of the reference spacetime, i.e., the term $\sigma\sigma
R^{\rm ref}$ in Eq.~(\ref{reference_QIM}).  Our reference embedding of
$(S,\sigma)$ into $(M^{\rm ref},g^{\rm ref})$ is required only to evaluate
the reference shear term, $(\tilde{k}^{2}-\tilde{l}^{2})^{\rm ref}$.
This `higher order correction'---which I have argued accounts for
angular momentum---is not present in Lau's reference subtraction term.  
Also, his additional `rest frame' assumption is not required here because 
the IQE is already naturally a `rest frame energy'.

Let us now return to Eq.~(\ref{total_action}).  We know that when space is
non-compact the boundary (or quasilocal) energy-momentum tensor $T_{\cal
B}^{ab}=-\Pi^{ab}/(8\pi)$ diverges in general as $\cal B$ is taken to
infinity.\footnote{As elsewhere in this paper, we use the symbol ${\cal
B}$ loosely to refer to either a timelike three-surface in the interior of
$M$, bounding a finite spatial region, or the boundary at infinity.  It's
meaning should be clear from the
context in which it is used. }  To render it finite, Brown and York
suggest the use of a reference subtraction term that involves an isometric
embedding of $({\cal B},\gamma)$ into a suitable reference
spacetime.  However, like their prescription to embed $(S,\sigma)$ into a
suitable three-dimensional reference space, this prescription suffers
from the drawback that such a codimension-one embedding does not always
exist.  Recently,
Balasubramanian and Kraus~\cite{BK} have proposed an alternative
procedure:  Since it is always possible to add to the action a local
functional of the intrinsic geometry of the boundary without affecting the
equations of motion or the symmetries (but of course this alters $T_{\cal
B}^{ab}$), their idea is to choose this functional such that its
divergences cancel those of the original $T_{\cal B}^{ab}$, rendering the
improved boundary energy-momentum tensor finite as $\cal B$ is taken to
infinity.  No recourse is made to a reference embedding.  This procedure
was first applied to spacetimes that are asymptotically AdS space, in
which case the required `counterterms' amounted to a simple finite
polynomial in the curvature invariants of $\cal B$~\cite{BK}.  This
idea is exactly analogous to the standard prescription for removing
ultraviolet divergences in quantum field theory by adding to the
Lagrangian a finite polynomial in the fields.  Moreover, the conjectured
AdS/CFT correspondence~\cite{Mald} implies that the two procedures
are not merely analogous, they are one and the same~\cite{BK}.

Now since flat spacetime is recovered from AdS space by taking $\ell$ to
infinity, one might expect that in this same limit the counterterms
found by Balasubramanian and Kraus would produce counterterms suitable for
asymptotically flat spacetimes.  It is not obvious that this is
so~\cite{BK}.  However, Mann~\cite{Mann1} has suggested the following
generalization of their counterterm action:
\beq
I_{\rm ct}={1\over{8\pi}}\int_{{\cal B}_\infty}\,d^{3}x\,\sqrt{-\gamma}\,
\sqrt{2\left[{{2}\over{\ell^2}}+R(\gamma)\right]},
\label{Mann_formula}
\eeq
where $R(\gamma)$ is the scalar curvature of $({\cal B},\gamma)$, and
${\cal B}_{\infty}$ indicates that we are to take the limit as $\cal B$
goes to infinity.  For small $\ell$ Mann's formula reduces to the one
given by Balasubramanian and Kraus, but in addition it 
has a smooth flat spacetime limit as $\ell\rightarrow\infty$.  Moreover,
Mann showed that in many explicit examples it leads to a cancellation of
all divergences, and the remaining finite part agrees with that obtained
using the reference spacetime procedure~\cite{Mann1,Mann2}.  While a
counterterm
action and a reference mass are not the same thing, the resemblance
between the expressions in Eqs.~(\ref{Mann_formula}) and (\ref{AdSQIMref}) is
nevertheless striking.\footnote{I am indebted to R.B.~Mann for pointing
out to me the significance of the $\sigma\sigma R^{\rm ref}$ term in ${\rm 
IQE}^{\rm ref}$, and emphasizing that it provides at least some measure
of geometrical motivation for his expression in Eq.~(\ref{Mann_formula}).}

To see that the connection between ${\rm IQE}^{\rm ref}$ in
Eq.~(\ref{AdSQIMref}) and the AdS/CFT-inspired counterterm action is
probably much deeper than mere resemblance, we now turn to recent
work done by Kraus {\it et al}~\cite{KLS}.  Besides providing an
independent derivation of Mann's formula, and its generalization to higher
dimensions, of most interest to us here is their geometrical argument
suggesting
what the counterterm for $\Pi^{ab}$ in Eq.~(\ref{total_action}) should be 
in order to cancel divergences.  Their result is an expansion in powers of
$\ell$.  Denoting their counterterm ($\tilde{\Pi}_{ab}$) as $\Pi_{ab}^{\rm
ct}$, and specializing their result to a three-dimensional boundary $\cal
B$, they find:
\begin{eqnarray}
\Pi_{ab}^{\rm ct}&=&-{2\over\ell}\gamma_{ab}+\ell\left(R_{ab}-{1\over
2}\gamma_{ab}R\right)
+\ell^{3}\left\{{1\over 2}\gamma_{ab}\left(R_{cd}R^{cd}-{3\over
8}R^{2}\right)\right.\nonumber\\
&&\left.+{3\over 4}RR_{ab}
-2R^{cd}R_{acbd}+{1\over 4}D_{a}D_{b}R-\Box R_{ab}+{1\over
4}\gamma_{ab}\Box R\right\}+O(\ell^{5}).
\label{refPi}
\end{eqnarray}
The curvatures and covariant derivatives in this expression all refer to
the induced timelike three-metric $\gamma_{ab}$ on $\cal B$.  

Now consider
our usual two-surface $(S,\sigma)$ in the physical spacetime $(M,g)$,
which is an asymptotically AdS space.  As we did at the end of Sec.~4,
suppose that $S$ is such
that $(k^{2}-l^{2})>0$ and $k>0$.  Then we can always find a timelike unit
vector $u^a$ normal to $S$ such that $l=0$, and so
$\sqrt{k^{2}-l^{2}}=k=\Pi_{ab}u^{a}u^{b}$ is the Brown-York energy surface
density (modulo the factor of $-1/(8\pi)$).  In other words, our
unreferenced IQE reduces to the unreferenced Brown-York CQE, which would
be called the `unrenormalized' energy in Ref.~\cite{KLS}.  The counterterm
required to renormalize the energy surface density is thus $\Pi_{ab}^{\rm
ct}u^{a}u^{b}$, which we will denote at $E_{\rm ct}$.  Hence, our task is
to compare $E_{\rm ct}:=\Pi_{ab}^{\rm ct}u^{a}u^{b}$ with the integrand of
${\rm IQE}^{\rm ref}$ in Eq.~(\ref{AdSQIMref}); we expect to see at least
some measure of agreement between the two.

This comparison will not be straightforward, however, because on the one
hand we expect the integrand of ${\rm IQE}^{\rm ref}$ to depend on $\cal
R$, $\cal F$, and their derivatives in $S$, as discussed previously,
whereas on the other hand $E_{\rm ct}$ depends on the three-metric
$\gamma_{ab}$.  Nevertheless, let us see how far we can get.  Let $\cal
B$ be a three-surface in $(M,g)$ passing through $S$ in a direction
tangent to $u^{a}$ on $S$.  Different choices of $\cal B$ satisfying these
conditions will lead to different induced metrics $\gamma_{ab}$, but
this ambiguity will not affect our considerations.  At least $\gamma_{ab}$
on $S$ is uniquely determined, and some information about $\gamma_{ab}$ in
the neighborhood of $S$ is determined by the condition $l=0$.  Our choice
of $\cal B$ means that $l_{ab}$ defined in Eqs.~(\ref{extrinsic_curvatures}) 
is the extrinsic curvature of $(S,\sigma)$ as embedded in $({\cal
B},\gamma)$, and so the corresponding codimension-one Gauss embedding
equation reads
\beq
{\cal P}_{a}^{e}{\cal P}_{b}^{f}{\cal P}_{c}^{g}{\cal P}_{d}^{h}
R_{efgh} = {\cal R}_{abcd} + (l_{ac}l_{bd}-l_{bc}l_{ad}).
\label{AdSGauss}
\eeq
This is just a truncated version of Eq.~(\ref{Gauss}), except here
$R_{efgh}$ is the Riemann tensor of $({\cal B},\gamma)$, not $(M,g)$.

Now let $E_{\rm ct}^{(n)}$ denote the term in $E_{\rm ct}$ of order
$\ell^n$.  Inspection of Eq.~(\ref{refPi}) shows that $E_{\rm
ct}^{(-1)}=2/\ell$.  The term $E_{\rm ct}^{(1)}$ can be written in terms
of $G_{ab}$, the Einstein tensor of $({\cal B},\gamma)$, and we have
\beq
E_{\rm ct}^{(1)}=\ell G_{ab}u^{a}u^{b}={\ell\over 2}\sigma^{ac}\sigma^{bd}
R_{abcd}={\ell\over 2}({\cal R}-\tilde{l}^{2}).
\label{E1}
\eeq
The second equality is an easily derived identity (valid in any
codimension-one setting) relating the $uu$ component of the Einstein
tensor to the sectional curvature of the hypersurface orthogonal to
$u^{a}$ (in this case the hypersurface $S$ in $\cal B$).  The third
equality follows from contracting Eq.~(\ref{AdSGauss}) with
$\sigma^{ac}\sigma^{bd}$, and using the fact that $l=0$ by our choice of
$u^{a}$.  (And as usual, $\tilde{l}^2$ is shorthand for
$\tilde{l}_{ab}\tilde{l}^{ab}$.)  Thus, to order $\ell$ we have
\beq
E_{\rm ct}={2\over\ell}+{\ell\over 2}({\cal R}-\tilde{l}^{2})+O(\ell^{3})
=\sqrt{2\left[{2\over{\ell^2}}+{\cal R}-\tilde{l}^{2}+O(\ell^{2})\right]}.
\eeq
Comparing the last expression with Eq.~(\ref{AdSQIMref}) suggests the
correspondence:
\beq
(\tilde{k}^{2}-\tilde{l}^{2})^{\rm ref}\longleftrightarrow
-\tilde{l}^{2}+O(\ell^{2}).
\label{c1}
\eeq
Immediately we see something odd: we are identifying a boost invariant
quantity with one that is not, i.e., it seems that a $\tilde{k}^{2}$ is
missing from the right hand side.  I will comment on this shortly.  Let 
us assume for the moment that the right hand side reads
$(\tilde{k}^{2}-\tilde{l}^{2})$, in which case Eq.~(\ref{c1}) seems
reasonable: it suggests that, if we solve the embedding equations
(\ref{ref_Gauss}-\ref{ref_Ricci}) for $(\tilde{k}^{2}-\tilde{l}^{2})^{\rm 
ref}$ we will find that, to lowest order in $\ell$, the reference shear
term is the same as the unreferenced shear term, the difference to be seen
at a higher order in $\ell$.  On the other hand, this seems like a
problem: Would it not mean, e.g., that the shear terms in
Eq.~(\ref{largeSQIM}) basically cancel, thus ruining the Bondi-Sachs
mass result in Eq.(\ref{BondiMass}), which depends so crucially on 
$(\tilde{k}^{2}-\tilde{l}^{2})$?  The answer is No, because
Eq.~(\ref{largeSQIM}) is valid in the asymptotically flat case, not the
asymptotically AdS case.  To make a statement that is valid in the
asymptotically flat case ($\ell\rightarrow\infty$) we need to know $E_{\rm
ct}$ to all orders in $\ell$, then sum the infinite series, and finally
take the limit $\ell\rightarrow\infty$.  So being at the `other end' of
the series, Eq.~(\ref{c1}) has nothing to say about the asymptotically
flat case.  But we also expected $(\tilde{k}^{2}-\tilde{l}^{2})^{\rm 
ref}$ to depend on $\cal R$, $\cal F$, and their derivatives.  Why do we
not see these quantities on the right hand side of Eq.~(\ref{c1})?  The
answer is, We will---we just have to calculate $E_{\rm ct}$ to the next
order in $\ell$.

But before doing so I will comment on the `missing' $\tilde{k}^{2}$ in
Eq.~(\ref{c1}).  Kraus {\it et al} \cite{KLS} have devised an algorithm to
compute the extrinsic geometrical quantity $\Pi_{ab}^{\rm ct}$ from
the intrinsic geometry of $\cal B$.  Insofar as $l_{ab}$ (and thus
$\tilde{l}^{2}$) depends only on the metric $\gamma_{ab}$, there is no
doubt that the $-\tilde{l}^{2}$ term in Eq.~(\ref{c1}) is `correct'.  On
the other hand, $k_{ab}$ (and thus $\tilde{k}^{2}$) depends on the
extrinsic geometry of $\cal B$, being just a certain projection of
$\Pi_{ab}$ into $S$.  The algorithm of Kraus {\it et al} relies on the
fact~\cite{FG} that the {\it divergent} part of the derivative of
$\Pi_{ab}$ in the direction normal to $\cal B$ can be expressed in terms
of just the intrinsic geometry of $\cal B$.  In essence, their algorithm
is designed precisely to compute the $\it divergent$ part of
$\Pi_{ab}$.  The `correctness' of the accompanying finite part is a
subtle issue.  In a slightly different context, they discuss two different
counterterm actions that both properly cancel divergences, but that lead
to different finite terms in the action.  Furthermore, they point out that
their algorithm, when carried to all orders in $\ell$, might imply
singularities in the bulk spacetime, but that this is of no concern
because they truncate their counterterm expressions to a finite number of
terms, enough at least to cancel the divergences.  In our case we have the
quasilocal idea in mind, and so are interested in {\it all} of the finite
terms---it matters what happens in the bulk.  But going further with this
discussion will take us beyond the scope set for this simple comparison.  
I will just conclude by saying that, insofar as the shear terms almost
certainly represent a finite contribution to the energy, we do not {\it
necessarily} expect the algorithm of Kraus {\it et al} to produce a
$\tilde{k}^{2}$ term on the right hand side of Eq.~(\ref{c1}).  Our goals
are slightly different, and it is too much to expect exact agreement
between $E_{\rm ct}$ and the integrand of ${\rm IQE}^{\rm ref}$.

Nevertheless, it is still instructive to proceed with the comparison to
the next order in $\ell$.  In light of my previous remarks, we will make
the simplifying assumption that the metric on $\cal B$ has a product
structure: $\gamma_{ab}\,dx^{a}\,dx^{b}=-N^{2}\,dt^{2}+\sigma_{ij}(x) 
\,dx^{i}\,dx^{j}$, where $x^a=(t,x^{i})$ are local coordinates on $\cal
B$, $N$ is a constant lapse, and $\sigma_{ij}(x)$ is the metric on any
$t={\rm constant}$ two-surface $S$.  The idea is that $E_{\rm ct}$ and the
integrand of ${\rm IQE}^{\rm ref}$ should agree {\it at least} in their
dependence on the intrinsic geometry of $(S,\sigma)$.  Assuming such a
product structure for $\gamma_{ab}$ is a convenient was to isolate this
dependence, and ignore everything else.

In this case the only nonvanishing components of the Riemann tensor of
$\gamma_{ab}$ are $R_{ijkl}={\cal R}_{ijkl}$, the Riemann tensor of
$\sigma_{ij}$.  And clearly $l_{ab}=0$.  Thus $E_{\rm ct}^{(1)}$ in
Eq.~(\ref{E1}) reduces to $\ell{\cal R}/2$, and it is a simple exercise to
work out $E_{\rm ct}^{(3)}$.  The net result is
\beq
E_{\rm ct}={2\over\ell}+{\ell\over 2}{\cal
R}-{{\ell^{3}}\over{16}}\left(
{\cal R}^{2}-4\,\triangle{\cal R}\right)
+O(\ell^{5})
=\sqrt{2\left[{{2}\over{\ell^{2}}}+{\cal R}-{{\ell^{2}}\over{2}}\,
\triangle{\cal R}+O(\ell^{4})\right]},
\label{E3}
\eeq
where $\triangle$ is the Laplacian in $(S,\sigma)$.
Comparing the last expression with Eq.~(\ref{AdSQIMref}) we now have the
higher order correspondence:
\beq
(\tilde{k}^{2}-\tilde{l}^{2})^{\rm ref}\longleftrightarrow 
-{{\ell^{2}}\over{2}}\,\triangle{\cal R}+O(\ell^{4}).
\eeq
Thus we begin to see how a solution to our embedding equations might yield
an expression for $(\tilde{k}^{2}-\tilde{l}^{2})^{\rm ref}$ in terms of
${\cal R}$, ${\cal F}$, and their derivatives, as we have expected all
along.

To conclude this section we make two general observations.  First, it is
especially clear from the higher order expression in Eq.~(\ref{E3}) that
the AdS/CFT-inspired counterterm energy is, in fact, the square root of
some quantity.  This is not surprising, since the algorithm of Kraus {\it
et al}~\cite{KLS} is a means of solving a Gauss embedding equation for
$\Pi_{ab}^{\rm ct}$, and this equation is quadratic in $\Pi_{ab}^{\rm
ct}$.  But it is significant.  Beginning simply with the definition of the
quasilocal energy-momentum tensor as the functional derivative of the
action with respect to the boundary metric~\cite{BY}, in which there is no
square root in sight, the counterterm energy required to cancel
divergences unmistakably involves a square root.  Moreover, it concurs
with the square root introduced here, in the context of the IQE, as the
general relativistic analogue of the special relativistic
formula: $m=\sqrt{E^{2}-\vec{p}^{\;2}}$.  I believe it is unlikely this is
a mere coincidence.  Given that it is nonanalytic, a square root is too
unusual an object to occur without good reason.

Secondly, under the square root (in our case) is $2/\ell^{2}+
{\cal R} + (\tilde{k}^2 -\tilde{l}^2 )^{\rm ref}$.  In the case of the
AdS/CFT-inspired counterterm energy~\cite{KLS}, it is $2/\ell^{2}+
{\cal R} + X$, where $X$ is an infinite series in increasing powers of
$\ell$.  That $X$ is clearly not zero lends strong support for our
additional term $(\tilde{k}^2 -\tilde{l}^2 )^{\rm ref}$, which is thus
seen to be a necessary generalization of Lau's suggestion~\cite{Lau3}.  I
have argued that its necessity is closely linked to the proper inclusion
of angular momentum in the energy.  Given that angular momentum is a
subtle notion in general relativity, especially so at the quasilocal level
we envision here, it is not surprising that our biggest difficulty lies in
evaluating $(\tilde{k}^2 -\tilde{l}^2 )^{\rm ref}$.  In light of the
algorithm given by Kraus {\it et al}~\cite{KLS}, work is currently
in progress to try to apply similar techniques to solve the embedding
equations (\ref{ref_Gauss}-\ref{ref_Ricci}).  Since these
embedding equations are manifestly boost invariant, I expect at least to
recover the `missing' $\tilde{k}^{2}$ term in Eq.~(\ref{c1}), and
hopefully the entire series.

\section{Summary and discussion}

In this paper I have introduced a new definition of quasilocal energy that
is a simple modification of the Brown-York quasilocal energy.  I just
replace their energy surface density $k$ with $\sqrt{k^{2}-l^{2}}$,
where $l$ is the radial momentum surface density.  (For ease of
exposition here I will omit the $-1/(8\pi)$ factors.)  The principle
motivation for doing this stems from an analogy with the
formula: $m=\sqrt{E^{2}-\vec{p}^{\;2}}$ in special
relativity.  Identifying $E$ with $k$ (which are both energies), and
$\vec{p}$ with $l$ (both momenta), identifies $m$ with
$\sqrt{k^{2}-l^{2}}$.  Like $m$, $\sqrt{k^{2}-l^{2}}$ is a boost invariant
quantity, and hence the integral of $\sqrt{k^{2}-l^{2}}$ over a spacelike
two-surface $S$ gives rise to an `invariant quasilocal energy', or
IQE.  In what follows I will refer to the Brown-York quasilocal energy
as the CQE---canonical quasilocal energy.

There are several important consequences of replacing $k$ with 
$\sqrt{k^{2}-l^{2}}$:
\begin{itemize}
\item\label{first_bullet}
While $k$ is always well defined for any spacelike two-surface $S$, 
$\sqrt{k^{2}-l^{2}}$ is not.  Roughly speaking, it is real when $S$ lies 
in the exterior region of a black hole, zero when it is on the horizon,
and imaginary in the black hole interior.  Thus (again roughly
speaking) the IQE asserts that energy is real only outside of a black
hole.
\item
Both the CQE and the IQE require a reference energy subtraction
procedure.  Since $k$ is associated with a spacelike three-surface
spanning $S$, the reference space into which $S$ is to be isometrically
embedded is inherently three-dimensional.  Such a codimension-one
embedding does not always exist, but when it does, it is essentially
unique.  This means the CQE, when it is defined, is unique.  In contrast, 
$\sqrt{k^{2}-l^{2}}$ makes no reference to a three-surface spanning $S$,
and so the reference space(time) is inherently four-dimensional.  Such
codimension-two embeddings (at least of a generic non-round sphere into
Minkowski space) always exist~\cite{Brin}, but are not unique.  However,
in this situation there are two curvatures associated with $S$: the scalar
curvature $\cal R$, and the curvature of the normal bundle, $\cal F$.  A
necessary condition for an isometric embedding is that ${\cal R}^{\rm ref}
={\cal R}$.  I argued that demanding also ${\cal F}^{\rm ref} ={\cal F}$
is both a means to make the embedding essentially unique, and at the same
time, a geometrically natural way to properly incorporate angular momentum
into energy at the quasilocal level.  Indeed, since angular momentum is
associated with rotational kinetic energy, it {\it should} contribute to
the energy in some way.
\item
While ${\rm CQE}^{\rm ref}$ is associated with a reference {\it energy}
density $k^{\rm ref}$, ${\rm IQE}^{\rm ref}$ is concerned with a
reference {\it shear} term, $(\tilde{k}^{2}-\tilde{l}^{2})^{\rm
ref}$.  ($\tilde{k}_{ab}$ and $\tilde{l}_{ab}$ are the trace-free parts of
the two extrinsic curvatures of $S$.)  In a certain sense, the IQE already
inherently contains the correct reference energy, without recourse to a
reference embedding.  The reference embedding is required only to
determine the reference shear term, which is a higher order correction to
the energy associated with angular momentum.
\item
The CQE is sensitive to the sign of $k$, whereas since it involves
$\sqrt{k^{2}-l^{2}}$, the IQE is not.  Thus one can easily construct
simple examples for which the two energies give different results, even
when $l=0$.  Thus the IQE is not simply the `rest energy' version of the
CQE.  Note: the IQE naturally assigns zero energy to {\it any}
two-surface in flat spacetime.  This is because the natural reference
spacetime in this case is the very same spacetime, namely flat
spacetime.  So obviously one can always reference-embed the two-surface
identically (up to Poincar\'{e} transformations) to the way it is embedded
in the physical spacetime, and get ${\rm IQE}=0$.  The only subtlety that
may arise is if the two conditions: `isometric embedding' and `${\cal
F}^{\rm ref}={\cal F}$' do not uniquely determine 
$(\tilde{k}^{2}-\tilde{l}^{2})^{\rm ref}$.  Then the flat spacetime result
(${\rm IQE}=0$) may be reduced to a choice, rather than a necessary
fact.  To properly address this subtlety requires an in depth
understanding of the embedding equations.  But in any case, the fact that
${\rm IQE}=0$ in flat spacetime is independent of the motion of the
observers.  In contrast, moving observers in flat spacetime could measure
nonzero energy in the Brown-York approach~\cite{Ivan}.  This is because
under a radial boost the Brown-York energy surface density dilates by a
Lorentz factor, as in special relativity, whereas the reference energy
surface density does not.  According to Ref.~\cite{BY} the latter depends
only on the intrinsic geometry of $S$, and therefore does not know about
the time derivative of this geometry.\footnote{I thank R.B. Mann and
I.S. Booth for this remark on the Brown-York case.}
\end{itemize}

We examined both the large and small sphere limits of the IQE, taking $S$
to be asymptotically round for simplicity.  In an asymptotically flat
spacetime, the large sphere limit of the IQE in a spatial direction yields
the ADM mass.  In the future null direction it reduces to the Bondi-Sachs
mass, provided the reference shear term is a total divergence.  Short of
solving the embedding equations, I gave a heuristic argument which shows
that is is.  It is significant that this argument relies on the condition 
${\cal F}^{\rm ref}={\cal F}$, since this provides evidence that the
curvature of the normal bundle {\it is} involved in quasilocal energy,
albeit its involvement in this simple example is minimal.

The quantity $\sqrt{k^{2}-l^{2}}$ is proportional to the mean curvature of
$S$ as a two-surface embedded in the physical spacetime, and so the IQE is
a natural geometrical invariant of $S$.  Since the Hawking
mass~\cite{Hawk} is constructed using $(k^{2}-l^{2})$, the IQE can
be thought of roughly as the square root of the Hawking mass.  In the
small sphere limit the square root disappears, and to leading order the
IQE reduces to the Hawking mass (but differs from it at higher
order).  Thus, when matter is present, the lowest order contribution to
the IQE gives the standard result: $(4\pi r^{3}/3)\,T_{ab}^{\rm
mat}\,u^{a}u^{b}$, i.e., the expected matter energy contained in a small
sphere of proper radius $r$.  Note that $u^{a}$ here is not necessarily
the four-velocity of any observer on $S$, since the IQE is boost
invariant, and so independent of the observers' velocities on $S$.  Rather,
$u^{a}$ is the four-velocity that observers would have {\it if} they were
in the rest frame determined by $S$.  More precisely, in the small sphere
limit we considered, namely a $t,r={\rm constant}$ two-sphere in Riemann
normal coordinates (with $t\propto r$), $u^{a}=(\partial/\partial t)^{a}$
evaluated at the center of the sphere.  In the limit $r\rightarrow 0$,
the four-velocity $(\partial/\partial t)^{a}$ corresponds to observers who
at each point on $S$ have zero radial momentum, i.e., $l=0$.  In general,
since the IQE is an energy rather than a mass, the question arises, In
whose rest frame is the energy measured?\footnote{I thank A. Ashtekar for
posing this question.}  The answer is, The `quasilocal rest frame'
determined by the condition $l=0$ at each point on $S$.  Whenever
$(k^{2}-l^{2})>0$, observers on $S$ can always achieve this state of
motion by appropriate local radial boosts.  This (or more precisely,
$(k^{2}-l^{2})\geq 0$) is the same condition required for the unreferenced
IQE to be well defined in the first place---refer to the text in the first
bullet on page~\pageref{first_bullet}.

Returning to the small sphere case, in vacuo the leading order contribution 
due to gravitational energy occurs at order $r^{5}$.  At this order the
IQE results are inconclusive because it is expected that the reference
shear term will play a significant role, and without a solution to the
embedding equations (which is an extremely difficult problem) this term
cannot be determined.  Nevertheless, it was possible to show that in the
small sphere limit, the Hawking mass, which in this case is closely
related to the IQE, can be understood as a measure of the gravitational
energy contained in $S$ by considering certain tangential `electromagnetic'
fields $\vec{E}$ and $\vec{B}$ induced on $S$ by the Weyl curvature $S$ is
embedded in.  In terms of $\vec{E}$ and $\vec{B}$, gravitational energy
and radiation are essentially identical in nature to their counterparts in
electromagnetism, except for one crucial difference: the density
$r(\vec{E}\cdot\vec{E}+\vec{B}\cdot\vec{B})/(8\pi)$ is integrated
over the {\it surface} $S$ to determine the gravitational energy contained
in the spatial volume that $S$ encloses.  Here $r$ can be thought of as
the areal radius of $S$, so the measurement is truly quasilocal.

The IQE was analyzed in the context of asymptotically anti-de Sitter
spacetimes.  The fact that ${\rm IQE}^{\rm ref}$ depends explicitly on the
Riemann tensor of the reference spacetime (naturally taken to be anti-de
Sitter space)  was seen to play a significant role.  A connection was
established between ${\rm IQE}^{\rm ref}$ and a certain counterterm energy
that has recently been proposed~\cite{KLS} in the context of the
conjectured AdS/CFT correspondence.  Two similarities are
striking: (i) Both energies involve a square root, and (ii) the two
leading terms under the square root match.  The remaining term under the
square root in our case is the reference shear term,
$(\tilde{k}^{2}-\tilde{l}^{2})^{\rm ref}$; in the case of Ref.~\cite{KLS}
it is an infinite series in increasing powers of $\ell$, where $\ell$ is
the radius of curvature of the AdS space.  It was shown that the first two
nontrivial terms of this series (i.e., to the highest order given in
Ref.~\cite{KLS}) can plausibly be identified with
$(\tilde{k}^{2}-\tilde{l}^{2})^{\rm ref}$.  This agreement is impressive
because ${\rm IQE}^{\rm ref}$ and the AdS/CFT-inspired counterterm energy
are independently motivated, and derived quite differently.  It might
be possible to use techniques developed in Ref.~\cite{KLS} to solve our
embedding equations for $(\tilde{k}^{2}-\tilde{l}^{2})^{\rm ref}$.  The
present lack of a solution to these equations is the main outstanding
obstacle to further understanding the nature of the IQE.

A final remark is in order.  Most definitions of quasilocal energy,
including the IQE, assume that energy is associated with a closed
spacelike two-surface, $S$.  Given such a two-surface one can always find
a timelike unit normal vector field $u^a$, which at each point on $S$ is
supposed to correspond to an observer's instantaneous four-velocity.  But
this may not be a general enough setting.  While a two-parameter family of
observers will always sweep out a timelike three-surface $\cal B$, the
two-surface elements orthogonal to their world lines in $\cal B$ are not,
in general, integrable.  Thus a shift in emphasis from $S$ to $\cal B$,
i.e., from Eulerian to Lorentzian observers~\cite{Lau2}, might lead to a
deeper understanding of quasilocal energy, in particular of gravitational
radiation at the quasilocal level.  This shift would also bring the
quasilocal energy idea closer in line with the conjectured AdS/CFT
correspondence.  Whether or not this is the right direction, the
results in Sec.~7 strongly suggest that this is the direction the IQE is
pointing in.  

\vspace{1.5ex}
\begin{flushleft}
\large\bf Acknowledgments
\end{flushleft}
\noindent
I would like to thank Joseph Samuel, Madhavan Varadarajan, and Sukanya
Sinha at the Raman Research Institute for their continued encouragement.  
I am also indebted to Robert Mann, Ivan Booth, and Luis de Menezes at the
University of Waterloo for many enlightening discussions, and Joseph
Samuel and Robert Mann for providing critiques of the manuscript.  I
thank Gabor Kunstatter and Steven Carlip for input during the early stages
of this work.  This work was financially supported in part by the Natural
Sciences and Engineering Research Council of Canada, and the Department of
Science and Technology, Government of India.

\end{document}